\documentclass[12pt]{article}

\usepackage{jheppub}

\usepackage{mathtools,amsfonts,dsfont,esint}
\usepackage{enumitem}
\usepackage{breqn}
\usepackage[utf8]{inputenc}
\usepackage[english]{babel}
\usepackage{float}
\usepackage{caption}
\usepackage{adjustbox}
\usepackage{tikz}
\usepackage{tikz-3dplot}
\usetikzlibrary{decorations.pathreplacing}

\usepackage{xspace}
\newcommand{\TTbar}{{\texorpdfstring{\ensuremath{T\bar{T}}}
{TTbar}}\xspace}
\newcommand{\nopdfstring}[1]{\texorpdfstring{#1}{}}
\newcommand{\pd}{\partial}

\newcommand{\del}{\nabla}
\newcommand{\im}{\mathrm{i}}
\newcommand{\dm}{\mathrm{d}}

\def\eqa{\begin{eqnarray}}
\def\eqae{\end{eqnarray}}
\def\eq{\begin{equation}}
\def\eqe{\end{equation}}
\def\be{\begin{equation}}
\def\ee{\end{equation}}
\def\bea{\begin{eqnarray}}
\def\eea{\end{eqnarray}}
\def\ba{\begin{array}}
\def\ea{\end{array}}

\title{Non-perturbative aspects of entanglement structures in $T\bar{T}$-deformed CFTs}
\author{Wen-Xin Lai, Huajia Wang, and Yongjiang Xu}

\affiliation{Kavli Institute for Theoretical Sciences, University of Chinese Academy of Sciences, \\ Beijing 100190, China}
\emailAdd{laiwenxin@ucas.ac.cn} 
\emailAdd{wanghuajia@ucas.ac.cn} 
\emailAdd{xuyongjiang23@mails.ucas.ac.cn}

\abstract{Turning on the $T\bar{T}$-deformation in a two-dimensional CFT provides a unique window to study explicitly how non-local features arise in the UV as a result of the deformation. A sharp signature is the dynamical emergence of an effective length-scale $\propto \sqrt{\mu}$  that separates the local and non-local regimes of the deformed theory, effectively serving as a UV cut-off for computing observables in the local regime. In this paper, we study this phenomenon through the entanglement structures of the deformed theory. We focus on computing the Renyi entropies of single-interval sub-regions in the deformed vacuum states. We pay particular attention to the interplay between the bare entanglement cut-off inherited from the CFT computation and the effects from the $T\bar{T}$ deformations. Applying the general replica trick to the string theory formulation of $T\bar{T}$-deformed CFTs, we derive an explicit representation of the deformed replica partition function as a weighted integral of the CFT results evaluated at a dynamical cut-off, which is integrated over. We computed in detail the kernel functions of the integral representation, and performed the saddle-point analysis in the semi-classical limit of small $\mu$. We found that in addition to the perturbative saddle-point which identifies the dynamical cut-off with the bare entanglement cut-off, there exists another non-perturbative saddle-point that identifies the dynamical cut-off with the $T\bar{T}$ length-scale $\propto \sqrt{\mu}$, but whose contribution is exponentially small. We discuss how these non-perturbative effects can shed lights on the mechanism through which the $T\bar{T}$ length-scale may eventually replace the bare counter-part and become the effective entanglement cut-off.     }

\begin{document}

\maketitle
\clearpage

\section{Introduction}
In two dimensional spacetimes, local quantum field theories (QFTs) will develop non-local features under the so-called $T\bar{T}$-deformation \cite{Zamolodchikov:2004ce,Smirnov:2016lqw}, thereby providing a unique and indispensable opportunity to study non-local physics explicitly \cite{Dubovsky:2012wk,Cavaglia:2016oda,McGough:2016lol,Callebaut:2019omt}. These are deformations that can be defined in terms of the action by the flow equation:
\be\label{eq:TT_def}
\partial_\mu S^{T\bar{T}}(\mu) = \frac{1}{2\pi} \int d^2 x\; \text{det}\; T^\mu_{\alpha\beta}(x)
\ee
where $T^{\mu}_{\alpha\beta}$ denotes the ``instantaneous" stress-energy tensor derived from the deformed action $S^{T\bar{T}}(\mu)$, and $\mu$ is the deformation parameter. The original QFT action serves as the initial condition for (\ref{eq:TT_def}) at $\mu=0$. 

The standard dimensional analysis dictates that such deformations are RG irrelevant, i.e.~they leave the IR behavior of the original QFT intact while drastically modifying it in the UV. The nature of the UV modification can be inferred from the spectral properties of the deformed theory, which indeed point to non-local features. This is possible due to the integrability of the $T\bar{T}$ deformation, which comes from the special property of the $\text{det}\; T_{\alpha\beta}$ operator that as a composite operator its terms factorize when evaluated in the energy eigen-states \cite{Zamolodchikov:2004ce,Dubovsky:2012wk,Smirnov:2016lqw,Cavaglia:2016oda}. As a result, when compactified on a circle of radius $R$, the deformed spectrum $E_n$ satisfies the inviscid Burgers' equation:
\begin{equation}
\label{eq:Burgers}
\frac{\pd}{\pd\mu} E_n(\mu, R)
= \pi E_n \frac{\pd}{\pd R} E_n(\mu, R)
+ \frac{\pi}{R^3} P_n^2,
\end{equation}
while the momenta $P_n$ remain unchanged due to their quantization condition on a circle. When restricting to the cases of conformal field theories (CFTs) regarding the initial spectrum at $\mu=0$, the spectral flow equation can be solved explicitly \cite{Smirnov:2016lqw,Cavaglia:2016oda,McGough:2016lol}: 
\begin{equation}
\label{eq:deformed_spectrum}
E_n(\mu) = \frac{R}{2\pi\mu}
\Bigl(\sqrt{1 + \frac{4\pi\mu E_n(0)}{R} + \frac{4\pi^2\mu^2 P_n^2}{R^4}} - 1\Big).
\end{equation}

The UV behavior of the deformed spectrum  (\ref{eq:deformed_spectrum}) has two distinct possibilities depending on the sign of the deformation parameter $\mu$: for $\mu>0$ the asymptotic density of states $g(E) = e^{S(E)}$ exhibits that of the Hagedorn growth: $S(E) \propto E$, which is typical in theories with stringy degrees of freedom; for $\mu<0$ the spectrum becomes complex above a certain UV threshold $E>E_c$, suggesting that a UV cut-off should exist in some way if the deformed theory were to remain unitary. Both scenarios are indicative of the non-local nature of the UV modification coming from the $T\bar{T}$ deformation. For either sign of $\mu$, the non-locality manifests through a characteristic length-scale $\epsilon_{T\bar{T}}\propto \sqrt{|\mu|}$ below which the deformed theory cease to behave like a local QFT: for $\mu>0$ a stringy description may take over; and for $\mu<0$ the theory may simply be trivial. 

The length-scale $\epsilon_{T\bar{T}}$ plays a more explicit role in the context of holography. For the $\mu<0$ case, in contrary to the usual AdS/CFT dictionary the bulk dual of a $T\bar{T}$-deformed holographic CFT was proposed to feature a finite boundary at $r\sim \epsilon_{T\bar T}^{-1}$ instead of the usual asymptotic boundary $r\to \infty$ \cite{McGough:2016lol}, where the boundary theory can be thought to ``live" in. For $\mu>0$, the bulk dual of the $T\bar{T}$-deformed holographic CFTs is less clear; some interesting suggestions based on the so-called glue-on surfaces were proposed in \cite{Apolo:2023vnm,Apolo:2023ckr}. An alternative bulk interpretation of holographic $T\bar{T}$-deformations in terms of the mixed boundary condition was proposed in \cite{Guica:2019nzm}, which was based on the observation that $\text{det}\;T_{\alpha\beta}$ takes the form of a double-trace deformation that implements in holography a change of dictionary for the boundary conditions. The mixed boundary condition interpretation holds for both signs of $\mu$, and it is equivalent to the finite cut-off surface picture for $\mu<0$ when only pure gravity is considered in the bulk. Other variants of the $T\bar{T}$ deformation, especially in relation to holography, have been proposed and studied extensively. Examples include the single-trace $T\bar{T}$ deformation \cite{Giveon:2017nie,Borsato:2018spz,Araujo:2018rho,Apolo:2019zai,Benjamin:2023nts,Dei:2024sct} and the $J\bar{T}$ deformation in the presences of additional conserved charges \cite{Guica:2017lia,Aharony:2018ics,LeFloch:2019rut,Chakraborty:2019mdf,Apolo:2019yfj,Anous:2019osb}.

The dynamical emergence of the length-scale $\epsilon_{T\bar{T}}$ is one of the most important phenomena that the $T\bar{T}$ deformation can cause in the deformed theory. It is therefore desirable to study observables that are directly sensitive to such effects. For example, one could study the modified behavior of the $T\bar{T}$-deformed correlation functions at short-distance or equivalent at large-momentum, see \cite{Kraus:2018xrn,Cardy:2019qao,He:2019vzf,He:2020udl,Kruthoff:2020hsi,Cui:2023jrb,Aharony:2018ics,He:2023kgq,Barel:2024dgv,Chen:2025jzb,Hirano:2025alr}. On the other hand, the deformed theory can be approximated at low energies by the original local QFT, now as an effective field theory (EFT). In this perspective, the length-scale $\epsilon_{T\bar{T}}$ provides a notion of the UV cut-off, and could therefore affect deformed quantities that are UV divergent in the original QFT. We emphasize that in these quantities the UV physics is not probed by direct kinematic arrangements, e.g.~taking the short-distance limit of operator insertions in the correlation functions -- but rather by the regulator of the UV divergences. An important class of such examples consists of the entanglement quantities, e.g.~von-Neumann entropies, Renyi entropies, etc. For generic subsystem size $\ell_A$, these entanglement quantities feature logarithmic UV divergences, or are logarithmically sensitive to the UV cut-off $\Lambda$, i.e. $S_A \sim \ln{\left(\ell_A \Lambda\right)}$.  Physically, they are produced by the short-distance entanglement across the boundary of the subsystem. A natural treatment is to simply use $\epsilon_{T\bar{T}}^{-1}$ as a practical definition of $\Lambda$ in the $T\bar{T}$-deformed results. This proposal seems especially tempting when the RT formula \cite{Ryu:2006bv,Hubeny:2007xt} for computing the holographic entanglement entropy is applied naively to the cut-off surface interpretation of the deformed bulk dual \cite{Donnelly:2018bef,Chen:2018eqk,Lewkowycz:2019xse,Park:2018snf,Banerjee:2019ewu,Murdia:2019fax,Jeong:2019ylz,Asrat:2020uib,Li:2020zjb,Apolo:2023ckr,Chang:2024voo} -- the minimal surface (geodesics in the case of $\text{AdS}_3$) now ends on the cut-off surface at $r\sim \epsilon_{T\bar{T}}^{-1}$, regulating the otherwise diverging geodesic length by such terms as  $\sim\ln{\left(1/\epsilon_{T\bar{T}}\right)}$. 

Usually, a ``bare" cut-off $\Lambda\sim \epsilon^{-1}$ already exists in the absence of the $T\bar{T}$ deformation -- to regulate the UV divergence in the entanglement computation of the original QFT. This is the case even for CFTs whose definition does not require a UV cut-off. In practice, it could be implemented in the form of a small hole about the entanglement boundary (of radius $\sim \epsilon$) removed from the path-integral, though it could be understood more rigorously from the perspectives of the von-Neumann algebra and algebraic QFT \cite{Witten:2018zxz,Goto:2025xxx}. Therefore, a careful study of the entanglement quantities in the $T\bar{T}$-deformed theory should proceed by imposing the bare cut-off as the ``actual" regulator -- the parameter $\mu$ only affects the deformed Lagrangian -- e.g.~when evaluating the action. In this way, the deformed quantities retreat continuously to the original QFT results in the limit of $\mu \to 0$. The replacement of the bare cut-off by that of the emergent length-scale $\epsilon_{T\bar{T}}$, if correct, should be observed as a dynamical phenomenon in the process of increasing $\mu$.  

Unfortunately, despite the great success it brings to studying the deformed spectrum, the special integrability of the $T\bar{T}$ deformation is not as helpful when computing the entanglement quantities in the $T\bar{T}$-deformed theories. In practice most of the explicit computations were carried out in the perturbative theory, i.e.~the results were expanded perturbatively in small $\mu$ \cite{Chen:2018eqk,He:2019vzf}. Attempts of non-perturbative computations were only successful in cases where convenient symmetries are present \cite{Donnelly:2018bef,Lewkowycz:2019xse,Apolo:2023vnm}. The implicit premise of performing the perturbative analysis in small $\mu$ is that among all possible scales/parameters in the set-up, $\mu$ is parametrically the smallest. Other scales/parameters could include the bare cut-off scale $\epsilon$;  thermal length-scale $\beta$ when considering the high temperature states; the inverse of the central charge $1/c$ in holography; or the subtracted Renyi index $(n-1)$ when taking the replica limit $n\to 1$. Subtleties and potential confusions related to the order of limits could arise when multiple scales/parameters are taken to be small. In addition, in the interest of observing a potential dynamical take-over of the effective cut-off scale between $\epsilon$ and $\epsilon_{T\bar{T}}$, perturbation theory alone is rarely helpful; the revelation often requires re-summation, i.e.~non-perturbative treatment of the computation. As a simple illustration of this, consider a toy model expression $1/(\epsilon + \lambda)$ that ``resolves" the $1/\epsilon$ divergence at $\lambda =0$ dynamically into $1/\lambda$ for $\lambda\gg\epsilon$. The mechanism for the resolution becomes obscure when expanded in small $\lambda$: 
\be
\frac{1}{\epsilon+\lambda} = \frac{1}{\epsilon}\left(1-\frac{\lambda}{\epsilon}+\frac{\lambda^2}{\epsilon^2}+...\right)
\ee
Formally, the series appears increasingly divergent in terms of the original cut-off $\epsilon$, and only upon re-summation is the resolution revealed. 

Therefore in order to understand the most interesting aspects in the entanglement structure of the $T\bar{T}$-deformed theory, such as the emergence of $\epsilon_{T\bar{T}}$ as a dynamical cut-off, it is crucial that we treat the computation in a framework that is non-perturbative in $\mu$. This is the main goal of this paper. We consider the case where the original theory is a CFT. Based on the motivation described so far, we pay particular attention to the bare cut-off $\epsilon$ that is specified in the original CFT: how it appears in the non-perturbative formulation of the $T\bar{T}$ deformed entanglement computation, and how it interacts with the dynamics of the deformation. To focus on this part of the $T\bar{T}$-deformed physics and not to be distracted by the presence of additional scales/parameters, we compute the Renyi entanglement entropies for a single-interval subregion $A$ of the infinite line in the vacuum state $\Omega_\mu$ of the deformed theory.

It is easy to realize that the definition of the $T\bar{T}$-deformation in the original form (\ref{eq:TT_def}) is unproductive for any meaningful attempts of the non-perturbative analysis. Among other things, even obtaining the non-perturbative form of the deformed action at finite $\mu$ is a highly non-trivial task. Closed-form expressions can only be obtained for very special theories, e.g.~the Nambu-Goto action (at the classical level) for the deformed free boson. Fortunately, alternative formulations of the $T\bar{T}$-deformation exist that are more convenient for our goal. It has been understood that the essential physics of the $T\bar{T}$-deformation can be summarized by a pair of dynamical coordinates $X^{\pm}$ \cite{Dubovsky:2012wk,Guica:2019nzm,McGough:2016lol,Callebaut:2019omt}. Roughly speaking, these are state-dependent coordinate systems in which the deformed theory looks un-deformed, and therefore should be identified with the original coordinates, i.e.~the ``physical" coordinates before turning on the deformation.  In one formulation, the $T\bar{T}$-deformation can be described as a minimal coupling between the original CFT and a dynamical flat JT gravity theory \cite{Dubovsky:2017cnj}. In the other formulation, the $T\bar{T}$-deformed CFT can be described as a string theory whose world-sheet CFT consists of the un-deformed CFT and a two-dimensional target-space CFT. The string theory could be critical or non-critical depending on whether the central charge of the original CFT is equal to 24 or not \cite{McGough:2016lol,Callebaut:2019omt}. In the critical case the action of the target-space CFT is genuinely free; in the non-critical case it is modified by an additional term, which yet does not affect the dynamics essentially because it can be gauged away. Both formulations are argued to be equivalent to (\ref{eq:TT_def}), e.g.~they can both reproduce the deformed spectrum (\ref{eq:deformed_spectrum}). In both formulations the dynamical coordinates $X^{\pm}$ arise naturally. In the string theory formulation, they are indeed the two-dimensional target-space fields $X^{\pm}$. Due to their relation to the original CFT coordinates, it is in fact the target-space manifold that is identified with the physical space-time that the deformed CFT lives in, while the world-sheet manifold is dynamical and to be integrated over. However, the mapping degree between the world-sheet manifold and the target-space manifold is usually restricted to one, i.e.~with the unit winding along both directions. 

These formulations are convenient for non-perturbative analysis because the deformation parameter $\mu$ appears only in the added sector, and simply as a coupling constant. 
In the flat JT-gravity formulation $\mu$ is related to the cosmological constant of the gravitational sector; in the string theory formulation $\mu$ is related to the string tension of the target-space sector in the worldsheet CFT. In the ideal scenarios, the deformed quantities may admit integral-representations, in which $\mu^{-1}$ appears as a pre-factor in the exponent of the integrand, i.e.~integrals of the general form: 
\be\label{eq:integral_ansatz}
A(\mu) \sim \int dx\; e^{-\frac{1}{\mu} K(x)} Q(x)
\ee
We then reach a productive set-up for the non-perturbative analysis, to which powerful semi-classical methods can be applied. Works of such kind have been performed for the deformed rational CFT torus partition function \cite{Dubovsky:2018bmo,Hashimoto:2019wct,Callebaut:2019omt,Gu:2024ogh}, in conjunction with the resurgence analysis of the perturbation series \cite{Gu:2024ogh,Gu:2025tpy}. 
Similar results have also been obtained in the deformed Maxwell and Yang-Mills theories \cite{Griguolo:2022xcj,Griguolo:2022hek}.
In this paper, we will work with the string theory formulation of the $T\bar{T}$-deformation, critical or non-critical. We will focus on the case of $\mu>0$ which has a more physical semi-classical limit as a string theory. Our key strategy is to seek integral representations of the general form (\ref{eq:integral_ansatz}) for the entanglement quantities of interest, using which the non-perturbative aspects of entanglement in the deformed CFT can be easily analyzed. 

This paper is organized as follows.  In section (\ref{sec:setup}) we quickly review the string theory formulation of the $T\bar{T}$ deformation, in which we set up the general framework of computing the Renyi entropy using the replica trick. In section (\ref{sec:infinite}) we apply the framework specifically to the case where the subsystem is the semi-infinite interval $[0,\infty)$, with cut-off holes removed about $z=0$ and $z=\infty$. We derive an integral representation of the result, from which we proceed with the saddle-point analysis to extract the non-perturbative aspects of the result. In section (\ref{sec:finite}) we perform a similar computation for the case where the subsystem is a finite interval, which is not equivalent to the semi-infinite line because the conformal symmetry relating the two cases is absent in the $T\bar{T}$-deformed CFT. We conclude the paper in section (\ref{sec:discuss}) with some discussions and outlooks for future investigations.


\subsection*{Notations and scales}

Here we collect some key notations and scales in this paper for clarity:

\begin{itemize}
\item $\epsilon$ is the target space cutoff around the boundary $\pd A$ of the entangling subregion~$A$. $L = -2\log\epsilon$ is a convenient reparametrization for $\epsilon$, not to be confused with the length of the interval $A$, which would be $(b-a)$ in the finite interval case.

\item $\epsilon'$ is the cutoff of the worldsheet annulus around the origin. $\ell = -2n\log\epsilon'$ is a convenient reparametrization for $\epsilon'$ and represents the length of the worldsheet cylinder after the plane to cylinder map. It is also the modular parameter of the annulus / cylinder worldsheet.

\item $\mu$ is the $\TTbar$ deformation parameter. $\lambda = 2\pi\mu / R^2$ is the dimensionless coupling where $R$ is the length scale of the target space: $ds^2 = R^2\, dX^+ dX^-$. 

\item $\epsilon_{T\bar{T}} \sim \sqrt{c\mu}$ is the proposed dynamical cutoff induced by the \TTbar deformation. Here $c$ is the central charge of the undeformed CFT.
\end{itemize}

\section{Entanglement and replica trick in \TTbar-deformed CFTs}\label{sec:setup}

In this section, we outline the general framework of the computation. To begin with, we assume that the deformed vacuum state $\Omega_\mu$ at the deformation parameter $\mu$ is described using the Euclidean path-integral by the Hartle-Hawking wave-functional with the deformed action: 
\be \label{eq:deformed_state}
\langle \varphi| \Omega_\mu\rangle = \int \left[\mathcal{D}\Phi\right]_\varphi\; \exp{\left[-S^{T\bar{T}}\left(\mu,\Phi,H^-\right)\right]}
\ee
where $\Phi$ denotes collectively the quantum fields in the original CFT, and $S^{T\bar{T}}\left(\mu,\Phi,H^-\right)$ is the deformed action at parameter $\mu$ on the lower-half-plane $H^-$. The functional argument $\varphi$ of the wave-functional specifies the boundary condition imposed for the path-integral $\left[\mathcal{D}\Phi\right]_\varphi$ at the $\tau=0$ time-slice. The deformed Renyi entropy is defined by: 
\be\label{eq:renyi_def}
S^n_A\left(\Omega_\mu\right) = \frac{1}{1-n}\ln{\left[\frac{\text{Tr}\rho^n_A\left(\Omega_\mu\right)}{\text{Tr} \rho_A\left(\Omega_\mu\right)^n}\right]}
\ee
Applying the standard replica trick to (\ref{eq:deformed_state}), evaluating (\ref{eq:renyi_def}) requires one to compute the following Euclidean path-integral:
\be \label{eq:replica_def}
\text{Tr} \rho^n_A\left(\Omega_\mu\right) = \int \mathcal{D}\Phi\, \exp{\left[-S^{T\bar{T}}\left(\mu,\Phi,\Sigma^A_n\right)\right]} = Z(\mu, \Sigma^A_n) 
\ee
where $S^{T\bar{T}}\left(\mu,\Phi, \Sigma^A_n\right)$ is now the $T\bar{T}$-deformed action defined on the Euclidean background manifold $\Sigma^n_A$ obtained by cyclically gluing $n$ copies of the complex-plane along the interval $A$. The conical singularities of $\Sigma^A_n$ located at the end points of $A$ are responsible for the UV divergence of $S^n_A$. To be concrete, we regulate the divergence with the standard cut-off procedure of removing holes of radius $\sim \epsilon$ around the conical singularities. As was discussed in the introduction, it is not realistic to evaluate (\ref{eq:replica_def}) using the original formulation (\ref{eq:TT_def}). Instead, we will formulate the computation of (\ref{eq:renyi_def}) in the framework where the $T\bar{T}$-deformed CFT is described as a string theory propagating in a two-dimensional target-space \cite{Callebaut:2019omt}. 

\subsection{\TTbar-deformed CFT as a string theory: a quick recap}
In this subsection, we provide a quick recap of the string theory formulation of the $T\bar{T}$-deformation, focusing only on the necessary ingredients that concern us. This was first proposed for the deformation of critical CFTs with $c=24$ \cite{McGough:2016lol}, and subsequently extended to non-critical CFTs \cite{Callebaut:2019omt}. For a $T\bar{T}$-deformed CFT on a background space-time $\mathcal{M}$, the corresponding string theory formulation is described by a world-sheet CFT consisting of two sectors: (i) the original CFT, now appearing as a world-sheet CFT; (ii) a two-dimensional target-space CFT described by $X^{\pm}$,  where the target-space manifold is given by the background space-time $\mathcal{M}$. The dynamics of the target-space CFT is described by the following action: 
\be\label{eq:target_onshell}
S_{\mathcal{M}}(X^{\pm},\mu) = \frac{1}{2\pi\mu}\int d^2z \left(\bar\partial X^+(z,\bar{z})\,\partial X^-(z,\bar{z}) -\frac{\kappa}{4}R \log{\left(\partial X^+ \bar{\partial} X^-\right)}\right)
\ee
where $\kappa = \frac{c}{24}-1$ for the original CFT of central charge $c$ and $R$ is the world-sheet curvature. The deformation parameter $\mu$ now serves as the coupling constant of the target-space CFT, or more physically as the string tension. The second term in (\ref{eq:target_onshell}) is there so that the central charge $c_X$ of the target-space CFT is $c_X = 26-c$. In this way the total Weyl anomaly cancels among the world-sheet CFTs including the ghost sector and is equal to zero.  Our focus in this paper deals only with the genus-one world-sheets, for which we can gauge-fix the world-sheet metric to be flat $R=0$. The second term of (\ref{eq:target_onshell}) is then absent and does not play any dynamical part -- it only modifies the stress tensor definition in (\ref{eq:target_onshell}).  When the world-sheet contains boundaries, one may include a Gibbons-Hawking-like term in addition to the second term in (\ref{eq:target_onshell}) such that the boundary condition for the world-sheet metric is consistent with the variational principle. Such a boundary term is proportional to the extrinsic curvature $K$ of the boundary. In this paper $K$ can again be set to zero by working in an appropriate gauge, e.g.~in which the world-sheet is the finite cylinder with flat edges. More details related to this can be found in the discussion section (\ref{sec:discuss}). We will therefore neglect the second term in (\ref{eq:target_onshell}) and the associated boundary term when performing explicit computations in subsequent analysis.  

Now the gauge-fixed string partition function can be written as:
\be\label{eq:deformation_overall}
Z(\mu,\mathcal{M}) = \int \frac{d\zeta_{\mathcal{M}'}}{\text{CKG}}\int \mathcal{D} X^{\pm} \mathcal{D}\Phi\mathcal{D}b\mathcal{D}c\; e^{-S_{CFT}(\Phi,\mathcal{M}')-S_{\mathcal{M}}(X^{\pm},\mu,\mathcal{M}')-S_{gh}(b,c,\mathcal{M}')}
\ee
We emphasize again that the background space-time $\mathcal{M}$ is recorded in the definition of the target-space sigma-model parametrized by $X^{\pm}$; while the world-sheet manifold is denoted by $\mathcal{M}'$, whose metric has locally been gauged-fixed to be flat and the resulting $bc$ ghost action $S_{gh}$ included. We have also included a potential integral over the moduli space $\zeta_{\mathcal{M}'}$, as well as the modding out by the conformal killing group (CKG), when they exist. In addition, the mapping degree between the $\mathcal{M}'$ and $\mathcal{M}$ is restricted to one. This restriction on topology may appear arbitrary from a string theory perspective. On the other hand, it is based on the same restriction that the original $T\bar{T}$-deformed spectrum is reproduced exactly from the string theory or topological gravity formulations \cite{Callebaut:2019omt, Dubovsky:2018bmo}. It is unclear what role the higher winding (or higher genus) mappings play in the $T\bar{T}$-deformed physics. 
Even if included, contributions from the higher winding mappings are exponentially suppressed for $\mu>0$, see Eq (\ref{eq:higher_winding}) in a later footnote. Due to these considerations, in this paper we will restrict our attention to the winding one sector.
Eq (\ref{eq:deformation_overall}) is the starting point of using the string theory formulation to evaluate the replica trick path-integral (\ref{eq:replica_def}) for the $T\bar{T}$-deformed CFT. 
   
\subsection{\TTbar-deformed replica trick: the stringy perspective}
Now we begin formulating the computation of (\ref{eq:replica_def}) in terms of the general string partition function  (\ref{eq:deformation_overall}). This is simply achieved by specifying the target-space manifold $\mathcal{M}$ to be the one obtained from performing the replica trick: the Euclidean manifold $\Sigma^A_n$ with $n$-fold conical-excesses at the boundaries $\partial A$. To simplify notations from now on we will stop labeling the subregion $A$, and will simply write $\Sigma_n$. Evaluating (\ref{eq:replica_def}) therefore simply amounts to inserting $\mathcal{M}=\Sigma_n$ in (\ref{eq:deformation_overall}):
\be\label{eq:deformation_trace}
\text{Tr} \rho_A^n\left(\Omega_\mu\right) = \int \frac{d\zeta_{\mathcal{M}'}}{\text{CKG}}\int \mathcal{D} X^{\pm} \mathcal{D}\Phi\mathcal{D}b\mathcal{D}c\; e^{-S_{CFT}(\Phi,\mathcal{M}')-S_{\Sigma_n}(X^{\pm},\mu,\mathcal{M}')-S_{gh}(b,c,\mathcal{M}')}
\ee
We pause here to again emphasize our treatment regarding the conical-excess singularities of the target space manifold $\Sigma_n$. As was mentioned, we will follow the usual QFT regularization of removing small holes around these singularities. In the string theory formulation of the $T\bar{T}$-deformed CFT, this treatment is then uplifted to the target-space -- now the target space manifold is $\Sigma_n$ but with the conical-excess singularities removed by small holes of radius $\sim \epsilon$. We denote by $\tilde{\Sigma}_n$ the target space manifold that is regulated in this way. 

It is worth comparing this treatment with the conventional string theory perspective, where the target-space geometry is dynamical and constrained by the on-shell condition. Stringy entanglement in this perspective has been extensively considered \cite{Dabholkar:1994ai,Dabholkar:2001if,Dabholkar:2014ema,He:2014gva,Prudenziati:2016dbc,Prudenziati:2018jcf,Witten:2018xfj,Dabholkar:2023ows}. The fact that the target space manifold $\Sigma_n$ contains conical-excess singularities means that it is not an on-shell background for string theory. In particular, instead of imposing a geometric cut-off like in QFTs, the conical singularity resulting from the replica trick is expected to resolve by itself via the back-reaction -- possibly describable in an off-shell framework of string theory \cite{Susskind:1994sm,Ahmadain:2022tew,Ahmadain:2022eso},  giving rise to an on-shell geometry that is smooth. Our treatment does not consider the back-reaction and instead simply removes the singularities. It reflects the perspective that the $T\bar{T}$-deformed CFT, despite having a stringy formulation, originates from a mundane CFT living on a fixed background space-time. In particular, this background manifold is specified at $\mu = 0$ as part of the initial condition, and should not evolve with $\mu$. Once we have decided to implement a geometric UV cut-off in the original CFT replica trick at $\mu=0$, e.g.~working with the regulated $\tilde{\Sigma}_n$, the target-space manifold is then fixed as $\tilde{\Sigma}_n$ at finite values of $\mu$. In the future it is interesting to explore whether and how our treatment regarding the $T\bar{T}$-deformed entanglement can be made sense of in the conventional string theory perspective. We remark that by working with this prescription we have not hidden the off-shell nature of the model -- it shows up when the boundary conditions are taken into consideration. We will come back to this point in the later sections, see in particular appendix (\ref{app:F0_limits}). 

From now on we restrict to the cases where $A$ is a single interval $A=[a,b]$ -- this is the main focus of this paper. The target space $\tilde{\Sigma}_n$ is now topologically a cylinder. Since the mapping degree is restricted to 1, the world-sheet manifold should match topologically, which we take to be the complex plane with two punctures. By the world-sheet conformal invariance, the locations of the punctures can be fixed at $z=0$ and $z=\infty$; the punctured holes can be arranged to have the radii $r=\epsilon'$ about $z=0$ and $r=1/\epsilon'$ about $z=\infty$.  The world-sheet manifold is therefore parametrized by the single modulus $\epsilon'$, and the string partition function  (\ref{eq:deformation_trace}) becomes: 
\be\label{eq:deformation_regulated}
\text{Tr} \rho_A^n\left(\Omega_\mu\right) = \int^\infty_0 \frac{d\ell}{2\pi\ell}\int \mathcal{D} X^{\pm} \mathcal{D}\Phi\mathcal{D}b\mathcal{D}c\; e^{-S_{CFT}(\Phi,\epsilon')-S_{\tilde{\Sigma}_n}(X^{\pm},\mu,\epsilon')-S_{gh}(b,c,\epsilon')}
\ee
where we have written the integration measure for $\epsilon' = e^{-\ell/(2n)}$, and divided $2\pi$ from the conformal Killing group of the world-sheet rotation about $z=0$. The factor of $n$ in the definition of $\epsilon'$ will become clear later. To simplify expressions in the subsequent discussion, we consider for the moment when the subregion $A$ is the semi-infinite half-line $A=[0,\infty]$. Once we have a good understanding regarding this case, the formulation can be easily extended to other cases, e.g.~when $A$ is the finite-interval. In this case, the regulated target-space $\tilde{\Sigma}_n$ contains $n$-fold conical-excess at $X^{\pm}=0$ and $X^{\pm}=\infty$, with a hole of radius $\epsilon$ removed about $X^{\pm}=0$; and a hole of radius $\epsilon^{-1}$ removed about $X^{\pm}=\infty$. In contrary to the dynamical world-sheet cut-off $\epsilon'=e^{-\ell/(2n)}$, the target-space cut-off $\epsilon=e^{-L/2}$ enters the string partition function (\ref{eq:deformation_regulated}) through $\tilde{\Sigma}_n$ as a fixed external parameter. From now on we will specify the $L$-dependence explicitly and write $\text{Tr} \rho^n_A\left(L,\Omega_\mu\right)$ for the corresponding string partition. 

\subsection{Boundary conditions for the world-sheet BCFT}

The computation in its current form amounts to calculating the partition function of a boundary CFT from the world-sheet perspective. A rigorous treatment of the world-sheet BCFT computation should proceed by specifying the boundary conditions for all the sectors. For the original CFT sector, a natural candidate for the boundary state in the closed string channel is the one that should have been imposed in the original CFT computation $|B_{CFT}\rangle$. Its overlap squared with the vacuum state $|\langle B_{CFT}|\Omega_{CFT}\rangle|^2$ is related to the CFT boundary entropy. For the target-space CFT, it is more convenient to directly specify the boundary condition for the path-integral. From a sigma-model point of view, the target-space is a manifold with fixed boundaries, which we denote as $\mathcal{C} = \partial\tilde{\Sigma}_n$, while the world-sheet has the boundary $\gamma=\partial \tilde{\Sigma}'_n$. The path-integral is over mapping functions from the world-sheet manifold to the target-space manifold, and the boundary condition is specified by the corresponding mapping function from $\gamma$ to $\mathcal{C}$. It is important to stress that this is not equivalent to the completely Dirichlet boundary condition, e.g, those of the form: 
\be\label{eq:bc_general}
X^{\pm}(\sigma) = \mathcal{C}^{\pm}\left(\theta^*(\sigma)\right),\;\; \sigma \in \gamma
\ee
and where $\theta^*(\sigma)$ is some fixed mapping from the intrinsic world-sheet coordinates $\sigma$ for $\gamma$ to the intrinsic target-space coordinate $\theta$ for $\mathcal{C}$. This boundary condition breaks the diffeomorphic invariance of the boundary and can lead to pathologies for the world-sheet metric path-integrals \cite{Alvarez:1982zi}. Instead, one should allow boundary variations $\delta X^{\pm}$ that are tangent to $\mathcal{C}$. In this sense, it is analogous to the D-brane boundary condition -- Dirichlet along some target-space directions (transverse) and Neumman along the others (longitudinal). What distinguishes our set-up from those of ordinary planar D-branes is that $\mathcal{C}$ is generally curved, so the boundary conditions are non-linear in terms of the target-space fields $X^{\pm}$. For the half-line example, the target-space boundary consists of the inner and outer circles, described by the boundary condition:
\be
\sqrt{X^+(z,\bar{z})\, X^-(z,\bar{z})}_{\gamma_1} = \epsilon,\quad\sqrt{X^+(z,\bar{z})\, X^-(z,\bar{z})}_{\gamma_2} = 1/\epsilon
\ee
where $\gamma_{1,2}$ are the inner and outer parts of world-sheet annulus boundaries. In terms of (\ref{eq:bc_general}) we should impose: 
\be \label{eq:bc_specific}
X^{\pm}(\epsilon' e^{i\sigma_1}) = \epsilon e^{\pm in\theta_1(\sigma_1)},\quad X^{\pm}\left(\frac{1}{\epsilon'} e^{i\sigma_2}\right) = \frac{1
}{\epsilon} e^{\pm in\theta_2(\sigma_2)}
\ee
but allow $\theta_{1,2}(\sigma)$ to vary over smooth monotonous functions from $[0,2\pi]$ to $[0,2\pi]$. They represent the freedom of the boundary reparametrization. The full partition function is then written as the integral over the boundary reparametrization: 
\be
\text{Tr}\rho^n_A\left(L,\Omega_\mu\right) = \int \mathcal{D}\theta_1 \int \mathcal{D}\theta_2\; Z(L,\theta_1,\theta_2)
\ee
where the integrand $Z(L,\theta_1,\theta_2)$ is computed by imposing (\ref{eq:bc_specific}) as the usual completely Dirichlet boundary condition for the target-space fields $X^{\pm}(z,\bar{z})$. In more details the integrand can be further written as: 
\bea\label{eq:Z_Dirichlet}
Z(L,\theta_1,\theta_2) &=& \int^\infty_0 \frac{d\ell}{2\pi \ell}\sum_{X^{\pm}_{cl}(\theta_1,\theta_2)} e^{-S_{\tilde{\Sigma}_n}\left(X^{\pm}_{cl}(\theta_1,\theta_2),\mu,\epsilon'\right)}\times \left[\int \mathcal{D}\delta X^{\pm}\;e^{-\frac{1}{2\pi\mu} \int d^2 z \delta X^+ \left(\partial\bar{\partial}\right) \delta X^-} \right]\nonumber\\
&\times &\left[\int \mathcal{D}b\mathcal{D}c\; e^{-S_{gh}(b,c,\epsilon')}\right] \times\left[\int \mathcal{D}\Phi\;e^{-S_{CFT}(\Phi,\epsilon')}\right]
\eea
where $X^{\pm}_{cl}(\theta_1,\theta_2)$ is a classical solution satisfying (\ref{eq:bc_specific}) for the fixed $\theta_{1,2}$, and $\delta X^{\pm}$ denotes the fluctuations on top of $X^{\pm}_{cl}(\theta_1,\theta_2)$ satisfying the (homogeneous) Dirichlet boundary condition. As was shown and analyzed in \cite{Rychkov:2002ni}, the boundary conditions for a non-geodesic $\mathcal{C}$ are in general not conformally invariant -- the quantum corrections from integrating the $\theta_{1,2}$ modes are in general sensitive to the UV cut-off $\Lambda$. The planar D-brane boundary condition is special in that such corrections are UV convergent, and thus conformal.  

\subsection{Conical excesses: from the target-space to the world-sheet}
Finally let us discuss in more precise terms how the conical excess properties of the target space $\tilde{\Sigma}_n$ affect the string partition function (\ref{eq:Z_Dirichlet}). Locally the target-space is still a flat manifold. The fact that it contains conical excesses is reflected in the restriction of the mapping $\delta X^{\pm}(z,\bar{z})$ in (\ref{eq:Z_Dirichlet}) to have winding number only in multiples of $n$ around $X^{\pm}=0$ or $X^{\pm}=\infty$ as the world-sheet coordinate winds once around $z=0$ or $z=\infty$. As a consequence, the path-integral is only over $\delta X^{\pm}$ with such restricted winding properties, not the full functional space on the world-sheet. To account for this, We can define the conformal coordinates $w=z^n, \bar{w}=\bar{z}^n$ on the world-sheet, then the path-integral over the mapping $\mathcal{D}\delta X^{\pm}(w,\bar{w})$, now as functions of $(w,\bar{w})$, is standard -- there is no restriction for their winding properties.
As a result, the one-loop factor, which is the only quantum correction to the partition function of the free target-space CFT, is simply given by the functional determinant of the Laplacian operator $\partial_w\partial_{\bar{w}}$ subject to the Dirichlet boundary condition. There are two factors of such one-loop determinants, coming from $\delta X^{\pm}$ respectively. On the other hand, the free $bc$ ghost CFT also consists of two bosons. After transforming to the $(w,\bar{w})$ coordinate frame, the one-loop determinants from the $bc$ ghost CFT will cancel those of the target-space CFT in the string partition function (\ref{eq:Z_Dirichlet}): 
\be\label{eq:cancellation}
\left[\int \mathcal{D}\delta X^{\pm}\;e^{-\frac{1}{2\pi\mu} \int d^2 w\; \delta X^+ \left(\partial_w\partial_{\bar{w}}\right) \delta X^-} \right]
\times \left[\int \mathcal{D}b\mathcal{D}c\; e^{-S^{w,\bar{w}}_{gh}(b,c,\epsilon')}\right] =1
\ee
We make some additional comments regarding the cancellation (\ref{eq:cancellation}). Since in our case the world-sheet contains boundaries, the functional determinant of the Laplacian depends on the boundary condition imposed on the world-sheet boundaries. For the fluctuations $\delta X^{\pm}$ on top of the classical solution $X^{\pm}_{cl}$, the Dirichlet boundary condition is composed. As will be discussed momentarily, the ghost fields satisfy boundary conditions (\ref{eq:ghost_bc}), which for the vector ghost field $c^i$ (whose boundary condition determines the ghost determinant) is equivalent to the Neumman and Dirichlet boundary conditions respectively for the longitudinal and transverse components. Despite this caveat, it can be shown that on the annulus the functional determinant of the Laplacian is identical for the Neumman and the Dirichlet boundary conditions \cite{Alvarez:1982zi}, so the cancellation remains valid. In Appendix (\ref{app:cancellation}) we provide some details for this.

We emphasize that working in the $(w,\bar{w})$ coordinates is crucial for arranging explicitly the cancellation between the one-loop factors --  in some sense the target-space winding is ``neutralized" by doing this. In performing the corresponding conformal transformation $z\to w=z^n,\;\bar{z}\to\bar{w}=\bar{z}^n$ to the target-space and the $bc$ ghost CFTs, a net Weyl anomaly appears. It is therefore more convenient to perform the conformal transformation for all three sectors of the world-sheet CFTs. By doing this, the total Weyl anomaly cancels on the world-sheet. We note that the world-sheet manifold contains boundaries in our analysis. In this case the 2d BCFT is free of Weyl anomaly as long as the conformal boundary conditions for the world-sheet CFT sectors are imposed in the form of a BRST invariant Cardy state $|B\rangle$ satisfying\,\footnote{In higher dimensional $(d\geq 3)$ BCFTs, the Weyl anomaly can contain boundary terms that depend on the so-called boundary central charges. They are properties of the conformal boundary conditions imposed \cite{Herzog:2017xha}. In this case, the absence of the total Weyl anomaly requires the cancellation of these boundary central charges. In 2d BCFT there is no counterpart for boundary central charges, and the absence of the total Weyl anomaly can be guaranteed by the bulk properties alone, see e.g.\ page 96 of \cite{Polchinski:1998rq}. We thank Matthias Gaberdiel and Rongxin Miao for useful discussions related to this subject.}: 
\be\label{eq:BCFT_BRST}
\left(\mathcal{Q}+\bar{\mathcal{Q}}\right) |B\rangle = 0
\ee
where in terms of the Virasoro generators (matter CFT and $\delta X^{\pm}$) and the ghost oscillators the BRST charges are: 
\bea\label{eq:BRST}
&& \mathcal{Q} = \sum_n {: \mathcal{L}_{-n} c_n :} - \frac{1}{2} \sum_{m,n}(m-n) :c_{-m} c_{-n} b_{m+n}:\nonumber\\
&& \bar{\mathcal{Q}} = \sum_n {: \bar{\mathcal{L}}_{-n} \bar{c}_n :} - \frac{1}{2} \sum_{m,n}(m-n) :\bar{c}_{-m} \bar{c}_{-n} \bar{b}_{m+n}:
\eea
For generic conformal boundary conditions imposed on the matter CFT and the Dirichlet boundary conditions for the target-space fluctuations $\delta X^{\pm}$, the Virasoro generators satisfy: $\left(\mathcal{L}_n-  \bar{\mathcal{L}}_{-n}\right)|B\rangle=0$, so (\ref{eq:BCFT_BRST}) is satisfied if we impose on the ghost sector the following boundary conditions \cite{Callan:1987px}: 
\be \label{eq:ghost_bc}
\left(c_n +\bar{c}_{-n}\right)|B\rangle = \left(b_n -\bar{b}_n\right) |B\rangle = 0
\ee
In this setup, the conformal breaking only comes from the integral over the boundary condition reparametrizations $\theta_{1,2}$. The original world-sheet $\mathcal{M}'$ is now transformed to a manifold, which we label as $\tilde{\Sigma}'_n$, that also contains $n$-fold conical-excess singularities at $w=0$ and $w=\infty$. They are now regulated by removing holes of radii $r=(\epsilon')^{n}=e^{-\ell/2}$ about $w=0$ and $r=(\epsilon')^{-n}=e^{\ell/2}$ cut about $w=\infty$. 

\subsection{General integral representation\nopdfstring{ of $\operatorname{Tr} \rho^n_A(L,\Omega_\mu)$}}
Assembling everything that we have discussed so far, and working in the conformal frame of $(w,\bar{w})$, the resulting string partition function can be written as:\,\footnote{The path-integral measure for $\theta_{1,2}$ inherits from that of the field-space metric for the target-space coordinate at the boundaries $X_{\text{inner}}^{\pm} = \epsilon e^{\pm in\theta_1},\;X_{\text{outer}}^{\pm} = \epsilon^{-1} e^{\pm in\theta_2}$: 
\be 
ds^2\propto \delta X_{\text{inner}}^+ \delta X_{\text{inner}}^- + \delta X_{\text{outer}}^+ \delta X_{\text{outer}}^- =n^2 \epsilon^2 \delta \theta_1^2 + n^2 \epsilon^{-2} \delta \theta_1^2
\ee
From this we obtain the linear path-integral measure in (\ref{eq:master}): 
\be
\mathcal{D}X^{\pm}_{\text{inner}} \mathcal{D}X^{\pm}_{\text{outer}} \rightarrow \mathcal{D}\theta_1 \mathcal{D}\theta_2
\ee
} 
\bea\label{eq:master}
\text{Tr}\rho^n_A(L,\Omega_\mu)&=& \int \mathcal{D}\theta_1 \int \mathcal{D} \theta_2 \int^\infty_0 \frac{d\ell}{2\pi \ell} \sum_{X^{\pm}_{cl}(\theta_1,\theta_2)} e^{-S_{\tilde{\Sigma}_n}\left(X^{\pm}_{cl}(\theta_1,\theta_2),\mu,\tilde{\Sigma}'_n\right)} \left(\int \mathcal{D}\Phi\, e^{-S_{CFT}\left(\Phi, \tilde{\Sigma}'_n\right)}\right) \nonumber\\
&=&\int^\infty_0 \frac{d\ell}{2\pi \ell}\, K(L,\ell,\mu)\; \text{Tr}\rho^n_A(\ell,\Omega_{CFT})
\eea
where in the second line we have written the deformed result as a weighted integral over the CFT replica result at a dynamical UV cut-off $\left(\epsilon'\right)^n=e^{-\ell/2}$ to be integrated over. We remind the readers that the explicit $A$-dependence on the LHS of (\ref{eq:master}) enters the RHS through the branch structure (with the tip excised) of the replica manifold $\tilde{\Sigma}_n$, which encodes the end-points of the interval $A$ (thus also the interval length); and the parameter $\epsilon = e^{-L/2}$, which encodes the regulator prescription of $\partial A$. The readers are also cautioned not to confuse $L$ with the length of $A$.

\begin{figure}[!ht]
\tdplotsetmaincoords{70}{0}
\begin{center}
\begin{adjustbox}{center}
\begin{tikzpicture}[tdplot_main_coords, scale=1]
  \useasboundingbox (-7,-7,0) rectangle (7,7,2);
  
  \draw[thick] (-6,0,0) arc[start angle=180, end angle=360, radius=2];
  \draw[->,thick,black,line width=1.3] (-4.05,-2,0) -- (-3.95,-2,0);
  \draw[dashed] (-2,0,0) arc[start angle=0, end angle=180, radius=2];
  \draw[thick] (-4,0,2) circle (0.5);
  \draw[->,thick,black,line width=1.3] (-4.05,-0.5,2) -- (-3.95,-0.5,2);
  \draw (-4.7,-0.6,2) node[above] {$\epsilon'$};
  \draw (-6.3,-2,0) node[above] {$1/\epsilon'$};
  \draw[thick] (-6,0,0) -- (-4.5,0,2);
  \draw[thick] (-2,0,0) -- (-3.5,0,2);
  \node[below] at (-4,-2.2,0) {\shortstack{World-sheet\\$(w,\bar{w})$\\$\sigma\in[0,2\pi n)$}};
  \node[above] at (-3.4,0.3,2) {$\sigma$};
  \node[above] at (-1.8,-1,0) {$\sigma$};

  \draw[draw=blue, thick,line width=1.3] (2,0,0) arc[start angle=180, end angle=360, radius=2];
  \draw[->,thick,blue,line width=1.3] (3.95,-2,0) -- (4.05,-2,0);
  \draw[draw=blue, dashed] (6,0,0) arc[start angle=0, end angle=180, radius=2];
  \draw[draw=red, thick,line width=1.3] (4,0,2) circle (0.5);
  \draw[->,thick,red,line width=1.3] (3.95,-0.5,2) -- (4.05,-0.5,2);
  \draw (3.3,-0.6,2) node[above] {$\epsilon$};
  \draw (1.7,-2,0) node[above] {$1/\epsilon$};
  \draw[thick] (2,0,0) -- (3.5,0,2);
  \draw[thick] (6,0,0) -- (4.5,0,2);
  \node[below] at (4,-2.2,0) {\shortstack{Target-space\\$\left(X^{+},X^{-}\right)$\\$\theta_{1,2}\in[0,2\pi n)$}};
  \node[above] at (4.6,0.3,2) {$\theta_1(\sigma)$};
  \node[above] at (6.5,-1,0) {$\theta_2(\sigma)$};
  
  \draw[->, thick] (-1.5,0) -- (1.5,0) node[midway, above] {$X^{\pm}_{cl}(w,\bar{w})$};
\end{tikzpicture}
\end{adjustbox}
\end{center}
\caption{A graphic illustration for the classical solution used to compute the kernel function $K(L,\ell,\mu)$. For better visualization, the conical singularities are shown as defects instead of excesses.}\label{fig:illustration of the Master Equation}
\end{figure}
The weight kernel function $K$ is computed by the on-shell target-space action of the classical mapping solutions, and with the boundary condition reparametrization integrated over, see Figure (\ref{fig:illustration of the Master Equation}): 
\be\label{eq:kernel}
K(L,\ell,\mu) = \int \mathcal{D}\theta_1 \int \mathcal{D} \theta_2\sum_{X^{\pm}_{cl}(\theta_1,\theta_2)} e^{-S_{\tilde{\Sigma}_n}\left(X^{\pm}_{cl}(\theta_1,\theta_2),\mu,\tilde{\Sigma}'_n\right)}
\ee
The dependence on $\ell$ and $L$ of (\ref{eq:kernel}) enters through the definition for the target-space and world-sheet manifolds $(\tilde{\Sigma}_n,\tilde{\Sigma}'_n)$.  Eq (\ref{eq:master}) and Eq (\ref{eq:kernel}) are the main results of this section.  They have translated the entanglement computation of our interest in the $T\bar{T}$-deformed CFT into a well-defined and concrete task. In the next section, we continue with the example of the half-line region $A$ and perform the computation of (\ref{eq:master}) explicitly. 

Before concluding this section, we extract the general principles behind (\ref{eq:master}) so that it can be extended to the single finite-interval case. We leave the more general cases, e.g.~multi-interval, non-vacuum states, etc, to the discussion section (\ref{sec:discuss}). The key ingredients can be summarized as follows: 
\begin{itemize}
\item The string partition function corresponding to $\text{Tr} \rho^n_A\left(L,\Omega_\mu\right)$ is computed by identifying the target-space manifold $\mathcal{M}$ as the replica background that enters the original QFT computation: the branched-manifold $\tilde{\Sigma}_n$ with n-fold conical excess singularities around $\partial A$ regulated by removing small holes of radius $\epsilon=e^{-L/2}$ -- the holes form the target-space boundaries. 

\item The restriction on the mapping degree between the world-sheet and the target-space manifolds being one dictates that the world-sheet manifold $\mathcal{M}'$ also contains boundaries that map to the target-space boundaries. They take the form of holes on the world-sheet. The radius $\epsilon'=e^{-\ell/(2n)}$ of the world-sheet holes is now part of the world-sheet moduli and needs to be integrated over.

\item As a world-sheet BCFT, the boundary condition for the original CFT sector can be specified by inheriting the conformal boundary conditions \cite{Cardy:1989ir,Ohmori:2014eia,Hung:2019bnq} imposed at $\mu=0$. For the target-space CFT sector, the boundary condition is Dirichlet in the transverse direction and Neumann in the longitudinal direction with respect to the target-space boundaries. As a result, the string partition function involves a path-integral $\mathcal{D}f$ over all boundary reparametrizations: $f: \partial \mathcal{M}' \to \partial \mathcal{M}$. For curved target-space boundaries, this path-integral is generally UV-divergent, indicating that the target-space boundary condition is not conformal.

\item A convenient conformal frame $(w,\bar{w})$ on the world-sheet is obtained in which the conical excess around the world-sheet holes matches with that of the target-space: $\mathcal{M}'=\tilde{\Sigma}'_n$. In this frame the one-loop determinants from the target-space fields fluctuations cancel against that of the $bc$ ghost CFT, leaving only the on-shell contribution from the target-space and ghost CFT sectors combined. The contribution from the original CFT sector computed in the frame $(w,\bar{w})$ gives the CFT result $\text{Tr} \rho^n_A\left(\ell,\Omega_{CFT}\right)$ regulated at the dynamical cut-off $(\epsilon')^n = e^{-\ell/2}$. The string partition function for $\text{Tr} \rho^n_A(L,\Omega_\mu)$ then naturally takes the general form of (\ref{eq:master}): a weighted integral over the CFT results at a dynamical cut-off, whose kernel is computed by the classical on-shell action of the target-space CFT, but with all reparametrized boundary conditions integrated over.  

\end{itemize}

\section{Renyi entropies on a semi-infinite line}\label{sec:infinite}

In this section we carry out the explicit computations of (\ref{eq:master}) for the case of half-line subregion $A$.  The key object is the kernel function $K(L,\ell,\mu)$ defined in (\ref{eq:kernel}). The $\mu$-dependence of the string partition function (\ref{eq:master}) comes entirely from the kernel function. As a result it encodes all aspects, both perturbative and non-perturbative, of the effect the $T\bar{T}$ deformation brings to the entanglement structure. Therefore we will begin the section by computing this object systematically. 

We recall that the kernel function is computed by performing the path-integral over the reparametrized boundary conditions: 
\be 
K(L,\ell,\mu) = \int \mathcal{D} \theta_1 \int \mathcal{D} \theta_2\; e^{-\frac{R^2}{2\pi \mu} F\left(\theta_1,\theta_2\right)},\;\;F\left(\theta_1,\theta_2\right) = \int_{\tilde{\Sigma}'_n} d^2 w \,\partial_{\bar w} X^{+}_{cl}\,\partial_{w} X^{-}_{cl}
\ee
where $F(\theta_1,\theta_2)$ is the on-shell action evaluated on the classical solutions $X^{\pm}_{cl}$ to the Laplace equation in $(w,\bar{w})$ that satisfy the corresponding Dirichlet boundary conditions specified by $(\theta_1,\theta_2)$. We have supplied a scale $R$ to restore the dimensionality, which represents the overall size of the target-space and combines with the $T\bar{T}$ scale $\mu$ to form a dimensionless coupling $\lambda^{-1} = R^2/(2\pi \mu)$.

As was discussed, the world-sheet manifold $\tilde{\Sigma}'_n$ is defined by the conformal frame $w=z^n$. It is a manifold with conical excesses around $w=0$ and $w=\infty$. For the purpose of computing the classical on-shell action $\partial X^{+}_{cl} \bar{\partial} X^{-}_{cl}$, which is conformal, it is more convenient to work in the original flat conformal frame $(z,\bar{z})$.  Now the classical solutions $X^{\pm}(z,\bar{z})$ satisfy the Laplace equation and the Dirichlet boundary conditions in the usual complex-plane $(z,\bar{z})$: 
\bea \label{eq:target_space_EOM}
&&\partial_z \partial_{\bar{z}} X^{\pm}_{cl}(z,\bar{z})=0\\
&& X^{\pm}\left(\epsilon' e^{i\sigma}\right) = \epsilon e^{\pm in\theta_1(\sigma)}\nonumber\\
&&X^{\pm}\left(1/\epsilon'e^{i\sigma}\right) = \epsilon^{-1} e^{\pm in\theta_2(\sigma)}\nonumber
\eea
We emphasize that the reparametrizations $\theta_{1,2}$ are constrained to have winding number one, i.e.~they are smooth and monotonous functions of $\sigma$ in the range $[0,2\pi]$. We also recall that in this paper we only work with $\epsilon$ that, being the physical cut-off in the target-space, is very small: $\epsilon = e^{-L/2}\ll 1$, namely $L\gg 1$. 

We now outline the steps to proceed. We will first compute the on-shell action $F(\theta_1,\theta_2)$ as a functional of $\theta_1$ and $\theta_2$. This is done by solving  (\ref{eq:target_space_EOM}) and evaluating the classical action on the solutions obtained. Having done this, we can view $F(\theta_1,\theta_2)$ as an effective action for the dynamical fields $\theta_{1,2}$ on the unit circle $\sigma\in [0,2\pi]$. The kernel function is then computed from the path-integral of this effective theory. The perturbative limit of small $\mu$ corresponds to its weak-coupling limit $\lambda\ll 1$. We can therefore follow the semi-classical approach of first finding the saddle-point configurations that minimize the effective action $F(\theta_1,\theta_2)$, and then computing the perturbative corrections in small $\lambda$. As a result, the full string partition function (\ref{eq:master}) can also be treated in the semi-classical framework.  

\subsection{Effective action \texorpdfstring{$F(\theta_1,\theta_2)$}{F} for the reparametrization fields}
We begin our analysis by first deriving the full quantum effective action $F(\theta_1,\theta_2)$ for the reparametrization fields $\theta_1,\theta_2$, whose path integral will then be computed perturbatively. It amounts to solving bulk equations of motion subject to the fixed boundary reparametrization $(\theta_1, \theta_2)$. 

The world-sheet in $(z,\bar{z})$ coordinate is an annulus with inner and outer radii $r=\epsilon$ and $r=\epsilon^{-1}$ respectively.  The classical solution $X^{\pm}_{cl}$ of the Laplace equation subject to the Dirichlet boundary condition can be obtained by the standard Green's function method: 
\be
X^{\pm}_{cl}(r,\sigma) = \int_{\partial\mathcal{M}'}X^{\pm}_{bc}\del' G(r',\sigma';r,\sigma)\cdot\hat{n}\dm s
\ee
where $G(r',\sigma';r,\sigma)$ is the Green's function on the annulus satisfying:
\be\label{eq:Greens_def}
{\del'}^2 G(r',\sigma';r,\sigma)=\delta(r'-r)\,\delta(\sigma'-\sigma)/r.
\ee
It can be solved in the form of a series expansion: 
\bea\label{eq:annulus_Greens}
G(r',\sigma';r,\sigma)&=&\frac{1}{2\pi}\left(\frac{\ln{\frac{r_<}{\epsilon'}\ln{\frac{{\epsilon'}^{-1}}{r_>}}}}{2\ln{\epsilon'}}+\sum_{k=1}^{\infty}\frac{\left(r_<^k-\frac{{\epsilon'}^{2k}}{r_<^k}\right)\left(r_>^k-\frac{{\epsilon'}^{-2k}}{r_>^k}\right)}{k({\epsilon'}^{-2k}-{\epsilon'}^{2k})}\cos{\left(k(\sigma-\sigma')\right)}\right)\nonumber\\
r_< &=& \text{min}\lbrace r, r' \rbrace,\;\; r_> = \text{max}\lbrace r, r' \rbrace
\eea
According to the divergence theorem in complex coordinates:
\begin{equation}
    \int_{\mathcal{M}}\dm^2 z(\partial v^z+\bar{\partial}v^{\bar{z}})=\im\oint_{\partial \mathcal{M}}(v^z \dm \bar{z}-v^{\bar{z}}\dm z).
\end{equation}
The action can be rewritten in terms of boundary conditions:
\begin{equation}
    \begin{aligned}
        F&=\int_0^{2\pi}\dm\sigma\int_{0}^{2\pi}\dm\sigma'\Bigg[X^+\left(\frac{1}{\epsilon'},\sigma\right)K\left(\frac{1}{\epsilon'},\sigma';\frac{1}{\epsilon'},\sigma\right)X^-\left(\frac{1}{\epsilon'},\sigma'\right)\\
        &-X^+\left(\frac{1}{\epsilon'},\sigma\right)K\left(\epsilon',\sigma';\frac{1}{\epsilon'},\sigma\right)X^-(\epsilon',\sigma')-X^+(\epsilon',\sigma)K\left(\frac{1}{\epsilon'},\sigma';\epsilon',\sigma\right)X^-\left(\frac{1}{\epsilon'},\sigma'\right)\\
        &+X^+(\epsilon',\sigma)K(\epsilon',\sigma';\epsilon',\sigma)X^-(\epsilon',\sigma')\Bigg]
    \end{aligned}
\end{equation}
Where:
\begin{equation}
    K(r',\sigma';r,\sigma)=\frac{rr'}{2}\left(\partial_r-\frac{\im}{r}\partial_\sigma\right)\partial_{r'}G(r',\sigma';r,\sigma)
\end{equation}
After some calculations, we can obtain the explicit form of $K(r',\sigma';r,\sigma)$. Substituting it back into the equation, we obtain the following:
\begin{equation}
    \begin{aligned}
        F&=\frac{1}{4\pi}\sum_{k=-\infty}^{\infty}\int_0^{2\pi}\dm \sigma\int_0^{2\pi}\dm\sigma'\frac{ke^{-\im k(\sigma-\sigma')}}{\sinh{k\ell/n}}\Bigg[{\epsilon'}^{2k}X^+\left(\frac{1}{\epsilon'},\sigma\right)X^-\left(\frac{1}{\epsilon'},\sigma'\right)\\
        &-X^+\left(\frac{1}{\epsilon'},\sigma\right)X^-(\epsilon',\sigma')-X^+(\epsilon',\sigma)X^-\left(\frac{1}{\epsilon'},\sigma'\right)+{\epsilon'}^{-2k}X^+(\epsilon',\sigma)X^-(\epsilon',\sigma')\Bigg].
    \end{aligned}
\end{equation}

Plugging in form of the boundary conditions in  (\ref{eq:target_space_EOM}), we therefore obtain the effective action $F(\theta_1,\theta_2)$ for the boundary reparametrizations $\theta_{1,2}$: 
\bea\label{eq:action}
        F(\theta_1,\theta_2)&=&\frac{1}{4\pi}\sum_{k=-\infty}^{\infty}\int_0^{2\pi}\dm \sigma\int_0^{2\pi}\dm\sigma'\;\frac{ke^{-\im k(\sigma-\sigma')}}{\sinh{(k\ell/n)}}
        \times \Big({\epsilon'}^{2k} \epsilon^{-2} e^{in(\theta_2(\sigma)-\theta_2(\sigma'))}\nonumber\\
        &-&e^{in(\theta_2(\sigma)-\theta_1(\sigma'))}
        - e^{in(\theta_1(\sigma)-\theta_2(\sigma'))} +{\epsilon'}^{-2k} \epsilon^2 e^{in(\theta_1(\sigma)-\theta_1(\sigma'))}\Big)
\eea
where the coefficients at $k=0$ are defined by taking the $k\to 0$ limit:
\begin{equation}
\lim_{k\rightarrow0}\left(\frac{k}{\sinh{(k\ell/n)}}\right)=\frac{n}{\ell}
\end{equation}

\subsection{Classical on-shell action: leading order analysis}
Now we can treat the path-integral that computes the kernel function in the perturbative theory of small $\lambda = 2\pi\mu/R^2$:  
\bea \label{eq:path-integral-1}
K(L,\ell,\lambda) &=& \int \mathcal{D}\theta_1 \int \mathcal{D}\theta_2\; e^{-\lambda^{-1} F(\theta_1,\theta_2)} \nonumber\\
&=& \exp{\left(- \lambda^{-1}\sum_{m\geq 0} \lambda^m S_m(L,\ell)\right)}
\eea
where $S_m(L,\ell)$ is the $m$-th order contribution in the perturbation theory -- each order a function of the target-space and world-sheet cut-off parameters $L$ and $\ell$. The leading order contribution $S_0(L,\ell)$ is determined by the saddle-point solution $\theta^*_{1,2}(\sigma)$ that minimizes the effective action $F(\theta_1,\theta_2)$. This is computed by solving the saddle-point equation of (\ref{eq:action}), which takes the form:
\begin{equation}\label{eq:half_line_EOM}
    \begin{aligned}
        e^{2\im n\theta_1(\sigma)} &= \frac{\sum\limits_{k=-\infty}^{\infty}\frac{ k e^{ik\sigma} }{\sinh{(k\ell/n)}}\left(\epsilon^2{\epsilon'}^{-2k}\alpha_{k}-\beta_{k}\right)}{\sum\limits_{k=-\infty}^{\infty}\frac{ k e^{-ik\sigma} }{\sinh{(k\ell/n)}}\left(\epsilon^2{\epsilon'}^{-2k}\alpha_{k}^*-\beta_{k}^*\right)},
        e^{2\im n\theta_2(\sigma)} = \frac{\sum\limits_{k=-\infty}^{\infty}\frac{ k e^{ik\sigma} }{\sinh{(k\ell/n)}}\left(\epsilon^{-2}{\epsilon'}^{2k}\beta_{k}-\alpha_{k}\right)}{\sum\limits_{k=-\infty}^{\infty}\frac{ k e^{-ik\sigma} }{\sinh{(k\ell/n)}}\left(\epsilon^{-2}{\epsilon'}^{2k}\beta_{k}^*-\alpha_{k}^*\right)}\\
        \alpha_{k} &= \frac{1}{2\pi}\int^{2\pi}_0 d\sigma e^{in\theta_1(\sigma)} e^{-ik\sigma},\;\;\beta_{k} =\frac{1}{2\pi}\int^{2\pi}_0 d\sigma e^{in\theta_2(\sigma)} e^{-ik\sigma}
    \end{aligned}
\end{equation}
where the solutions $\theta^*_{1,2}$ are constrained to be real functions that are smooth and monotonous between $0$ and $2\pi$. Despite the complicated non-linear form of (\ref{eq:half_line_EOM}), it is easy to verify that the simplest guess: 
\be\label{eq:half_line_saddle}
\theta^*_1(\sigma) = \theta^*_2(\sigma) = \sigma
\ee
solves (\ref{eq:half_line_EOM}) while satisfying the constraint. By plugging the saddle-point configuration (\ref{eq:half_line_saddle}) into (\ref{eq:action}), the leading-order action $S_0$ of the path-integral (\ref{eq:path-integral-1}) is given by:\,\footnote{It is straight-forward to re-derive (\ref{eq:S_0}) for the higher winding sector $w=m>1$, and the leading order action is simply given by replacing $n\to mn$ in (\ref{eq:S_0}): 
\be\label{eq:higher_winding}
\lambda^{-1}S_0 \sim \frac{4\pi n m}{\lambda}\left(\frac{\sinh^2\left(\frac{\ell-L}{2}\right)}{\sinh{(\ell)}}\right)
\ee
We therefore see that the winding $w$ map gives a higher action for $m>1$ and thus for $\mu>0$ contributes exponentially smaller had they been included.} 
\be\label{eq:S_0}
\lambda^{-1}S_0(L,\ell) = \frac{4\pi n}{\lambda} \left(\frac{\sinh^2{(\frac{\ell-L}{2})}}{\sinh{(\ell)}}\right)
\ee
The $\ell$-integral based on (\ref{eq:S_0}) is convergent only for $\mu>0$ due to the positive divergence of $S_0$ in the $\ell\to 0$ region. For this reason we only focus our analysis on the $\mu>0$ case, as was mentioned in the introduction. A proper treatment for $\mu<0$ involves performing an analytic continuation of $\mu$ in the complex plane from the positive to the negative axis. In the semi-classical analysis, this can be systematically done by tracking the saddle-points and their corresponding Lefschetz thimbles for the complexified $\ell$ integral as $\mu$ rotates in the complex plane, paying particular attention to the Stokes' phenomenon in this process. This was done in \cite{Gu:2024ogh, Gu:2025tpy} for the case of $T\bar{T}$-deformed torus partition functions. We leave the task of analytically continuing $\mu$ for future investigations, and focus only on the $\mu>0$ case in this paper. 

At the leading order, the kernel function is approximately a Gaussian distribution for $\ell$ that is highly peaked at $\ell^*$: 
\be\label{eq:gaussian_leading}
K(L,\ell,\lambda) \approx e^{-\frac{1}{2\delta}(\ell-\ell^*)^2}
\ee
where $\ell^*$ is the the minimum of $S_0$\,:\,\footnote{We discard the other solution  $\ell^*=-L$ of (\ref{eq:ordinary_saddle}), see the subsection (\ref{sec:saddle}) for more explanations.}
\bea\label{eq:ordinary_saddle}
\partial_\ell \left(\frac{\sinh^2\left(\frac{\ell-L}{2}\right)}{\sinh{(\ell)}}\right)_{\ell^*} = \frac{n \sinh{\left(\frac{\ell^*-L}{2}\right)}\sinh{\left(\frac{\ell^*+L}{2}\right)}}{\sinh^2{(\ell^*)}}= 0
\quad\Longrightarrow\quad
\ell^*=L 
\eea
and the variance $\delta$ is give by:
\be\label{eq:variance}
\delta = \lambda/S^{''}_0(\epsilon^*,\epsilon) =\frac{\lambda \sinh{(L)}}{2\pi n}
\ee
In other words, at the leading order in small $\lambda$ the kernel function alone creates a strong attraction for the dynamical (world-sheet) cut-off ${\epsilon'}^n=e^{-\ell/2}$ to match with the physical (target-space) cut-off $\epsilon=e^{-L/2}$. It follows that by treating (\ref{eq:master}) as a Gaussian integral of $\ell$ about the peak $\ell^*$ and including higher order corrections to (\ref{eq:path-integral-1}), one naturally produces the perturbation series of $\text{Tr}\rho^n_A\left(L, \Omega_\lambda\right)$ starting from the original CFT result $\text{Tr}\rho^n_A\left(L, \Omega_{CFT}\right)$. We remark that these statements are valid only for the sign of $\mu>0$. For the opposite sign of $\mu<0$, the distribution from (\ref{eq:S_0}) is pathological along the integration contour $\ell\in[0,\infty)$. 

\subsection{Quantum corrections from reparametrizations: one-loop analysis}

In this section, we study the quantum corrections $S_{m\geq 1}(L,\ell)$ to the kernel function in the perturbative expansion (\ref{eq:path-integral-1}). This amounts to integrating over small fluctuations near the saddle-points $\theta^*_{1,2}(\sigma) = \sigma$: 
\be 
\theta_{1,2}(\sigma) = \sigma + \delta \theta_{1,2}(\sigma)
\ee
where the fluctuations $\delta \theta_{1,2}(\sigma)$ are now restricted to real and periodic (i.e.~having zero winding number) functions of $\sigma$. The path-integral for the kernel function can now be written in the form: 
\bea\label{eq:fluctuations}
K(L,\ell,\lambda) &=& \int \mathcal{D} \delta \theta_1 \int \mathcal{D} \delta \theta_2\;e^{-\lambda^{-1} F\left(\sigma + \delta \theta_1(\sigma), \sigma+\delta \theta_2(\sigma)\right)} \nonumber\\
&=& e^{-\lambda^{-1}S_0(L,\ell)} \int \mathcal{D} \delta \theta_1 \int \mathcal{D} \delta \theta_2 \; e^{-I(\delta \theta)}\nonumber\\
I(\delta \theta)&=&\int d\sigma \int d\sigma' \delta \theta_i(\sigma) \,D^{ij}(\sigma,\sigma')\, \delta \theta_j(\sigma')+ \mathcal{O}(\delta \theta^3)
\eea
where in the last line we have truncated the effective action $I$ for $\delta \theta_{1,2}$ at the quadratic order. Due to the non-linear relation $X^{\pm} \sim e^{\pm in\theta_{1,2}(\sigma)}$ between the target-space fields $X^{\pm}$ and the boundary reparametrizations $\theta_{1,2}$, the effective action $I$ is an interacting theory for the fluctuations $\delta \theta_{1,2}$ despite coming from a free theory for $X^{\pm}$. In this paper, we focus only on the one-loop correction. The effects of the higher order corrections in analogous contexts were studied in \cite{Rychkov:2002ni}, and we will leave such analysis of (\ref{eq:fluctuations}) to future investigations.  

\subsubsection{Explicit expansion of \texorpdfstring{$I$}{I} to all orders}
We begin by finding the ``kinetic matrix" $D^{ij}(\sigma,\sigma')$ for the fluctuations $\delta \theta_{1,2}$. In fact, in our case it is possible to write down the interacting action $I$ explicitly to all orders of $\delta \theta$, thanks to the free theory nature of the original target-space CFT. We will work out this expression, even though the subsequent analysis is confined to the one-loop order. By introducing a matrix, the expression for 
$F(\theta_1,\theta_2)$ can be rewritten in the following more compact form.
\begin{equation}
    \begin{aligned}
        F(\theta_1,\theta_2)&=\frac{1}{4\pi}\sum_{i,j}\sum_{k=-\infty}^{\infty}\int_0^{2\pi}\dm \sigma\int_0^{2\pi}\dm\sigma'e^{\im n\theta_i(\sigma)}\alpha_k^{ij}e^{-\im k(\sigma-\sigma')}e^{-\im n\theta_j(\sigma')}\\
        \alpha_k^{ij}&=\begin{pmatrix}
                    \frac{k{\epsilon'}^{-2k}\epsilon^{2}}{\sinh{(k\ell/n)}} & -\frac{k}{\sinh{(k\ell/n)}} \\
                    -\frac{k}{\sinh{(k\ell/n)}} & \frac{k{\epsilon'}^{2k}\epsilon^{-2}}{\sinh{(k\ell/n)}}
                 \end{pmatrix}
    \end{aligned}
\end{equation}
Then we can perform the expansion of the boundary conditions.
\begin{equation}
    F=\sum_{i,j}\sum_{k=-\infty}^{\infty}\sum_{r,s=0}^{\infty}\frac{\im^{r-s}n^{r+s}}{r!s!}\frac{\alpha_k^{ij}}{4\pi}\int_0^{2\pi}\dm\sigma\int_0^{2\pi}\dm\sigma'\delta\theta_i^r(\sigma)\,e^{\im(n-k)\sigma}\delta\theta_j^s(\sigma')\,e^{\im(k-n)\sigma'}
\end{equation}
To proceed, we notice that the non-perturbative effective action $F(\theta_1, \theta_2)$, even though non-local in the spatial coordinates $\sigma$ and $\sigma'$, still enjoys translation invariance. It can therefore be simplified by performing the Fourier transformation on a circle $\sigma \in [0,2\pi]$: 
\begin{equation}
    \delta\theta_i(\sigma)=\sum_{p=-\infty}^{\infty}b_{ip}e^{\im p\sigma},\quad b_{ip}=\frac{1}{2\pi}\int_{0}^{2\pi}\dm\sigma\delta\theta_i(\sigma)e^{-\im p\sigma}
\end{equation}
This leads us to the final expression:
\begin{equation}
    F=\pi\sum_{i,j}\sum_{r,s=0}^{\infty}\sum_{p=q}\frac{\im^{r-s}n^{r+s}}{r!s!}\alpha_{p+n}^{ij}b^r_{i\vec{p}}b^{s*}_{j\vec{q}}.
\end{equation}
where $p=\sum_{i=1}^{r}p_i$, $q=\sum_{i=1}^{s}q_i$, $b^r_{i\vec{p}}=b_{ip_1}\dotsb b_{ip_r}$ and $b^s_{j\vec{q}}=b_{jq_1}\dotsb b_{jq_s}$. The linear term is absent in the expression above, which is consistent with the fact that we are expanding around a classical saddle.
Consequently, the expression for $I(\delta\theta)$ is:
\begin{equation}
    I(\delta\theta)=\frac{\pi}{\lambda}\sum_{i,j}\sum_{r,s=1}^{\infty}\sum_{p=q}\frac{\im^{r-s}n^{r+s}}{r!s!}\alpha_{p+n}^{ij}b^r_{i\vec{p}}b^{s*}_{j\vec{q}}.
\end{equation}
To second order, the result takes the following form:
\begin{equation}
    I^{(2)}(\delta\theta)=\frac{\pi n^2}{\lambda}\sum_{i,j}\sum_{p=-\infty}^{\infty}\left(\alpha^{ij}_{p+n}b_{i,p}b_{j,-p}-\alpha_n^{ij}b_{i,p}b_{i,-p}\right).
\end{equation}

\subsubsection{One-loop: UV divergence and RG flow}
Now we focus on computing the one-loop correction $S_1$, which amounts to calculating the log-determinant of the kinetic matrix $D^{ij}(\sigma,\sigma')$. As we shall see, the result is UV divergent and therefore requires the addition of counter-terms to make sense of. We can view such counter-terms as producing an RG flow of the effective action $F(\theta_1,\theta_2)$. We stress that despite the appearance of UV divergences, the path-integral is still independent of the Weyl scale factor of the world-sheet, i.e.~had we chosen to rescale the world-sheet annulus such that its boundaries become: 
\be
r_{in} = R'\epsilon', \;\; r_{out}= R'/\epsilon'
\ee
the effective action (\ref{eq:action}) remains the same. 
The UV divergence is only associated with the angular cut-off in the effective theory of boundary reparametrizations \eqref{eq:master}, and is therefore dimensionless. 

In the momentum representation, the determinant of $D^{ij}(\sigma,\sigma')$ is factorized into a product over distinct momenta: 
\be
\text{det} D^{ij}(\sigma,\sigma') = \prod_{k} \text{det}D^{ij}_k 
\ee
where the determinant for each of the factors is given by:
\bea
\text{det}D^{ij}_k &=& \frac{8(n^2-k^2)\sinh^2{(k\ell/n)}}{\cosh{(2\ell)}-\cosh{(2k\ell/n)}} + \frac{8n^2(1-\cosh{(\ell-L)})}{\sinh^2{(\ell)}}\nonumber\\
&-& \frac{8n(n+k)\left(\cosh{(k\ell/n)}-\cosh{\left(k\ell/n+\ell-L\right)}\right)}{\cosh{\left((2+k/n)\ell\right)}-\cosh{(k\ell/n)}}\nonumber\\
&-& \frac{8n(n-k)\left(\cosh{(k\ell/n)}-\cosh{\left(k\ell/n-\ell+L\right)}\right)}{\cosh{\left((2-k/n)\ell\right)}-\cosh{(k\ell/n)}}
\eea
We begin by noticing that $\text{det} D^{ij}_k=0$ for $k=0$. This implies that among the eigen-modes of the fluctuations, there exists a zero-mode with $k=0$. Physically the zero-mode comes from the invariance of the effective  action (\ref{eq:action}) under equal translation in $\theta_{1,2}$ along both the inner and outer boundaries:
\be \label{eq:zero_mode}
\theta_{i}(\sigma) \to \theta_{i}(\sigma)+ a,\;\;\;i=1,2 
\ee
It simply performs a global translation of the saddle-point solution. In contrast, performing equal but opposite translations: 
\be 
\theta_{i}(\sigma) \to \theta_{i}(\sigma)+ (-1)^i a,\;\;i=1,2
\ee
corresponds to a massive mode with the eigenvalue $d_0 = \frac{2n}{\sinh{(\ell)}}$ in the $k=0$ sector. Unlike the massive modes, the zero-mode needs to be removed from the Gaussian integrals of fluctuations, and instead serves as a collective coordinate, e.g.~labeled by the amount of translation $a\in [0,2\pi]$ in (\ref{eq:zero_mode}). It needs to be integrated separately:
\be 
K(L,\ell,\lambda)= e^{-\lambda^{-1}S_0(L,\ell)} \int^{2\pi}_0 da N(a) \int \mathcal{D}\overline{\delta \theta_1} \mathcal{D}\overline{\delta \theta_2}\,e^{-S'_1+\mathcal{O}(\lambda)}
\ee
where $S'_1$ denotes the quadratic term excluding the zero-mode, and the path-integral $\mathcal{D}\overline{\delta \theta_{1,2}}$ has also removed the zero-mode component. The integration measure $N(a)$ is determined by the metric of the configuration space, whose standard definition takes the form:
\be\label{eq:metric}
\langle \delta \psi, \delta \varphi\rangle = \frac{n^2}{\lambda}\left(\int^{2\pi}_0 d\sigma \,\delta \psi^{*}_1(\sigma)\,\delta \varphi_1(\sigma)+\int^{2\pi}_0 d\sigma \,\delta \psi^{*}_2(\sigma)\,\delta \varphi_2(\sigma) \right)
\ee
where $\delta \psi = (\delta \psi_1,\delta \psi_2)$ and $\delta \varphi = (\delta \varphi_1,\delta \varphi_2)$ are two tangent vectors of  the configuration space. The pre-factor in (\ref{eq:metric}) is the coupling of the effective action, which enters the metric of the configuration space.
The integration measure $N(a)$ for the constant zero-mode is then given by: 
\be
N(a) = \langle \mathds{1}, \mathds{1}\rangle^{1/2} = \frac{n}{\sqrt{\lambda}}
\ee

Integrating first over the zero-mode, and then over the remaining massive modes gives:
\bea\label{eq:S1_bare}
K(L,\ell,\lambda) &=& \frac{2\pi n}{\sqrt{\lambda}} e^{-\lambda^{-1}S_0(L,\ell)} \times \exp{\left(- \frac{1}{2}\ln \prod_{d_i\neq 0} d_i +\mathcal{O}(\lambda)\right)}\nonumber\\
 \Longrightarrow\quad
S_1(L,\ell)&=& \frac{1}{2}\ln{\left(\frac{\lambda d_0}{4\pi^2n^2}\right)}+\frac{1}{2} \sum_{k\geq 1} \ln{\text{det} D^{ij}_k}
\eea
This is potentially UV divergent due to the infinite sum over the momenta $k\to \infty$. To isolate the divergence, we compute the large $k$ expansion of $\text{det} D^{ij}_k$:
\bea\label{eq:large_k_det} 
\text{det} D^{ij}_k  &=& 4k^2 \left(1-\frac{\Delta S}{k} + ...\right)
\quad\longrightarrow\quad \ln{\text{det} D^{ij}_k} = \ln{(4k^2)}- \frac{\Delta S}{k}+ ...\nonumber\\
\Delta S &=&\frac{4n\sinh{\left(\frac{\ell-L}{2}\right)}\sinh{\left(\frac{\ell+L}{2}\right)}}{\sinh{(\ell)}}
\eea
where $...$ represent terms that decrease sufficiently fast such that they converge upon summing over $k\to \infty$. Such terms include both power-law decays, e.g.~$\sim 1/k^2$ and exponential decays, e.g.~$\sim e^{-k\ell/n}$. The terms kept in (\ref{eq:large_k_det}) give rise to the following divergent terms:
\be\label{eq:S1_divergences}
\frac{1}{2}\sum^{\Lambda}_{k=1}\ln{\text{det}D^{ij}_k} \sim \left(\Lambda+\frac{1}{2}\right)\ln{(2\Lambda)}-\Lambda -\left(\frac{\Delta S}{2}\right)\ln{\Lambda} 
\ee
where $\Lambda$ is the UV cut-off for the discrete momentum $k$. To cancel these divergences, one needs to introduce a set of counter-terms. They appear by splitting the ``bare" effective action (\ref{eq:action}) into a sum of the ``renormalized" part and a set of $\Lambda$-dependent counter-terms:
\be
\lambda^{-1}F(\theta_1,\theta_2) = \lambda_{ren}^{-1}F_{ren}(\theta_1,\theta_2) + \text{counter-terms}
\ee
such that UV divergences from the loop integral/summation of the renormalized part are canceled by the counter-terms. From now on we only work with the renormalized part, and so will omit the ``ren" subscripts. 

The first two terms in (\ref{eq:S1_divergences}) come from the sum over the $\ln{(4k^2)}$ term in (\ref{eq:large_k_det}). They can be absorbed into a one-loop counter-term $\delta F_0$ for the ``cosmological constant" $F_0$ of the effective action $F$:
\be
\delta F_0  =\lambda \Lambda - \lambda\left(\Lambda+\frac{1}{2}\right)\ln{(2\Lambda)} + ... 
\ee
where ... denotes the finite-part of $\delta F_0$ that will be discussed later. 

The remaining log divergence in (\ref{eq:S1_divergences}) is more interesting. From the perspective of the full effective action $F(\theta_1,\theta_2)$, the expansion (\ref{eq:path-integral-1}) is formally analogous to the perturbative expansion about the ``instanton" background (\ref{eq:half_line_saddle}), with the winding number one playing the role of the topological number. To extend the analogy further, one can imagine relaxing the constraints on the winding number, then there are more saddle-points coming from the other winding sectors. For example, the constant solution $\theta^*_{1,2}(\sigma)=\theta_0$ can be viewed as the ``vacuum" sector  saddle-point.  

In a renormalizable theory, the log divergence arising from the one-loop correction in the instanton background has the same effect as renormalizing the coupling constants that appear in the leading-order instanton-action. The renormalization is consistent with the standard RG analysis. e.g. via perturbation theory in the vacuum sector. For example, in computing the QCD instanton partition function the quantum correction from the one-loop determinant on the instanton background gives rise to a log divergence that is proportional to the classical action, effectively renormalizing the coupling constant in the classical instant action: 
\be
\frac{1}{g_{bare}}S_0+\left(\beta_g\ln{\Lambda}\right)S_0 + ... \to \frac{1}{g(\rho^{-1})}S_0 + ...
\ee
where $\beta_g$ coincides with the one-loop QCD beta function, and $g(
\rho^{-1}
)$ is the running coupling evaluated at the IR cut-off of the instanton background, in this case the inverse of the instanton size $\rho$. 

If the effective action $F(\theta_1,\theta_2)$ is renormalizable, then the one-loop divergence $\frac{\Delta S}{2}\ln{\Lambda}$ in (\ref{eq:S1_divergences}) can be similarly absorbed by the counter-terms of the coupling constants in the effective action. While due to the apparent non-local form (\ref{eq:action}) it is unclear whether $F(\theta_1,\theta_2)$ constitutes a renormalizable theory, we will proceed under the assumption that it does. On the one hand, the translation-invariance of (\ref{eq:action}) means that the effective action is not very different from the usual renormalizable theories when analyzed in the momentum space; on the other hand, the renormalizability of a very similar effective action was explicitly illustrated in the more careful analysis of \cite{Rychkov:2002ni}, thus giving us more confidence about the assumption. We leave a systematic examination of the renormalizability of (\ref{eq:action}) to the future. Following the analogy with the QCD instanton calculation, we expect that: 
\be\label{eq:log_div_counter}
\frac{\Delta S}{2}\ln \Lambda = -\lambda^{-2} \delta \lambda\; S_0 + \lambda^{-1}\delta L\left( \partial_L S_0\right)
\ee
where $\delta \lambda$ and $\delta L$ are the counter-terms for the effective coupling $\lambda$ as well as the target-space cut-off $\epsilon = e^{-L/2}$, now also viewed as a coupling in the effective theory. From the expressions for $S_0$ and $\Delta S$ we can solve the counter-terms in (\ref{eq:log_div_counter}) to be:
\be
\delta \lambda = -\frac{\lambda^2}{2\pi}\cosh{(L)} \ln{\Lambda} + \cdots,\quad\delta L =-\frac{\lambda}{2\pi } \sinh{(L)}\ln{\Lambda} + \cdots
\ee
where ${\cdots}$ again denotes the finite terms. The divergent terms of the counter-terms dictate the one-loop $\beta$ functions for the RG flows of the renormalized couplings:
\bea\label{eq:RG}
\beta_{\lambda} &=& E\,\partial_{E} \lambda(E) = -\frac{\lambda^2}{2\pi} \cosh{L} \nonumber\\
\beta_{L} &=& E\,\partial_{E} L(E) = -\frac{\lambda}{2\pi } \sinh{L}
\eea
where $E$ is the sliding scale.
From the RG perspective, one can easily read out from (\ref{eq:RG}) an important property of the effective theory $F(\theta_1,\theta_2)$, that it is ``asymptotically free" \footnote{This is true only for the positive sign of the $T\bar{T}$ deformation parameter $\mu>0$; for the opposite sign $\mu<0$, the RG behaviors in the UV and IR simply switch. }, i.e. the effective couplings $\lambda$ flows to zero in the UV $(E\to \infty)$ together with $L$.
On the other hand, the effective theory flows to the strong-coupling regime in the IR $(E\to 0)$, possibly resulting in a phase transition where $1/\lambda = R^2/(2\pi \mu) \to 0$. The scale at which this happens depends on the renormalization conditions. From the point of view of the target-space, now as a manifold that changes with the RG scale, the phase transition can be depicted as the boundaries of target-space annulus ``pinching off" when $R\to 0$. 

From these discussions we conclude that the log divergences $(\ln{\Lambda})\Delta S/2$ in the one-loop correction amounts to replacing the parameters $(\lambda,\epsilon = e^{-L/2})$ in the leading order term $\lambda^{-1}S_0(L,\ell)$ by the running couplings according to (\ref{eq:RG}). To complete the analogy with the QCD instanton example we need to determine the physical scale at which to evaluate these running couplings. In the QCD instanton example this is set by the IR scale $\rho^{-1}$ from the size $\rho$ of the instant background. It is not difficult to see that in (\ref{eq:action}) there are two such IR scales. In our case these scales are dimensionless numbers because they concern the angular variables $\sigma$. One of them is determined by the finite ``angular size" of the circle $\sigma\in [0,2\pi]$, or equivalently by that the sum over momenta $k$ is discrete and only includes integers, so we can effectively treat this IR scale as simply 1. The other scale is given by $\ell^{-1}$, which controls the exponential corrections $e^{-k\ell/n}$ to the large $k$  approximation (\ref{eq:large_k_det}) \footnote{In this paper, we do not consider cases where $n$ is parametrically large.}. In some sense $\ell^{-1}$ acts like a temperature, which is a standard IR cut-off in finite temperature QFTs. Based on these, we propose to set the physical scale $E$ at which we evaluate the running parameters $\lbrace \lambda(E),L(E)\rbrace$ by the larger one of the two scales just mentioned: 
\be
E_{\ell} = \text{max}\lbrace 1, \ell^{-1}\rbrace
\ee

\subsubsection{One-loop: finite parts and renormalization conditions}
While the $\Lambda$-dependent parts of the counter-terms serve to remove the UV-divergences of the loop summation, the explicit form of the renormalized expression for $S_1(L,\ell)$ is fixed by the finite parts of the counter-terms, which we now determine. 

In the standard RG analysis, the finite parts of the counter-terms are fixed by imposing a set of renormalization conditions. These usually take the form of specifying the values of the running couplings at a particular physical scale $M$. In this way they fix the integration-constants of the RG flow equations, which are essentially the finite parts of the counter-terms.   

In our case, there are three renormalized parameters: the cosmological constant $F_0$, the effective couplings $\lambda$, and the target-space cut-off $\epsilon=e^{-L/2}$. Based on the discussions so far, they enter the renormalized kernel function as: 
\be\label{eq:kernel_ren}
K(L,\ell,\lambda) = \exp{\left(-\frac{F_0(\ell)}{\lambda(E_\ell)} - \frac{4\pi n}{\lambda(E_{\ell})} \frac{\sinh^2{\left(\frac{\ell-L(E_{\ell})}{2}\right)}}{\sinh{(\ell)}}\right)}
\ee
We shall impose the renormalization conditions by specifying their values at a particular scale of (\ref{eq:kernel_ren}), in this case represented by $\ell$. A very natural choice is based on the connection between  the string partition function (\ref{eq:master}) and the un-deformed CFT results in the limit of $\lambda \to 0$: 
\be\label{eq:CFT_limit}
\lim_{\lambda \to 0} \text{Tr}\rho^n_A \left(L, \Omega_\lambda\right) \to \text{Tr} \rho^n_A\left(L,\Omega_\mu\right)
\ee
where we stress that $(\lambda,L)$ in (\ref{eq:CFT_limit}) are the physical parameters in contrast to their running counter-parts in (\ref{eq:kernel_ren}). This is equivalent to imposing the following condition for the kernel function: 
\be\label{eq:ren_cond_1}
\lim_{\lambda\to 0} K(L,\ell,\lambda) \to 2\pi L\;\delta(\ell-L)
\ee
This fits well with the leading-order result $K(L,\ell,\lambda) \approx e^{-\lambda^{-1}S_0(L,\ell)}$, which as analyzed in (\ref{eq:gaussian_leading}) is approximately a sharp Gaussian distribution near $\ell^*=L$. Expressing the $\delta$-function as a limit of Gaussian distributions, the condition (\ref{eq:ren_cond_1}) requires that the renormalized kernel function near $\ell = L$ should take the form: 
\be\label{eq:ren_cond_2} 
K(L\approx \ell,\lambda) = L\sqrt{\frac{2\pi}{\delta}}\; \exp{\left[-\frac{(\ell-L)^2}{2\delta} + \mathcal{O}(\lambda)\right]},\quad \delta = \frac{\lambda \sinh{(L)}}{2\pi n}
\ee
While the higher order terms in (\ref{eq:ren_cond_2}) are not fixed by (\ref{eq:ren_cond_1}), the terms up to the one-loop order are uniquely fixed. In terms of the renormalized constants, (\ref{eq:ren_cond_1}) amounts to imposing a renormalization condition for the running cosmological constant $F_0(\ell)$ that fixes its value at $\ell=L$:
\be\label{eq:ren_cond_3}
F_0(L) = \frac{\lambda}{2}\ln{\left(\frac{\lambda \sinh{(L)}}{4\pi^2L^2n}\right)},\;\;\lambda(E_L)=\lambda,\;\;L(E_L)=L 
\ee
In other words, the running parameters $\lbrace\lambda(E),L(E)\rbrace$ are required to match the physical parameters $(\lambda,L)$ at $E=E_L$; while the cosmological constant $F_0(\ell)$ at $\ell=L$ produces the correct normalization of the Gaussian distribution near $\ell\approx L$ so that (\ref{eq:ren_cond_1}) holds.

We remark that while (\ref{eq:ren_cond_1}) may appear natural for the renormalization condition that fixes the finite part of the running cosmological constant $F_0(L)$, it is only a choice. It only affects the overall normalization of $K$ and consequently the replica partition function $\text{Tr} \rho_A^n$. The normalization factor itself does not appear in physical observables. For example, we can modify the choice by adding an arbitrary constant term to $F_0$ that is proportional to the Renyi index $n$: 
\be\label{eq:cc_density}
F_0(L) \to F_0(L) + n \times \left(\text{const}\right)
\ee
Note that the angular range is proportional to $n$, so physically (\ref{eq:cc_density}) amounts to a shifting of the angular density for the cosmological constant. It is easy to see that such modifications (\ref{eq:cc_density}) cancel out in physical quantities, e.g.~the Renyi entropy: 
\be
S_n \propto \ln{\text{Tr}\rho_A^n}-n \ln{\text{Tr} \rho_A}
\ee

Based on these renormalization conditions, the finite parts in the counter-terms $\delta \lambda$ and $\delta L$ can be specified implicitly by requiring that the trajectory of the RG flow (\ref{eq:RG}) satisfies (\ref{eq:ren_cond_3}). The finite part in the cosmological counter-term $\delta F_0$ requires more work. As a counter-term it should not depend on the world-sheet scale $\ell$ and instead only on $(\lambda,L)$. To impose the renormalization condition (\ref{eq:ren_cond_3}) we need to compute the $\ell$-dependent quantum corrections (with the UV-divergences removed) to the cosmological constant $F_0$, analogous to the thermal correction to the free energy in finite temperature QFTs. To proceed, we split the sum over $k$ in (\ref{eq:S1_bare}) into two parts: those of small $k$ and those of large $k$: 
\bea 
&&S_1(L,\ell)= \frac{1}{2}\ln{\left(\frac{\lambda d_0}{4\pi^2n^2}\right)}+\frac{1}{2} \sum_{k\geq 1} \ln{\text{det} D^{ij}_k}\nonumber\\
&=& \left(\frac{1}{2}\ln{\left(\frac{\lambda d_0}{4\pi^2n^2}\right)}+\frac{1}{2}\sum_{k\leq k_c} \ln{\text{det} D^{ij}_k}\right) + \frac{1}{2} \sum_{k\geq k_c} \ln{\left[4k^2 \left(1-\frac{\Delta S}{k} + ...\right)\right]}
\eea
where we recall that $d_0 = \frac{2n}{\sinh{(\ell)}}$ is the other non-zero eigenvalue in the $k=0$ sector. The split is such that for $k\leq k_c$, the behavior of $\text{det} D^{ij}_k$, due to its $\ell$-dependence, is qualitatively distinct from the large $k$ expression of (\ref{eq:large_k_det}). As can be analyzed in the appendix (\ref{app:F0_limits}), in the $\ell\gg1$ limit $k_c$ is sharply specified at $k_c = n$; while in the $\ell \ll 1$ limit it is only parametrically given by the scaling $k_c \sim n \ell^{-1}$. In analogy with finite temperature QFTs, the $k\leq k_c$ contributions represent the ``thermal" modifications to the constant part of an otherwise ``zero-temperature" result. The modifications are well-defined at ``low temperatures" $\ell\gg 1$ (modulo exponentially suppressed terms $\propto e^{-n\ell}$) due to the sharpness of $k_c$; while as is often the case when describing thermal effects at ``high temperatures", they are only parametrically defined for $\ell\ll 1$. Fortunately, the parametric ambiguity of the results in the $\ell \ll 1$ regime is not important for our subsequent saddle-point analysis -- only the $\ell \gg 1$ regime is relevant. In practice we can identify the thermal corrections with the corrections to the cosmological constant: 
\be\label{eq:cc_finite}
\lambda^{-1}F_0(\ell) = \frac{1}{2}\ln{\left(\frac{\lambda d_0}{4\pi^2n^2}\right)}+\frac{1}{2}\sum_{k\leq k_c} \ln{\text{det} D^{ij}_k} + \lambda^{-1}\left(\delta F_0\right)_{finite}
\ee
At the one-loop order, the difference between the running coupling $\lambda(E_\ell)$ and the physical coupling $\lambda$ is of higher-order, so we have simply written $\lambda$ in (\ref{eq:cc_finite}).  Analyzing (\ref{eq:cc_finite}) is difficult for general $\ell$. It is sufficient for our purpose to obtain the leading order results of (\ref{eq:cc_finite}) in the ``high temperature" limit of $\ell \ll 1$ and the ``low temperature" limit of $\ell \gg 1$. We leave the details of the analysis to the appendix (\ref{app:F0_limits}) and simply write down the expression for the renormalized kernel function at one-loop order: 
\be\label{eq:kernel_oneloop}
K(L,\ell,\lambda) \approx \exp{\left(-\frac{4\pi n \sinh^2{\left(\frac{\ell-L}{2}\right)}}{\lambda \sinh{(\ell)}}\right)} \times
\begin{cases}
e^{\frac{n+1}{2}(\ell-L)-\frac{L}{2}},\;\;\;\,\;\;\;\;\; \;\ell \gg L\\
e^{n(\ell-L) -\frac{L}{2}},\;\;\;\;\;\;\;\;\;1\ll \ell \ll L\\
e^{-\frac{nL}{2\ell}},\;\;\;\;\;\;\;\;\;\;\;\;\;\;\;\;\;\;\;\;\ell\ll 1
\end{cases}
\ee
where in the exponent we have omitted all sub-leading terms that are not important for our subsequent analysis regarding non-perturbative effects, e.g.~in addition to the various constant terms, we have disregarded the finite counter-term in the $\ell \ll 1$ case and the $\ln{(\ell)}$ terms in the $\ell\gg 1$ cases. 

We make the observation that the one-loop correction from the reparametrization modes exhibits a duality $\ell \to 1/\ell$ between the $\ell\gg 1$ and $\ell\ll 1$ limits, albeit not as a self-duality. This behavior resembles that of the free energy in a 2d BCFT under modular $S$ transformation \cite{Cardy:1989ir,Blumenhagen:2009zz}, suggesting the possibility of a more standard BCFT formalism for what we have studied regarding the target-space CFT. We leave this as an interesting follow-up direction to pursue in the future.  

We wrap up by discussing the behavior of the running parameters $\lbrace\lambda(E_{\ell}), L(E_{\ell})\rbrace$ in (\ref{eq:kernel_ren}). It is easy to realize that for $\ell\gg 1$, the physical scale $E$ at which to evaluate these running parameters is frozen: 
\be
E_{\ell \gg 1} = \text{max}\lbrace 1, \ell^{-1}\rbrace = 1
\ee
In fact, we can conclude from here that the $\ell$-dependent running couplings appearing in the string partition function (\ref{eq:master}) remain in the weakly coupled region. For $\ell \gg 1$, they are frozen and stay the same with physical ones imposed at the renormalization scale $\ell = L \gg 1$, which were taken to be weakly-coupled $\lambda \ll 1$: 
\be
\lambda (E_{\ell}) = \lambda, \;\;\; L(E_{\ell})=L,\;\;\; \text{for all $\ell \gg 1$.}
\ee
Recall that in a generic RG flow of (\ref{eq:RG}) there may be a strong-coupling phase transition towards the IR. It does not occur in the string partition function of our interest because the renormalization condition is imposed at a scale $\ell = L \gg 1$ where the running has already effectively frozen. For the other limit of $\ell \ll 1$, the physical scale $E_{\ell} = \ell^{-1}$ moves towards the UV, and the couplings run further towards the weak-coupling limit than the physical ones: 
\be
\lambda (E_{\ell})^{-1} \sim \lambda^{-1}- \left(\frac{e^{L}}{4\pi}\right)\ln{\ell},\;\;\; L(E_{\ell}) \sim L +\left(\frac{\lambda e^{L}}{4\pi}\right)\ln{\ell}
\ee
Though the RG flow generates dependence on $\ell$ for the running parameters in the $\ell \ll 1$ limit, it is only logarithmic. In the absence of phase-transitions we do not expect such mild $\ell$-dependence to play any significant part in the non-perturbative aspects of the string partition function (\ref{eq:master}). For this reason when performing the non-perturbative analysis later in the section, we will neglect the running of the parameters $\lbrace\lambda(E_\ell),L(E_{\ell})\rbrace$ and only refer to their physical values whenever they appear in the string partition function. 

\subsubsection{Summary of renormalization prescriptions}
As it turns out, the one-loop analysis depends on renormalization treatments that may appear overwhelming at first pass. For the convenience of the readers, we end the one-loop analysis by gathering and summarizing various statements regarding the renormalization prescriptions we have adopted in this section. 
\begin{itemize}
\item The one-loop correction $S_1(L,\ell)$ is divergent due to the UV  contributions in the angular frequency.  
\item The divergences can be absorbed by adding the counter-terms $(\delta \lambda, \delta L, \delta F_0)$ to the effective couplings $\lambda, L$ and the ``cosmological constant" to the effective action $F_0$, making them RG-dependent. 
\item The beta functions of the running couplings are described by (\ref{eq:RG}), which reveals the asymptotically free nature of the underlying RG dynamics.
\item We fix the finite parts of the running parameters using the following renormalization conditions imposed at the physical scale $\ell= L$: (i) the running couplings $\lbrace\lambda(E),L(E)\rbrace$ at $E_{\ell}$ are equal to the physical values $\lbrace \lambda, L\rbrace$; (ii) the cosmological constant term $F_0$ is set such that the kernel approaches the $\delta$-function $\delta(\ell-L)$ with the correct normalization in the $\lambda \to 0$ limit. 
\item The renormalization condition on $F_0$ only affects the normalization of the replica partition function, and is not of direct physical relevance. It can be relaxed in ways that do not affect physical observables. 
\item For $L\gg1 $, the renormalization condition is set at a scale where the couplings $\lbrace\lambda(E),L(E)\rbrace$ have effectively stopped running due to the finite-size effect of the circular boundaries. In terms of computing the partition function, this saves us from the IR instabilities in an otherwise asymptotically free theory.    
\item When evaluating the action $S_1(\ell, L)$, we simply set the running couplings $\lambda(E),L(E)$ to the physical values. This is sufficient at the one-loop order. 
\end{itemize}

\subsection{String partition function: non-perturbative analysis}\label{sec:saddle}
We now perform the final integration in (\ref{eq:master}) over the world-sheet cut-off of $\left(\epsilon'\right)^n  = e^{-\ell/2}$ to evaluate the replica string partition function at the physical cut-off, i.e. the target-space cut-off of $\epsilon = e^{-L/2}$: 
\be\label{eq:master2}
\text{Tr}\rho^n_A\left(L,\Omega_\lambda\right) =  \int \frac{d\ell}{2\pi \ell}\; e^{-\lambda^{-1} \sum_{m\geq 0} \lambda^m S_m(L,\ell)}\; \text{Tr}\rho^n_A\left(\ell,\Omega_{CFT}\right)
\ee
where the kernel function is evaluated via the perturbative expansion (\ref{eq:path-integral-1}). We are mainly interested in analyzing the non-perturbative aspects of (\ref{eq:master2}),  which can be formally organized into the so-called trans-series \cite{ecalle1981fonctions,Marino:2012zq,Sauzin:2014qzt,Aniceto:2018bis}: 
\be\label{eq:trans-series} 
\text{Tr}\rho^n_A\left(L,\Omega_\lambda\right) = \sum_{i} e^{-\lambda^{-1}\mathcal{K}_i(n,L)} \left(\sum_{j\geq 0} a^j_i(n,L) \lambda^{j} \right) 
\ee
The index $i$ in the trans-series label the non-perturbative sectors of (\ref{eq:master2}), usually represented by distinct saddle-points of the $\ell$ integral in (\ref{eq:master2}); for each of the non-perturbative sectors $i$, there is a corresponding perturbative expansion whose coefficients we write as $a^j_i(n,L)$. The different sectors and their perturbative expansions are not completely independent, they can be related to one another by the powerful resurgence analysis \cite{Marino:2012zq,Sauzin:2014qzt,Aniceto:2018bis}, see \cite{Gu:2024ogh,Gu:2025tpy} for such results related to the $T\bar{T}$-deformed torus partition functions, and \cite{Hirano:2025alr} for the two-point functions.

In this paper we focus on the non-perturbative effects at their leading perturbative orders -- we only include the leading-order terms from each of the non-perturbative sectors in the trans-series. For this purpose, it is sufficient to truncate the expansion of (\ref{eq:path-integral-1}) to the one-loop order: 
\be
K(L,\ell,\lambda) \approx e^{-\lambda^{-1}S_0(L,\ell)
- S_1(L,\ell)}
\ee
The higher order terms vanish in the limit of $\lambda\to 0$, and are expected to only affect the perturbative coefficients of the trans-series. Plugging in the one-loop expression (\ref{eq:kernel_oneloop}), we land on the following explicit integral: 
\bea\label{eq:final_integral_halfline}
\text{Tr}\rho^n_A\left(L,\Omega_\lambda\right) &\sim &  \int^\infty_0 \frac{d\ell}{2\pi \ell}\; e^{-\frac{4\pi n}{\lambda}\left(\frac{\sinh^2\left(\frac{\ell-L}{2}\right)}{\sinh{(\ell)}}\right)}\;e^{-S_1(L,\ell)}\; \text{Tr}\rho^n_A\left(\ell,\Omega_{CFT}\right) \nonumber\\
e^{-S_1(L,\ell)} &\sim &
\begin{cases}
e^{\frac{n+1}{2}(\ell-L)-\frac{L}{2}},\;\;\;\;\;\;\; \;\ell \gg L\\
e^{n(\ell-L) -\frac{L}{2}},\;\;\;\;\;\;\;\;\;1\ll \ell \ll L\\
e^{-\frac{n L}{2\ell}},\;\;\;\;\;\;\;\;\;\;\;\;\;\;\;\;\;\;\ell\ll 1
\end{cases}
\eea

Our final goal in this section is to perform a saddle-point analysis for the integral (\ref{eq:final_integral_halfline}). Notice that both $e^{S_1(L,\ell)}$ and $\text{Tr}\rho^n_A\left(\ell,\Omega_\lambda\right)$ are independent of $\lambda$. A standard saddle-point analysis therefore proceeds by identifying and extremizing the exponent of the integrand that is leading order in small $\lambda$, in this case $\lambda^{-1}S_0(L,\ell)$. As was computed in (\ref{eq:ordinary_saddle}), solving the corresponding saddle-point equation gives: 
\be
\partial_\ell S_0(L,\ell) =0 \to \ell^* = \pm L
\ee
The solution $\ell^*=-L$ is excluded. It occurs at a negative and thus un-physical value. Mathematically it does not lie on the Lefschetz thimble associated with the integration contour $\ell \in [0,\infty)$, which for $\lambda>0$ is the integration contour itself. As a result, the saddle-point $\ell^*=-L$ does not contribute to the string partition function for $\lambda>0$. Expanding the integral near the physical saddle-point $\ell^* = L$  produces the perturbative sector of the trans-series: 
\be\label{eq:perturbation_Series}
\text{Tr} \rho^n_A\left(L, \Omega_\lambda\right) \sim \text{Tr} \rho^n_A \left(L, \Omega_{CFT}\right)\left(1+ \mathcal{O}(\lambda)\right) + ...
\ee
The perturbation series is independently computable using purely field theoretical techniques, e.g.~by applying the conformal perturbation theory to the CFT replica trick computation \cite{Chen:2018eqk}. In principle one can use the perturbative results for the task of benchmarking, e.g.~by comparing the perturbation series (\ref{eq:perturbation_Series}) using (\ref{eq:master}) with those from using the conformal perturbation theory. Doing this exercise successfully will greatly enhance the confidence in the validity of (\ref{eq:master}). In practice, this requires specifying the higher-order corrections to the kernel function in (\ref{eq:path-integral-1}). As we have seen for the one-loop term, due to the divergences and RG flow these corrections in general depend on the renormalization conditions. On the field theory side, the conformal perturbation theory computation encounters divergences too, and thus also requires regularization that is equivalent to imposing renormalization conditions. In other words, both computations (stringy v.s.\ conformal perturbation theory) are plagued with loose ends from treating the divergences, and it is unclear at this point how to relate them. In addition, the field theory computations often do not treat the bare cut-off as explicitly as in this paper, making the comparison even more obscure.\footnote{We thank Peng-Xiang Hao for helpful exchanges regarding possible comparisons with \cite{Chen:2018eqk}.} We therefore acknowledge that benchmarking (\ref{eq:master}) with field theory results is beyond the scope of the current investigation. This is the case even for simple CFTs such as the free boson, because the main subtleties lie in the target-space sector. We postpone this as an important future project to carefully translate our treatment into explicit field theory prescriptions so that a systematic benchmarking with (\ref{eq:master}), either perturbatively or numerically, can be performed. 

The key question is whether there are other saddle-points in (\ref{eq:final_integral_halfline}). Naively it is easy to see that we have exhausted the saddle-points thanks to the simplicity of the expression for $S_0(L,\ell)$. We point out more subtle mechanisms for saddle-points to arise. Let us write the integral (\ref{eq:final_integral_halfline}) more compactly as: 
\bea\label{eq:Saddle Point Analysis}
F(\lambda) &=& \int \frac{d\ell}{2\pi \ell}\; e^{-\lambda^{-1}A_0(\ell)-A_1(\ell)}\nonumber\\[.5ex]
A_0(\ell)&=& S_0(L,\ell), \;\;A_1(\ell) = S_1(L,\ell)-\ln{\text{Tr} \rho^n_A\left(\ell,\Omega_{CFT}\right)}
\eea
In addition to the standard way of first solving the leading-order saddle-point equation $A_0'(\ell^*)=0$ and then computing perturbative corrections to it, a more general way proceeds by finding $y$ where:
\be\label{eq:new_saddle_key}
\lim_{\ell\to y}A'_1(\ell)/A'_0(\ell)\to\infty
\ee 
Other than when $A'_0(y)=0$, the condition (\ref{eq:new_saddle_key}) can also be satisfied in other ways. For example it can occur at a singularity $y$ of the sub-leading term when $A'_1(y)\to \infty$. Alternatively (\ref{eq:new_saddle_key}) can be satisfied asymptotically, e.g.:
\be 
\lim_{\ell\to \lbrace\infty,0\rbrace}A'_1(\ell) \to \text{const},\;\;\lim_{\ell \to \lbrace\infty,0\rbrace} A'_0(\ell)\to 0
\ee
In any case, when (\ref{eq:new_saddle_key}) holds, a competition between the leading order term $\lambda^{-1}A_0'(\ell)$ and the sub-leading order term $A_1'(\ell)$ can be maintained near $y$ and gives rise to a solution $\ell^*$ to the saddle-point equation that satisfies: 
\be
A'_1(\ell^*)/A'_0(\ell^*) = -\lambda^{-1},\;\;\;\lim_{\lambda \to 0}\ell^* \to y 
\ee

Now let us explore whether for the string partition function  (\ref{eq:final_integral_halfline}) the condition (\ref{eq:new_saddle_key}) holds in addition to the solutions $y = \pm L$. In the absence of phase-transitions for both $S_1(L,\ell)$ and $\ln{\text{Tr} \rho^n_A\left(\ell,\Omega_\mu\right)}$, which is what will cause singularities for $A'_1(\ell)$, we are naturally led to examine the asymptotic behaviors of $A'_0(\ell)$ and $A'_1(\ell)$. Those of the $A'_0(\ell)$ and the $S'_1(L,\ell)$ term in $A'_1(\ell)$ can be readily obtained from (\ref{eq:kernel_oneloop}):
\bea 
&&\lim_{\ell\to 0} A'_0(\ell)  \sim -\left(\frac{\pi n e^{L}} {\ell^2}\right),\;\;\;\lim_{\ell\to \infty} A'_0(\ell) \sim 4\pi n e^{-\ell}\nonumber\\
&&\lim_{\ell \to 0} S'_1(L,\ell) \sim -\left(\frac{n L}{2\ell^2}\right),\;\;\lim_{\ell \to \infty} S'_1(L,\ell) \sim -\left(\frac{n+1}{2}\right)
\eea
where we have used the expression of $S_1(\ell,L)$ in the regime of $\ell \gg L$ -- we will check later if it holds self-consistently for the saddle-point solution. Regarding the CFT term $\ln{\text{Tr} \rho^n_A\left(\ell,\Omega_{CFT}\right)}$, the $\ell\to \infty$ limit is given universally by the leading order result of the single-interval Renyi entropy:
\be\label{eq:CFT_renyi_1}
\lim_{\ell \to \infty} \ln{\text{Tr}\rho^n_A\left(\ell,\Omega_{CFT}\right)} \sim \frac{c \ell}{12 n} 
\ee
where $c$ is the central charge of the un-deformed CFT. The other limit $\ell \to 0$ is rarely discussed in the context of entanglement structure because it corresponds to the cut-off hole eating up the full complex-plane. However, viewed as a BCFT partition function, $\text{Tr}\rho^n_A\left(\ell,\Omega_{CFT}\right)$ at small $\ell$ corresponds to the low-temperature limit in the open-string channel: 
\be\label{eq:CFT_renyi_2}
\lim_{\ell \to 0}\text{Tr}\rho^n_A\left(\ell,\Omega_{CFT}\right) = \sum_{h_i} e^{-\frac{h_i n}{\ell}} \approx e^{-h_0 n/\ell}
\ee
where $h_i$ denotes the spectrum of the BCFT Hamiltonian and $h_0$ is the ground-state energy. The value of $h_0$ is not universal and depends on the conformal boundary condition imposed at the cut-off surface. Gathering everything we obtain that: 
\be 
\lim_{\ell \to 0}\left[\frac{A'_1(\ell)}{A'_0(\ell)}\right] \sim \left(\frac{L+2h_0}{2\pi e^{L}}\right),\;\; \lim_{\ell \to \infty }\left[\frac{A'_1(\ell)}{A'_0(\ell)}\right] \sim -\left(\frac{6n(n+1)+c}{48\pi n^2}\right)e^{\ell} \to \infty
\ee

We conclude that (\ref{eq:new_saddle_key}) holds asymptotically as $\ell\to \infty$ so there is indeed another saddle-point that approaches $y = \infty$ in the limit of $\lambda \to 0$. 
Solving the saddle-point equation $A'_1(\ell^*)/A'_0(\ell^*)=-\lambda^{-1}$ consistently in this regime gives: 
\be\label{eq:new_saddle} 
\ell^*_{n.p.} = -\ln{\left(\lambda \kappa_n\right)},\;\;\;\kappa_n=\left(\frac{6n(n+1)+c}{48\pi n^2}\right) 
\ee
The new saddle point $\ell_{n.p.}^*$ lies on the Lefschetz thimble and therefore does contribute to the string partition function (\ref{eq:final_integral_halfline}). It indeed satisfies the self-consistency condition $\ell^*_{n.p.} \gg L$ as long as $\lambda \ll e^{-L}=\epsilon^2$. Its contribution is given by plugging (\ref{eq:new_saddle}) into (\ref{eq:final_integral_halfline}). The result is non-perturbative in small $\lambda$, and is given by: 
\be\label{eq:half_line_np}
\text{Tr} \rho^n_A\left(L,\Omega_{\mu}\right)_{n.p.} \sim P_\lambda(n)\; \text{Tr} \rho^n_A\left(\ell^*_{n.p.},\Omega_{CFT}\right),\;\;P_\lambda(n)=e^{-2\pi n\epsilon^2/\lambda}\left(\lambda \kappa_n\right)^{-\frac{n+1}{2}}e^{-\left(\frac{n}{2}+1\right)L}
\ee
where we have only kept in the classical action the leading order term in small $\epsilon$. The non-perturbative contribution (\ref{eq:half_line_np}) is proportional the CFT replica result computed in the $(w,\bar{w})$ coordinates, now regulated by a pair of cut-off holes, with the radii given by $r_{n.p.}=e^{-\ell^*_{n.p.}/2}$ about $w=0$ and $r_{n.p.}^{-1}$ about $w=\infty$. It is multiplied by a weight factor that is exponentially small $P_\lambda(n)\propto e^{-2\pi n\epsilon^2/\lambda}$. Moreover, this non-perturbative cut-off radius $r_{n.p.}$ is independent of the original cut-off $\epsilon$, and instead is of the same order as the effective length-scale $\epsilon_{T\bar{T}}$ predicted to emerge from the $T\bar{T}$-deformation: 
\be
r_{n.p.}= e^{-\ell^*_{n.p.}/2}= \sqrt{\lambda \kappa_n} \propto \sqrt{\mu} \propto \epsilon_{T\bar{T}}
\ee
The proportional constant $\kappa_n$ can be written as: 
\be
\kappa_n=\frac{c+c_{\text{eff}}}{48\pi n^2}
\ee
where $c_{\text{eff}} = 6n(n+1)$ can be viewed as the effective $n$-dependent shift to the CFT central charge, produced by the boundary reparametrization modes of the target-space CFT. In the limit of the holographic CFTs $c\gg c_{\text{eff}}$, the non-perturbative cut-off scale is given by: 
\be
r_{n.p.} \approx \frac{1}{4n}\sqrt{\frac{c \lambda}{3\pi}}
\ee
which has been suggested for $T\bar{T}$-deformed holographic CFTs \cite{Dubovsky:2012wk,Chakraborty:2018kpr,Apolo:2023ckr}. 

Finally, writing $\text{Tr} \rho^n_A\left(\ell,\Omega_{CFT}\right)$ simply as $Z_n(\ell)$ and including (\ref{eq:half_line_np}) into the full string partition function, we can compute the non-perturbative correction to the $T\bar{T}$-deformed Renyi entropy: 
\bea\label{eq:renyi}
&& S^n_A(L,\Omega_{\mu}) = \frac{1}{1-n}\ln{\left[\frac{\text{Tr} \rho^n_A\left(L,\Omega_{\mu}\right)}{\text{Tr}\rho^1_A\left(L,\Omega_{\mu}\right)^n}\right]} \approx  \frac{1}{1-n}\ln{\left[\frac{Z_n(L)+P_\lambda(n)Z_n(\ell^*_{n.p.})}{\left(Z_1(L) + P_\lambda(1) Z_1(\ell^*_{n.p.})\right)^n}\right]}\nonumber\\
&\approx & S^n_A(L,\Omega_{CFT}) + \frac{1}{1-n}\left[\frac{ P_\lambda(n)Z_n(\ell^*_{n.p.})}{Z_n(L)}- \frac{n P_\lambda(1) Z_1(\ell^*_{n.p.})}{Z_1(L)}\right]
\eea
we remind that the non-perturbative saddle-points $\ell^*_{n.p.}$ is $n$-dependent so they take different values in $Z_1$ and $Z_n$. Both $L$ and $\ell^*_{n.p.}$ occur in the range of validity for (\ref{eq:CFT_renyi_1}), plugging it in gives that: 
\bea
S^n_A(L,\Omega_{\mu}) &\approx & S^n_A(L,\Omega_{CFT}) + \left[\frac{e^{-g(n)}-ne^{-g(1)}}{1-n}\right]\nonumber\\
g(n) &=& \frac{2\pi n\epsilon^2}{\lambda}+\frac{(c+6n^2+12n)L}{12n}+4\pi n\kappa_n\ln{(\lambda \kappa_n)}
\eea
Now we can take the replica limit $n\to 1$ and obtain the corresponding non-perturbation correction to the $T\bar{T}$-deformed von-Neumann entropy at the leading order in small $\lambda$: 
\be
S^{vN}_A(L,\Omega_{\mu}) = S^{vN}_A(L,\Omega_{CFT}) + \frac{2\pi\epsilon^2}{\lambda} e^{-\frac{2\pi\epsilon^2}{\lambda}} \left[\frac{48\pi}{(c+12)\lambda}\right]^{1+\frac{c}{12}} e^{-\frac{(c+18)L}{12}}
\ee

\begin{figure}[!t]
\centering
\mbox{\hspace{0em}\includegraphics[width=.75\linewidth]{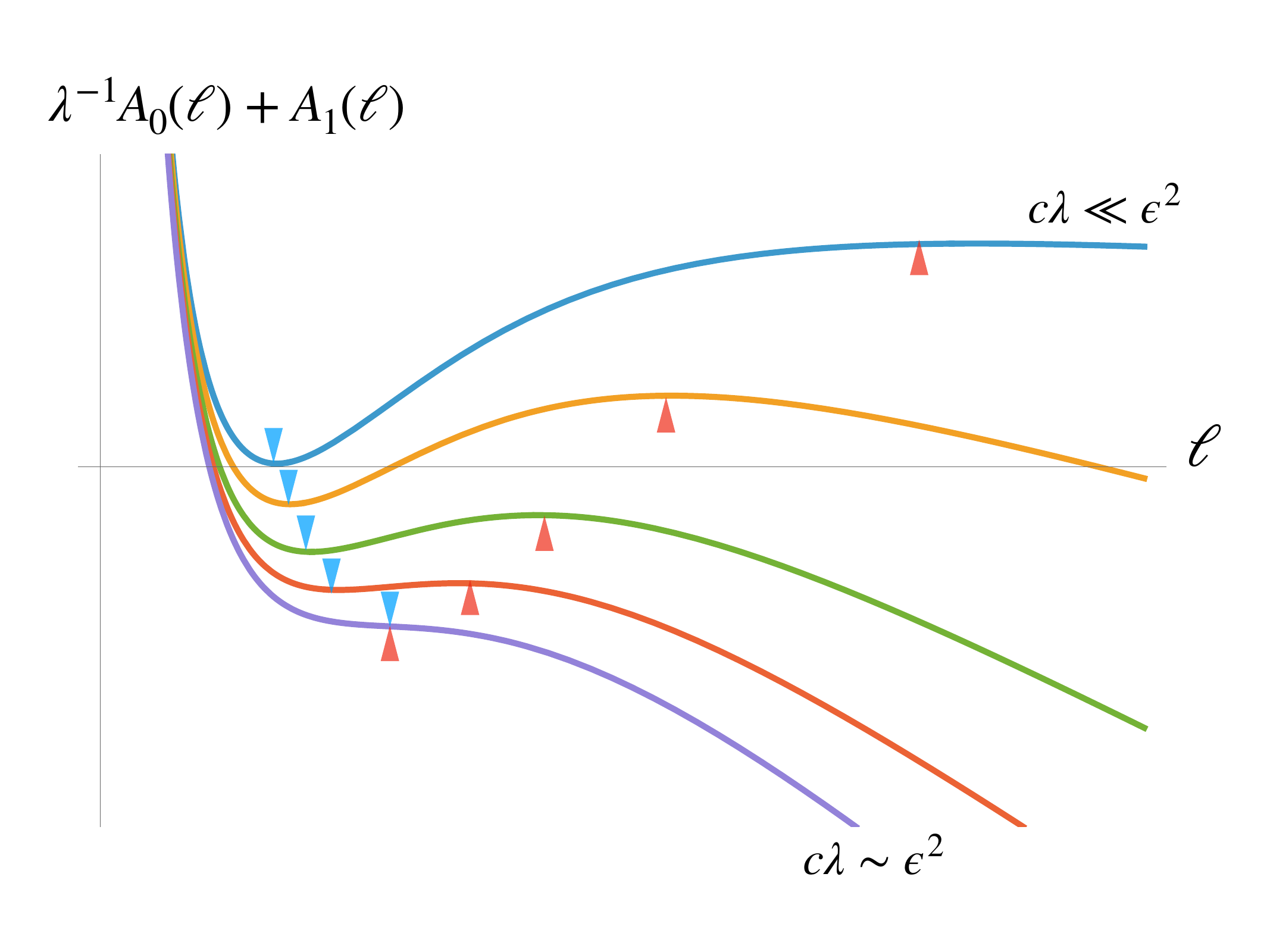}}
\caption{Schematic plot of the action $\lambda^{-1}A_0(\ell) + A_1(\ell)$, i.e.~the exponent of the integrand in \eqref{eq:Saddle Point Analysis}.
The perturbative saddle is marked with blue wedges while the non-perturbative saddle is marked with red wedges. The colored curves correspond to increasing values of $\lambda$ from top to bottom. As $\lambda$ increases, the two saddles approach each other and merge into a single saddle when $c\lambda \sim \epsilon^2$.%
}\label{fig:np-saddle}
\end{figure}

To summarize, our computation of the kernel function $K(L,\ell,\lambda)$ up to one-loop corrections allows us to perform a saddle-point analysis of the $\ell$-integral over the world-sheet moduli in the string partition function (\ref{eq:final_integral_halfline}). In addition to the ordinary saddle-point $\ell^*=L$ corresponding to the perturbative sector, we found a new saddle-point $\ell^*_{n.p.} \sim \ell_{T\bar{T}} \sim -\ln \epsilon_{T\bar T}$ via the competition between the leading order and sub-leading order terms in the asymptotic limit $\ell \to \infty$. 

We end this section with the following discussion. The non-perturbative contribution from the new saddle-point $\ell^*_{n.p.}$ is doing exactly what we had hoped to understand in the introduction -- it replaces \textit{dynamically} the original physical cut-off $\epsilon$ with the effective length-scale $\epsilon_{T\bar{T}}$. On the other hand, the effects from this new saddle-point are exponentially small, so it is not yet a full answer to the question of how the effective length-scale $\epsilon_{T\bar{T}}$ eventually takes over the physical cut-off $\epsilon$ in the final results. The new saddle-point we have found is more like the non-perturbative origin of the dynamical take-over that eventually takes place. 

A full understanding would naturally require us to go beyond the semi-classical limit, i.e.~not assuming that $\lambda$ is the smallest parameter and allowing it to become comparable with the cut-off scale $\epsilon^2$. We can have a glimpse of what happens then by approaching this limit from below, i.e.~keep increasing $\lambda$ towards $\epsilon^2$. Among other things, let us see what happens to the two saddle-points we have found in this process. Restricting the saddle-point equation in the large $\ell\geq L$ regime but not necessarily assuming that $\ell^*$ occurs asymptotically, we solve the following equation: 
\bea\label{eq:saddle_colliding}
&&\partial_\ell \left(\frac{4\pi n \sinh^2{\left(\frac{\ell-L}{2}\right)}}{\lambda\sinh{(\ell)}}\right) \approx \frac{4\pi n}{\lambda}\left(x-e^{L}x^2\right)= 4\pi n \kappa_n,\;\;\;x=e^{-\ell}\nonumber\\
&\rightarrow & x^2-e^{-L}x+\lambda e^{-L}\kappa_n =0\; \rightarrow \; x_{\pm} = \frac{e^{-L}}{2}\left(1\pm \sqrt{1-4\lambda e^{L}\kappa_n}\right)
\eea
Expanding $x_{\pm}$ in small $\lambda$ gives the two saddle-points $x_+ \approx e^{-L}$ and $x_-\approx\lambda \kappa_n$. However, as we increase $\lambda$ the two saddle-points $x_{\pm}$ approach each other. Taking the solution (\ref{eq:saddle_colliding}) at the face value, when $\lambda$ increases across $\lambda_c = \epsilon^2/(4\kappa_n)$ the two saddle-points $x_{\pm}$ collide and annihilate each other -- they leave the integration contour $\ell\in [0,\infty)$ and move into the complex-plane. This often signals a corresponding phase transition at $\lambda = \lambda_c$, which is the actual dynamical reason for the take-over between $\lambda_{T\bar{T}}$ and $\epsilon$. We will come back to this point in the discussion section (\ref{sec:discuss}).

\section{Renyi entropies on a finite interval}\label{sec:finite}
In this section, we switch to the case of a finite interval $A=[a,b]$, and compute the corresponding $T\bar{T}$-deformed Renyi entropies $S^n_A(\Omega_\mu)$. Unlike in the un-deformed CFT,  this is not trivially related to the half-line computation performed in the last section~(\ref{sec:infinite}). In the CFT, the finite interval $A=[a,b]$ is conformally equivalent to the half-line $A=[0,\infty)$ via the conformal transformation: 
\be
z=\left(\frac{z'-a}{b-z'}\right),\;\;\bar{z}=\left(\frac{\bar{z}'-a}{b-\bar{z}'}\right)
\ee
Upon turning on the $T\bar{T}$-deformation, the theory is no longer a CFT. As a result, the usual conformal transformations are not the symmetries of the theory, though one may still understand them as non-linear (i.e.~state-dependent) extensions of the symmetries in the deformed theory \cite{Guica:2019nzm,Guica:2020uhm,Kraus:2021cwf,Guica:2022gts,Georgescu:2022iyx,Du:2024bqk,Monten:2024efe,Chen:2025jzb}. In the string theory formulation of the $T\bar{T}$-deformed theory, the conformal transformations $\tilde{x}^{\pm}\to f^{\pm}(x^{\pm})$ in the original CFT are now lifted into the field re-definitions $\tilde{X}^{\pm}(z,\bar{z}) \to f^{\pm}\left(X^{\pm}\left(z,\bar{z}\right)\right)$ of the target-space CFT. They are not the symmetries of the target-space CFT except for the Poincare transformations, which happen to be its global symmetries.

For this reason, the finite interval case constitutes an independent computation. Nevertheless, the general procedure for performing the replica trick is almost identical to the half-line case. The only change is that now the target-space $\mathcal{M}=\tilde{\Sigma}_n$ is given by gluing $n$-copies of the punctured complex-plane along the subregion $A=[a,b]$. Again the conical excess singularities at $X^\pm=a$ and $X^\pm=b$ resulting from the gluing are regulated by removing two small holes about them, now both of small radius $\epsilon = e^{-L/2}$. This changes the boundary conditions for the world-sheet CFT of the target-space sector:\,\footnote{We comment that when dealing with a finite interval in a CFT, the following choice of the geometries for the cut-off holes: 
\be 
x^{\pm}_1(\theta) = \frac{a+b\epsilon e^{\pm in \theta}}{1+ \epsilon e^{\pm in\theta}},\;\; x^{\pm}_2(\theta) = \frac{a+b\epsilon^{-1} e^{\pm in \theta}}{1+ \epsilon^{-1} e^{\pm in\theta}}
\ee
are sometimes used to align with the underlying conformal isometries of the set-up. This is used later for the world-sheet. We do not bother with doing this in the target-space, since the conformal symmetries are no longer present. 
}
\be\label{eq:bc_finite}
X^{\pm}_{\gamma_1}(\sigma)=a+(b-a)\,\epsilon\, e^{\pm i n \theta_1(\sigma)},\;\;\;X^{\pm}_{\gamma_2}(\sigma)=b-(b-a)\,\epsilon\, e^{\mp i n \theta_2(\sigma)}
\ee
where $\gamma_{1,2}$ are the world-sheet boundaries. Correspondingly, in the conformal frame $(w,\bar{w})$ in which the one-loop factors from the bulk fluctuations in the target-space and the ghost sectors cancel, the world-sheet manifold $\mathcal{M}'$ also takes the from $\mathcal{M}'=\tilde{\Sigma}'_n$, which is obtained by gluing $n$-copies of the world-sheet complex-planes along a branch-cut, which we can also choose to be $[a,b]$, and with the world-sheet conical-excess singularities regulated by removing holes whose boundaries are described by: 
\be\label{eq:boundary_fintie_1}
w_1(\sigma)=\frac{a+b \left(\epsilon' e^{i\sigma}\right)^n}{1+ \left(\epsilon' e^{i\sigma}\right)^n},\;\; w_2(\sigma)=\frac{a+b \left({\epsilon'}^{-1} e^{i\sigma}\right)^{n}}{1+ \left({\epsilon'}^{-1} e^{i\sigma}\right)^{n}},\;\;\bar{w}_{1,2}(\sigma) =w_{1,2}(\sigma)^*
\ee
where $\left(\epsilon'\right)^n = e^{-\ell/2}$ is again the world-sheet cut-off scale on $\tilde{\Sigma}'_n$. In the small $\epsilon'$ limit (\ref{eq:boundary_fintie_1}) describes approximately two circular holes of radius $r'=(b-a)(\epsilon')^n$ about $a$ and $b$ respectively. These cut-off boundaries are integral curves of the conformal isometries that the world-sheet still enjoys. Having specified these modifications, we end with the same form of the string partition function:
\bea
\text{Tr}\rho^n_A(L,\Omega_\mu)&=& \int \mathcal{D}\theta_1 \int \mathcal{D} \theta_2 \int^\infty_0 \frac{d\ell}{2\pi \ell} \sum_{X^{\pm}_{cl}(\theta_1,\theta_2)} e^{-S_{\tilde{\Sigma}_n}\left(X^{\pm}_{cl}(\theta_1,\theta_2),\mu,\tilde{\Sigma}'_n\right)} \left(\int \mathcal{D}\Phi\, e^{-S_{CFT}\left(\Phi, \tilde{\Sigma}'_n\right)}\right) \nonumber\\
&=&\int^\infty_0 \frac{d\ell}{2\pi \ell} K(L,\ell,\mu)\; \text{Tr}\rho^n_A(\ell,\Omega_{CFT})
\eea

Let us sketch the plan for the remainder of this section. We will first compute the effective action $F(\theta_1, \theta_2)$, which requires solving the Laplace equations under the modified boundary conditions (\ref{eq:bc_finite}) and computing the on-shell action. We will then compute the leading order contribution to the kernel function $K(L,\ell,\mu)$ by finding the saddle-point $\theta^*_{1,2}(\sigma)$ that minimizes $F(\theta_1,\theta_2)$. While we have easily guessed and verified the exact saddle-point in the half-line case, doing the same becomes difficult for the finite-interval case. We will instead only find approximate solutions in various parametric limits of $\ell$ and $L$. As can be expected, it is significantly more difficult than in section (\ref{sec:infinite}) to compute the one-loop corrections. For this reason, in this section we will be satisfied with only using the leading-order result of $K(L,\ell,\lambda)$ for the subsequent non-perturbative analysis. As we have learned in the last section, including the one-loop corrections from the target-space sector only amounts to effectively shifting the CFT central charge $c\to c+c_{\text{eff}}$. While this shift does not alter the non-perturbative analysis in any qualitative way, it also becomes negligible when considering holographic CFTs, which we can restrict to if we want the results to be more accurate. We expect that the one-loop corrections have the same properties in the finite interval case.  

\subsection{Kernel function at the leading order}
We begin with solving the classical equations of motion for the target-space CFT: 
\be
\partial_w\partial_{\bar{w}}X^{\pm}(w,\bar{w}) = 0
\ee
with the boundary conditions specified at the world-sheet boundaries (\ref{eq:boundary_fintie_1}):
\begin{equation*}
X^{\pm}\left(w_1(\sigma),\bar{w}_1(\sigma)\right)=a+(b-a)\,\epsilon\, e^{\pm i n \theta_1(\sigma)},\;\;X^{\pm}\left(w_2(\sigma),\bar{w}_2(\sigma)\right)=b-(b-a)\,\epsilon\, e^{\mp i n \theta_2(\sigma)}
\end{equation*}

Similar to the last section, for the purpose of solving the classical equation we can work in the flat conformal frame $(z,\bar{z})$: 
\be
z=\left(\frac{w-a}{b-w}\right)^{1/n},\;\;\;\bar{z}=\left(\frac{\bar{w}-a}{b-\bar{w}}\right)^{1/n}
\ee
such that the world-sheet $\mathcal{M}'=\tilde{\Sigma}'_n$ becomes an annulus with holes of radii $\epsilon'$ and $1/\epsilon'$ removed about $z=0$ and $z=\infty$ respectively. Naively we can then simply use the Green's function on the annulus (\ref{eq:annulus_Greens}) and apply it to the modified boundary conditions: 
\be 
X^{\pm}\left(\epsilon' e^{i\sigma}\right)=a+(b-a)\,\epsilon\, e^{\pm i n \theta_1(\sigma)},\;\;\;X^{\pm}\left(1/\epsilon'e^{i\sigma}\right)=b-(b-a)\,\epsilon\, e^{\mp i n \theta_2(\sigma)}
\ee
to obtain the solution. However, there is a caveat coming from the non-compactness of target-space manifold $\tilde{\Sigma}_n$. The classical solution in the original conformal frame represents a mapping from $\mathcal{M}'=\tilde{\Sigma}'_n$ into $\mathcal{M}=\tilde{\Sigma}_n$, both of which contain $n$ copies of complex infinities. We need to specify how the infinities are mapped. Since we require that the solution be of the mapping degree one, a natural choice is that the world-sheet infinities be mapped into the target-space infinities in a way that respects the cyclic order of both $\tilde{\Sigma}'_n$ and $\tilde{\Sigma}_n$. In addition, the mapping is asymptotically linear towards these infinities\,\footnote{More generally we could also put in an anti-holomorphic term: 
\be 
\lim_{w,\bar{w}\to \infty} X^+(w,\bar{w}) \propto w + \alpha \bar{w},\;\;\lim_{w,\bar{w}\to \infty} X^-(w,\bar{w}) \propto \bar{w} +\bar{\alpha} w
\ee
For $\alpha \neq 0$ the on-shell action is divergent, which is fine and represents the expected IR-divergence from the non-compact target-space. Thanks to the special form of the target-space action $\sim \partial X^+ \bar{\partial} X^-$,  choosing $\alpha=0$ allows us to get rid of the IR-divergence. It therefore appeals naturally for us to make this choice.   
}
\be 
\lim_{w,\bar{w}\to \infty} X^+(w,\bar{w}) \propto w,\;\;\lim_{w,\bar{w}\to \infty} X^-(w,\bar{w}) \propto \bar{w}
\ee

In the conformal frame $(z,\bar{z})$, the $n$ copies of the world-sheet infinities in $(w,\bar{w})$ are mapped onto the unit circle and are located at $\lbrace z_i=e^{i\pi/n},\;i=1,...,n\rbrace$. The mapping condition regarding the infinities now takes the form of requiring that the classical solution $X^{\pm}_{cl}(z,\bar{z})$ contains $n$ holomorphic/anti-holomorphic simple poles at $\lbrace z_i \rbrace$ with the same residues. So we can write: 
\be
X^{+}_{cl}(z,\bar{z}) = \frac{p}{1+z^n}+\tilde{X}^{+}_{cl}(z,\bar{z}),\;\;\;X^-_{cl}(z,\bar{z})=\frac{\bar{p}}{1+\bar{z}^n}+\tilde{X}^{-}_{cl}(z,\bar{z})
\ee
where $\tilde{X}^{\pm}_{cl}(z,\bar{z})$ are regular harmonic functions on the annulus. We can solve them using the Greens function (\ref{eq:annulus_Greens}) with the modified boundary conditions: 
\bea
\tilde{X}^+_{cl}(\epsilon' e^{i\sigma}) &=& \tilde{X}^-_{cl}(\epsilon' e^{i\sigma})^* = a + (b-a)\,\epsilon\, e^{in\theta_1(\sigma)} - \frac{p}{1+{\epsilon'}^n e^{in\sigma}} \nonumber\\
\tilde{X}^+_{cl}(1/\epsilon' e^{i\sigma}) &=& \tilde{X}^-_{cl}(1/\epsilon' e^{i\sigma})^* = b -(b-a) \,\epsilon\, e^{-in\theta_2(\sigma)} - \frac{p}{1+{\epsilon'}^{-n}e^{in\sigma}}
\eea

Plugging the resulting solution back into the target-space action, we can obtain the effective action $F(\theta_1,\theta_2)$  in terms of the boundary reparametrizations $\theta_{1,2}$: 
\bea\label{eq:action2}
 F(\theta_1,\theta_2)&=&\frac{1}{8\pi^2}\sum_{k=-\infty}^{\infty}\int_0^{2\pi}\dm \sigma\int_0^{2\pi}\dm\sigma'\frac{ke^{-\im k(\sigma-\sigma')}}{\sinh{(k\ell/n)}} \times \Bigg[{\epsilon'}^{2k}Y^+_2(\sigma) Y^-_2(\sigma') 
-Y^+_2(\sigma) Y^-_1(\sigma') \nonumber\\
&-&Y^+_1(\sigma)Y^-_2(\sigma')+{\epsilon'}^{-2k}Y^+_1(\sigma)Y^-_1(\sigma')\Bigg]\nonumber\\
Y^{+}_1(\sigma) &=& Y^{-}_1(\sigma)^*= \tilde{X}^+_{cl}(\epsilon'e^{i\sigma}),\;\;\;
Y^{+}_2(\sigma) = Y^{-}_2(\sigma)^*=\tilde{X}^+_{cl}(1/\epsilon' e^{i\sigma})
\eea

There appears to be a free parameter $p$ floating in the expression of $F(\theta_1,\theta_2)$.
However, for generic values of $p$ the solutions will contain a $\ln{(z\bar{z})}$ term, e.g.:
\bea
\tilde{X}^+_{cl}(z,\bar{z})
&=&\left[\frac{(a-b)(1-\epsilon\alpha_0-\epsilon\beta_0)-p}{4\ln{\epsilon'}}\right]\ln{(z\bar{z})}+\left(\text{Polynomial terms in $z$, $\bar{z}$, $z^{-1}$ and $\bar{z}^{-1}$}\right)\nonumber\\
\alpha_{0}&=&\frac{1}{2\pi}\int_0^{2\pi}\dm\sigma e^{\im n\theta_1(\sigma)}, \quad\beta_0=\frac{1}{2\pi}\int_0^{2\pi}\dm\sigma e^{-\im n\theta_2(\sigma)}
\eea
In the presence of the $\ln{(z\bar{z})}$ terms, the corresponding mapping between the world-sheet and the target-space will self-intersect, e.g.~non-intersecting radial contours $|z|=const$ on the world-sheet will be mapped to target-space contours that intersect each other. Imposing the absence of the $\ln{(z\bar{z})}$ terms then fixes the value of $p$ to be:
\begin{equation}
    p=(a-b)\left(1-\epsilon\alpha_0-\epsilon\beta_0\right) 
\end{equation}

\subsubsection{Approximate saddle-points\nopdfstring{ $\theta^*_{1,2}$}}
\begin{figure}[!ht]
\centering
\mbox{\hspace{3em}\includegraphics[width=.85\linewidth]{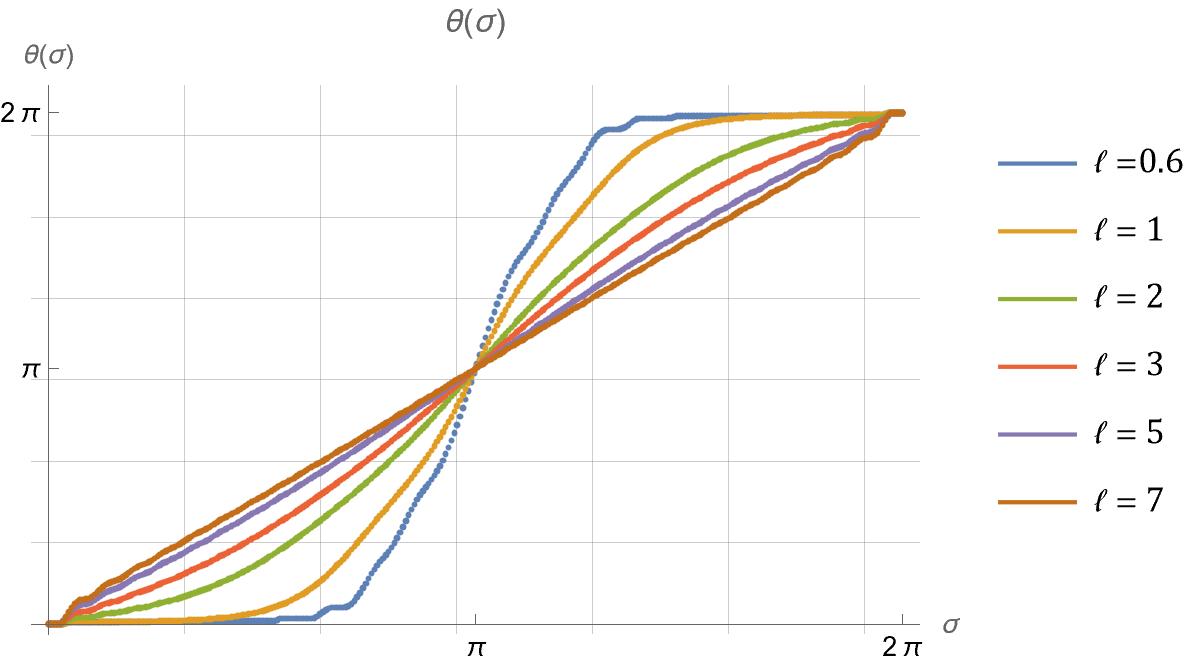}}
\caption{The full solution of $\theta(\sigma)$ with $a=1,\; b=2$ and $\epsilon=0.01$, extrapolates from $\theta(\sigma)=2\pi \Theta(\sigma-\pi)$ for small $\ell$ to $\theta(\sigma)=\sigma$ for large $\ell$. }\label{fig:finite_saddle}
\end{figure}

We now proceed to solve for the saddle-point solution $\theta^*_{1,2}(\sigma)$ that determines the leading order contribution $S_0(L,\ell)$. This is computed by solving the saddle-point equation of (\ref{eq:action2}), which after some rearrangement takes the form:
\begin{eqnarray}
    e^{2\im n\theta_1(\sigma)} &=& \frac{p\sum\limits_{j=1}^\infty\frac{(-1)^j (nj)}{\sinh{(j\ell)}}\left(e^{-j\left(\ell/2+in\sigma\right)}+e^{j\left(\ell/2+in\sigma\right)}\right)-\sum\limits_{k\neq 0}\frac{(b-a)\epsilon k e^{ik\sigma} }{\sinh{(k\ell/n)}}\left(\beta_{k}+e^{k\ell/n}\alpha_{k}\right)}{p^*\sum\limits_{j=1}^\infty\frac{(-1)^j (nj)}{\sinh{(j\ell)}}\left(e^{-j\left(\ell/2-in\sigma\right)}+e^{j\left(\ell/2-in\sigma\right)}\right)-\sum\limits_{k\neq 0}\frac{(b-a)\epsilon k e^{-ik\sigma} }{\sinh{(k\ell/n)}}\left(\beta_{k}^*+e^{k\ell/n}\alpha_{k}^*\right)}\nonumber\\
    e^{2\im n\theta_2(\sigma)} &=& \frac{p^* \sum\limits_{j=1}^\infty\frac{(-1)^j (nj)}{\sinh{(j\ell)}}\left(e^{-j\left(\ell/2+in\sigma\right)}+e^{j\left(\ell/2+in\sigma\right)}\right)-\sum\limits_{k\neq 0}\frac{(b-a)\epsilon k e^{-ik\sigma} }{\sinh{(k\ell/n)}}\left(\alpha_{k}^*+e^{-k\ell/n}\beta_{k}^*\right)}{ p\sum\limits_{j=1}^\infty\frac{(-1)^j (nj)}{\sinh{(j\ell)}}\left(e^{-j\left(\ell/2-in\sigma\right)}+e^{j\left(\ell/2-in\sigma\right)}\right)-\sum\limits_{k\neq 0}\frac{(b-a)\epsilon k e^{ik\sigma} }{\sinh{(k\ell/n)}}\left(\alpha_{k}+e^{-k\ell/n}\beta_{k}\right)}\nonumber\\
    \alpha_{k} &=& \frac{1}{2\pi}\int_{0}^{2\pi}\dm \sigma e^{\im n\theta_1(\sigma)}e^{-\im k\sigma}, \quad\beta_{k}=\frac{1}{2\pi}\int_{0}^{2\pi}\dm \sigma e^{-\im n\theta_2(\sigma)}e^{-\im k\sigma}.
\end{eqnarray}
where we remind that $p$ is fixed self-consistently as: 
\begin{equation}
    p=(a-b)\left(1-\epsilon\alpha_0-\epsilon\beta_0\right).
\end{equation}
It is in fact self-consistent to assume that the saddle-point solution satisfies $\theta_1(\sigma)= \theta_2(\sigma)$. In this case, we can check that:
\begin{eqnarray}
    \alpha^*_{k} = \beta_{-k},\quad\beta_{k}^* = \alpha_{-k},\;\;p=p^*
\end{eqnarray}
Then the RHS for the equations for both $\theta_1(\sigma)$ and $\theta_2(\sigma)$ become identical upon rewriting $k\to-k$ in one of the $k$-sums. As a result, we only need to solve for only a single unknown function $\theta_1(\sigma )=\theta_2(\sigma)=\theta(\sigma)$ satisfying: 
\begin{eqnarray}\label{eq:saddle_finite_2}
    e^{2\im n\theta(\sigma)} &=& \frac{p\sum\limits_{j=1}^\infty\frac{(-1)^j (nj)}{\sinh{(j\ell)}}\left(e^{-j\left(\ell/2+in\sigma\right)}+e^{j\left(\ell/2+in\sigma\right)}\right)-\sum\limits_{k\neq 0}\frac{(b-a)\epsilon k e^{ik\sigma} }{\sinh{(k\ell/n)}}\left(\alpha_{-k}^*+e^{k\ell/n}\alpha_{k}\right)}{p\sum\limits_{j=1}^\infty\frac{(-1)^j (nj)}{\sinh{(j\ell)}}\left(e^{-j\left(\ell/2-in\sigma\right)}+e^{j\left(\ell/2-in\sigma\right)}\right)-\sum\limits_{k\neq 0}\frac{(b-a)\epsilon k e^{-ik\sigma} }{\sinh{(k\ell/n)}}\left(\alpha_{-k}+e^{k\ell/n}\alpha_{k}^*\right)}\nonumber\\
    \alpha_{k}&=&\frac{1}{2\pi}\int^{2\pi}_0 d\sigma e^{in\theta(\sigma)} e^{-ik \sigma},\quad p=(a-b)\left(1-\epsilon\alpha_0-\epsilon\alpha_0^*\right). 
\end{eqnarray}
This remains a non-linear integral equation that is difficult to handle. To proceed, we first require that $\epsilon=e^{-L/2}\ll 1$, and then consider solving the saddle-point equation (\ref{eq:saddle_finite_2}) approximately in a few parametric limits of $\epsilon'=e^{-\ell/2n}$ that may be relevant for our purpose. We leave the details of the analysis to the appendix (\ref{app:saddle_approx}). At the leading order, the approximate solutions take the following simple forms: 
\begin{equation}\label{eq:finite_saddle}
    \theta^*(\sigma)=
    \begin{cases}
    0 +\cdots,\;\;\;\;\;\;\;\; {\epsilon'}^n\sim1\\
    \sigma +\cdots,\;\;\;\;\;\;\;\; {\epsilon'}^n\ll 1\\
    \end{cases}
\end{equation}
where ${\cdots}$ denotes higher order corrections in the corresponding limits. It may appear that the winding number constraint is violated by the constant form of the leading order solution (\ref{eq:finite_saddle}) in the limit ${\epsilon'}^n\sim 1,\; \epsilon \ll 1$. It is indeed satisfied by the full solution after including the sub-leading corrections. The full solution looks more like a smoothened step function $\theta(\sigma)=2\pi \Theta(\sigma-\pi)$,  see Figure (\ref{fig:finite_saddle}). Although we will not study the fluctuations about these saddle-point solutions, it is worth mentioning that for the finite-interval case there does not exist a zero-mode corresponding to a continuous symmetry group, e.g.~the global rotation of the target space in the half-line case. Nonetheless, there remains a discrete set of symmetries: 
\be 
\theta^*(\sigma) = \theta^*(\sigma) + \frac{2\pi m}{n},\;\;\;m = 1,...,n
\ee
They simply shuffle the $n$ replicas cyclically. Summing over the images gives rise to a pre-factor of $n$ for the path-integral, which is not important for subsequent analysis and will be ignored. 

\subsubsection{Leading order expressions for \texorpdfstring{$K(L,\ell,\mu)$}{K}}
The approximate forms of the saddle-point functions (\ref{eq:finite_saddle}) may now be used to evaluate the action \eqref{eq:action2}. In the limit of $\epsilon' = e^{-\ell/2n}\ll 1$, by plugging the approximate saddle-point solution $\theta^*(\sigma)\approx \sigma$, the leading-order contribution to the action is given by:
\begin{equation}
    \begin{aligned}
        \mu^{-1}S_0(L,\ell) &\approx  \frac{1}{\mu}\left(\frac{n(b-a)^2\epsilon^2{\epsilon'}^{-2n}}{\sinh{\ell}}-\frac{2n(b-a)^2\epsilon{\epsilon'}^{-n}}{\sinh{\ell}}+\sum_{m=1}^{\infty}\frac{nm(b-a)^2}{\sinh{m\ell}}\right)\\
        &= \frac{2 n(b-a)^2}{\mu}\left(e^{-L/2}-e^{-\ell/2}\right)^2 +...
    \end{aligned}
\end{equation}
where ... denotes sub-leading terms of order $\mathcal{O}(e^{-2\ell})$. Based on this, the leading order form of the kernel function $K(L,\ell,\mu)$ takes the form:
\bea\label{eq:finite_kernel_1}
    K(L,\ell,\mu)&\approx & \exp{\left[-\frac{2 n(b-a)^2}{\mu}\left(e^{-L/2}-e^{-\ell/2}\right)^2\right]}\nonumber\\
   &\approx & \exp{\left[-\frac{2 n}{\mu}(r'-r)^2\right]} 
\eea
where we have identified in this regime the dynamical (world-sheet) cut-off radius $r'$, read out from its original form (\ref{eq:boundary_fintie_1}), and the physical (target-space) cut-off radius defined by (\ref{eq:bc_finite}): 
\be\label{eq:finite_rs}
r' \approx (b-a)\,e^{-\ell/2},\;\;\;r=(b-a)\,e^{-L/2}
\ee
Similar to the case of half-line subregion, the kernel function (\ref{eq:finite_kernel_1}) in the semi-classical limit $\mu \ll 1$ describes approximately a Gaussian distribution of $r'$ peaked at $r$. 

For the other limit of ${\epsilon'}^n =e^{-\ell/2} \sim 1$, or equivalent the limit of $\ell \to 0$, by plugging in the approximate saddle-point solution $\theta(\sigma)=0$, the leading-order action can be estimated using the small $\ell$ expansion:  $\sum^\infty_{m=1} m/\sinh{(m\ell)} = \pi^2/(4\ell^2)+...$
\bea
    \mu^{-1}S_0(L,\ell)&\approx & \frac{n(b-a)^2}{\mu}\left(\frac{2e^{-L}}{\ell}+\sum_{m=1}^{\infty}\frac{m}{\sinh{m\ell}}\right)\nonumber\\
    &\approx & \frac{n(b-a)^2}{\mu}\frac{\pi^2}{4\ell^2}+ \cdots
\eea
where ${\cdots}$ denotes higher order terms in small $\ell$. The leading order expression of $K(L,\ell,\mu)$ is therefore given by:
\begin{equation}
    K(L,\ell,\mu)\approx\exp{\left[-\frac{n(b-a)^2}{\mu}\frac{\pi^2}{4\ell^2}\right]}
\end{equation}
The exponent of the kernel diverges as a double pole $\sim \ell^{-2}$ towards $\ell\to 0$. We can compare this with the small $\ell$ behavior of the half-line case, which diverges as a simple pole $\sim \ell^{-1}$. The distinction can be traced to the non-compactness of the target-space, technically it arises because of the poles included in the solution $X^{\pm}$. The $\ell \to 0$ behavior of the kernel is irrelevant for the purpose of our subsequent saddle-point analysis. For the full $\ell$ integral, the higher degree of $\ell\to 0$ divergence (double-pole v.s. simple-pole) in the exponent indicates that for $\mu>0$ the kernel suppresses the small $\ell$ contribution more in the finite interval case than in the half-line case.

\subsection{Non-perturbative analysis}
We proceed to perform a non-perturbative analysis of the finite-interval case of the string-partition function in analogy with the treatment of the half-line case. In this case, we only use the leading order results of the kernel function just obtained.

The analysis proceeds almost identically as the half-line case, so we shall simply state the results. There is again the simple saddle-point: 
\be
\ell^*\approx L
\ee
that identifies the dynamical (world-sheet) cut-off radius $r'$ with the physical (target-space) cut-off radius $r$ in (\ref{eq:finite_rs}). Expanding the string partition function near this saddle-point in small $\mu$ then generates the perturbation series starting from the CFT result. Based on the asymptotic behavior of $K(L,\ell,\mu)$ in the limits of $\ell \to \infty$ and $\ell\to 0$ which we have just estimated, we can similarly conclude that only the $\ell\to \infty$ regime can admit additional saddle-point of the nature discussed in section (\ref{sec:saddle}). The non-perturbative saddle-point can then be solved approximately as: 
\begin{equation}\label{eq:finite_lnp}
    \ell^*_{n.p.} \approx -2\ln{\left(\mu \kappa_n\right)},\;\;\;\kappa_n=\left(\frac{ce^{L/2}}{24 n^2(b-a)^2}\right) 
\end{equation}
The approximate expression for the solution is accurate for $c\mu \ll (b-a)^2\epsilon^2 \sim r^2$. Including its contribution, the string partition function for the deformed replica result across $A=[a,b]$, evaluated by removing holes of the physical cut-off radius \mbox{$r=(b-a)\,e^{-L/2}$} about $a$ and $b$, can be summarized as follows:
\bea\label{eq:finite_interval_np}
    \text{Tr} \rho^n_{[a,b]}\left(r,\Omega_{\mu}\right) &\sim & \text{Tr} \rho^n_{[a,b]}(r,\Omega_{CFT}) + e^{-\frac{2 n(b-a)^2\epsilon^2}{\mu}} \operatorname{Tr} \rho^n_{[a,b]}\left(r_{n.p.},\Omega_{CFT}\right) +...  
\eea
where for comparison we have expressed the RHS implicitly in terms of the CFT formula. The first term represents the perturbative contribution, while the second term represents the non-perturbative contribution and is again exponentially small. However, interpreted as a CFT result it is now evaluated by removing holes of the non-perturbative cut-off radius given by (\ref{eq:finite_lnp}): 
\be\label{eq:finite_rnp}
r_{n.p.} = (b-a)\,e^{-\ell^*_{n.p.}/2} \sim \sqrt{c \mu} \left(\frac{\sqrt{c\mu}}{r}\right) \sim \epsilon_{T\bar{T}}\left(\frac{\epsilon_{T\bar{T}}}{r}\right)
\ee

We make some comments regarding (\ref{eq:finite_rnp}). In contrast to the half-line case, for a finite interval $A=[a,b]$, the non-perturbative cut-off radius $r_{n.p.}$ is not simply given by the predicted $T\bar{T}$ effective length-scale $\epsilon_{T\bar{T}}$. There is an additional factor of the ratio $(\epsilon_{T\bar{T}}/r)$ between the $T\bar{T}$ length-scale and the physical cut-off radius. We have performed computations based on improved approximations, e.g.~by including sub-leading terms in the kernel function that come from the higher order corrections to the saddle-point solution (\ref{eq:finite_saddle}). The expression (\ref{eq:finite_rnp}) remains valid. The implication of (\ref{eq:finite_rnp}) therefore appears puzzling to us. From a conservative point of view, (\ref{eq:finite_rnp}) is inferred from an exponentially small non-perturbative correction computed in the limit $\mu \ll r^2$, and it is likely that one simply cannot extrapolate the scaling of (\ref{eq:finite_rnp}) to reflect the dynamical entanglement cut-off that actually emerges when $\mu \gtrsim r^2$. This is especially the case given a possible phase-transition at $\mu \sim r^2$ as discussed at the end of section (\ref{sec:saddle}), which the finite interval case also exhibits by doing more detailed saddle-point analysis.\,\footnote{
We point out that the phase transition is still controlled by the comparison between $\epsilon_{T\bar T}$ (not~$r_{n.p.}$) and the bare entanglement cut-off $r$.
} 

\begin{figure}[!ht]
\begin{center}
\begin{adjustbox}{center}
\begin{tikzpicture}
  \useasboundingbox (-7,-7) rectangle (7,3);
  

  \draw[thick] (-4,0) circle (1.7);
  \draw[->] (-4,0) -- ({-4-1.7* cos{45}},{-1.7*sin{45}});
  \draw[thick] (-4,0) circle (0.4);
  \draw[->] (-4,0) -- ({-4-0.4*cos{45}},{0.4*sin{45}});
  \draw (-4.5,0.3) node[above] {$\epsilon$};
  \draw (-5.5,-1.9) node[above] {$1/\epsilon$};
  \node[below] at (-4,-2.2) {Target-space};

  \draw[thick] (4,0) circle (2.2);
  \draw[->] (4,0) -- ({4-2.2* cos{45}},{-2.2*sin{45}});
  \draw[thick] (4,0) circle (0.25);
  \draw[->] (4,0) -- ({4-0.25* cos{45}},{0.25*sin{45}});
  \draw (4,0.3) node[above] {$r_{n.p.}\sim\sqrt{\mu}$};
  \draw (2.3,-2.3) node[above] {$1/r_{n.p.}$};
  \node[below] at (4,-2.2) {World-sheet};

  \draw[->,very thick] (-1.5,0) -- (1.5,0) node[midway, above] {n.p. saddle};

  \draw[thick] (-5.5,-5) circle (0.25);
  \draw[->] (-5.5,-5) -- ({-5.5-0.25*cos{45}},{-5-0.25*sin{45}});
  \draw (-6,-5.2) node[below] {$(b-a)\epsilon$};
  \draw[thick] (-2.5,-5) circle (0.25);
  \draw[->] (-2.5,-5) -- ({-2.5-0.25*cos{45}},{-5-0.25*sin{45}});
  \draw (-3,-5.2) node[below] {$(b-a)\epsilon$};
  \draw[-,thick] (-5.25,-5) -- (-2.75,-5);
  \node[below] at (-4,-6) {Target-space};
  \draw[decorate, decoration={brace}] (-5.5,-4.75) -- (-2.5,-4.75) node [midway, above] {$b-a$};

  \draw[-,very thick] (-1.5,-5) -- (0.5,-5);
  \node[above] at (-0.5,-5) {n.p. saddle};
  \draw[->,very thick] (0.5,-5) -- (1.5,-4);
  \node[above] at (1,-4) {picture 1};
  \draw[->,very thick] (0.5,-5) -- (1.5,-6);
  \node[below] at (1,-6) {picture 2};
  
  \draw[thick] (2.5,-4) circle (0.15);
  \draw[->] (2.5,-4) -- ({2.5-0.15*cos{45}},{-4-0.15*sin{45}});
  \draw (2.5,-4.1) node[below] {$r_{n.p.}$};
  
  \draw[-,thick] (2.65,-4) -- (5.35,-4);
  \draw[decorate, decoration={brace}] (2.5,-3.85) -- (5.5,-3.85) node [midway, above] {$b-a$};
  
  \draw[thick] (5.5,-4) circle (0.15);
  \draw[->] (5.5,-4) -- ({5.5-0.15*cos{45}},{-4-0.15*sin{45}});
  \draw (6.5,-3.85) node[below] {$r_{n.p.} \sim\sqrt{\mu}\left(\frac{\sqrt{\mu}}{r}\right)$};
  
  \draw[thick] (2.5,-6) circle (0.15);
  \draw[->] (2.5,-6) -- ({2.5-0.15*cos{45}},{-6-0.15*sin{45}});
  \draw (2.5,-6.1) node[below] {$r_{n.p.}$};
  
  \draw (4,-6.5) node[below] {World-sheet};
  \draw[-,thick] (2.65,-6) -- (5.35,-6);
  \draw[decorate, decoration={brace}] (2.5,-5.85) -- (5.5,-5.85) node [midway, above] {$(b-a)\frac{r}{\sqrt{\mu}}$};
  
  \draw[thick] (5.5,-6) circle (0.15);
  \draw[->] (5.5,-6) -- ({5.5-0.15*cos{45}},{-6-0.15*sin{45}});
  \draw (6,-6) node[below] {$r_{n.p.} \sim\sqrt{\mu}$};
  
\end{tikzpicture}
\end{adjustbox}
\end{center}
\caption{A schematic figure contrasting the saddle points between the half-line and finite-interval cases.}\label{fig:compare}
\end{figure}

These being said, it remains tempting to make some efforts in interpreting (\ref{eq:finite_rnp}) at face value. To begin with, since the CFT expression only depends on the combination: 
\be
\left[\frac{b-a}{r_{n.p.}}\right] = \left[\frac{(r/\epsilon_{T\bar{T}})(b-a)}{\epsilon_{T\bar{T}}}\right]
\ee
one can interpret (\ref{eq:finite_rnp}) as suggesting the following: when expressing the $T\bar{T}$-deformed result at $\mu \gtrsim r^2$ in terms of the CFT formula, in addition to replacing the bare entanglement cut-off radius $r$ by the $T\bar{T}$ length-scale $\epsilon_{T\bar{T}}$, one also needs to work with a re-scaled notion of the interval length: $(b-a)_{CFT} = (b-a)(r/\epsilon_{T\bar{T}})$. In this picture, one may explain the absence of the rescaling in the half-line case by its infinite interval length. We illustrate this in Figure (\ref{fig:compare}) by also comparing with the half-line case. We should point out that the scaling of the effective cut-off $\propto \epsilon_{T\bar{T}}$ in the half-line result cannot be recovered by taking additional limits of the finite interval case, particularly not from taking the large interval limit: $(b-a)\to \infty$. Part of this can be traced to the fact that in the un-deformed CFT, there is a topological gap between the two set-ups: they differ by a conformal inversion. We refrain from making further speculations, and leave the clarification of (\ref{eq:finite_rnp}) to the future. 

We end this section by evaluating the non-perturbative correction to the Rényi entropy in the presence of the $T\bar{T}$-deformation according to \eqref{eq:renyi}\,\footnote{We stress that since we are missing the one-loop correction to the kernel function, the sub-leading in $\lambda$ dependence of the results are to be taken seriously (at best) only for CFTs with large $c$.}. Taking into account the result for $Z_n$ and $Z_1$ in the finite-interval case, the expression for Rényi entropy can be simplified to the following form:
\begin{eqnarray}
    S^n_A(L,\Omega_{\mu}) &\approx & S^n_A(L,\Omega_{CFT}) + \left[\frac{e^{-g(n)}-ne^{-g(1)}}{1-n}\right]\nonumber\\
    g(n) &=& \frac{2 n(b-a)^2\epsilon^2}{\mu}+\frac{cL}{12n}+\frac{c}{6n}\ln{(\mu \kappa_n)}
\end{eqnarray}
After taking the replica limit $n\to 1$, we can obtain the corresponding non-perturbation correction to the $T\bar{T}$-deformed von-Neumann entropy at the leading order in small $\mu$:
\begin{equation}
    S^{vN}_A(L,\Omega_{\mu}) = S^{vN}_A(L,\Omega_{CFT}) + \frac{2(b-a)^2\epsilon^2}{\mu}e^{-\frac{2(b-a)^2\epsilon^2}{\mu}}\left[\frac{24(b-a)^2}{ce^{L/2}\mu}\right]^{\frac{c}{6}}e^{-\frac{cL}{12}}
\end{equation}

\section{Discussions and outlooks}\label{sec:discuss}
In this paper, we studied the non-perturbative aspects of entanglement structures in the $T\bar{T}$-deformed CFTs. Our main motivation was to make progress in understanding the emergence of the effective length-scale $\epsilon_{T\bar{T}} \propto \sqrt{\mu}$ that is predicted to replace the bare entanglement cut-off $\propto \epsilon=e^{-L/2}$ of the original CFT, which we keep and hold fixed in the deformed theory. We focused on the simpler case of one-interval subregion $A$, and computed the Renyi-entropy in the vacuum state $\Omega_\mu$ of the deformed theory. We considered both the half-line and the finite-interval cases, which are no longer conformally related in the deformed CFT. The computation was facilitated by the string theory formulation of the $T\bar{T}$-deformed CFT. The deformed replica-trick computation becomes a string partition function including an additional target-space CFT sector. 

By working with the appropriate world-sheet conformal frame, the string partition function takes the form of (\ref{eq:master}). It describes a weighted integral of $\ell$ over the CFT results, evaluated at the dynamical cut-off $\propto (\epsilon')^n = e^{-\ell/2}$ . The kernel function $K(L,\ell, \mu)$ of the integral relates the physical (target-space) and dynamical (world-sheet) cut-offs. It is obtained by computing the on-shell classical action of the target-space CFT, but the boundary condition is allowed to vary under boundary reparametrizations, which is then path-integrated. Working in the limit of small dimensionless coupling $\lambda = 2\pi \mu/R^2$, where $R$ is the target-space scale, one can perform a semi-classical computation of $K(L,\ell,\lambda)$ by finding the optimal reparametrization configuration and computing the loop corrections. For physical (target-space) cut-off boundaries that are non-geodesic, the loop-corrections are UV-divergent and thus render the computation RG-dependent. 

For the half-line case, we computed the semi-classical kernel function up to the one-loop correction. The result was used in evaluating the $\ell$-integral of the string partition function (\ref{eq:master}) via the saddle-point analysis. We found two saddle-points. One of them simply identifies the dynamical (world-sheet) cut-off $\ell^*$ with the physical (target-space) cut-off $L$, expanding about which produces the perturbation series about the CFT result. There is also a non-perturbative saddle-point $\ell^*_{n.p.}$ that coincides parametrically with the predicted $T\bar{T}$ cut-off $e^{-\ell^*_{n.p.}/2} \sim \sqrt{\mu} \sim \epsilon_{T\bar{T}}$. This saddle-point arises from a dynamical interplay between the kernel function and the CFT factor, which are of different orders in small $\lambda$ but balanced near the asymptotic limit $\ell \to \infty$. We computed the exponentially small non-perturbative corrections to the Renyi and von-Neumann entropies associated with the saddle-point $\ell^*_{n.p.}$. For the finite-interval case, due to the technical difficulties associated with finding the optimal reparametrization configuration exactly, our computation stopped at the leading-order for the kernel function. Using only the leading-order approximation of the kernel function, we performed the non-perturbative analysis and also found two saddle-points, one generating the perturbation series and the other suggesting a non-perturbative notion of the entanglement cut-off. In this case the identification between the non-perturbative saddle-point and the $T\bar{T}$ length-scale $\epsilon_{T\bar{T}}$ is less clear than in the half-line case.  

We conclude this paper by discussing a few open questions. For clarity we focus on the half-line results. As was commented in section (\ref{sec:saddle}), though we have identified the saddle-point $\ell^*_{n.p.}$ that is clearly related to the emergence of the effective $T\bar{T}$ length-scale as a candidate scale for the entanglement cut-off, a full understanding of how it dynamically replaces the bare entanglement cut-off $\epsilon$ is still far from clear based on the semi-classical computations in this paper. The effects from the saddle-point $\ell^*_{n.p.}$ is non-perturbatively small for $\lambda \ll \epsilon^2$. The $T\bar{T}$-deformation in this regime is a harmless perturbation on top of everything else, including the effective entanglement cut-off. Metaphorically, what we have found is the ``seed" of the transition that eventually takes place as $\lambda$ increases to $\lambda \gtrsim \epsilon^2$, lurking exponentially small in the semi-classical regime. We can examine what happens when extrapolating our analysis to the regime $\lambda \gtrsim \epsilon^2$. Firstly, as we have demonstrated before, the perturbative and non-perturbative saddle-points approach each other as $\lambda$ increases towards a critical value $\lambda \to \lambda_c\sim \epsilon^2$, across which both saddle-points first collide and then leave the integration contour $\ell \in [0,\infty)$ into the complex-plane. Secondly, in this process the curvatures of the integrand exponent at both saddle-points approach zero, implying that the magnitudes of the quantum fluctuations for $\ell$ about these saddle-points are diverging towards the collision, at which point the semi-classical approximation breaks down. 

What to make of the results beyond the collision point $\lambda_c$ is an interesting open question. We make some comments regarding the possibilities. In one scenario, the semi-classical approximation breaks down only near the collision point, and the results can still be approximated by the saddle-points, now analytically continued onto the complex plane, on the other side of but far away from $\lambda_c$. There are now two saddle-points being the complex conjugates of each other. The string partition function receives contributions from both of them. We can estimate what happens then by working with the following approximate integral representing the half-line case: 
\be
\text{Tr} \rho^n_A(L,\Omega_\lambda) \sim \int \frac{dx}{x\ln{x}} \exp{\left[\frac{4\pi n}{\lambda}\left(x-\frac{e^L}{2} x^2\right)-4\pi n \kappa_n \ln{x}\right]},\;\;\;x=e^{-\ell} 
\ee
The saddle-point solutions $x_{\pm}$ are given by: 
\be
x_{\pm} = \frac{e^{-L}}{2}\left(1\pm \sqrt{1-4\kappa_n e^{L}\lambda}\right)
\ee
Beyond the collision, i.e.~for $\lambda> \lambda_c = 1/(4\kappa_n e^{L})$, the saddle-points form a pair of complex-conjugates, so are their contributions to the string partition function. Including both and using the fact that $x_{\pm} = \sqrt{e^{-L}\kappa_n \lambda}\;e^{\pm i\tan^{-1}\sqrt{\lambda/\lambda_c -1}}$, we can obtain: 
\bea
\text{Tr} \rho^n_A\left(L,\Omega_\lambda\right) &\sim & e^{4\pi n \kappa_n\left(\frac{1}{2} + \frac{\lambda_c}{\lambda}\right)} \cos{\left[4\pi n \kappa_n\left(\frac{\lambda_c}{\lambda}\sqrt{\frac{\lambda}{\lambda_c}-1}+\tan^{-1}\sqrt{\frac{\lambda}{\lambda_c}-1}\right)\right]}\nonumber\\
&\times & \left(\sqrt{e^{-L}\kappa_n \lambda}\right)^{4\pi n \kappa_n}
\eea
where we should require $\lambda$ to be far above the critical value $\lambda_c$. This result can be unpacked as the CFT replica result evaluated at the cut-off radius $r = \left(e^{-L}\kappa_n \lambda\right)^{1/4} \sim \sqrt{\epsilon_{T\bar{T}} \cdot  \epsilon}$, multiplied by a weight factor that is oscillatory in $\lambda$. According to this scenario, there is only one effective entanglement cut-off for $\lambda>\lambda_c$, which is given by the geometric mean of the bare cut-off $\epsilon = e^{-L/2}$ and the $T\bar{T}$ length-scale $\epsilon_{T\bar{T}} \sim \sqrt{\lambda}$. It follows as a consequence that the bare entanglement cut-off $\epsilon$ does not decouple from the effective description even for $\lambda \gg \lambda_c$ in this scenario. This is odd from the IR perspective of the deformed theory. 

It therefore calls for another scenario that may be more realistic, in which the semi-classical approximation simply breaks down for all $\lambda \gtrsim \lambda_c$. This aligns with the suggestion that there is a phase transition at $\lambda \sim \lambda_c$. A distinct semi-classical framework, if exists, may emerge on the other side $\lambda>\lambda_c$. It is interesting to speculate about the nature of this phase-transition. In its most essential terms, the phase transition arises as a competition between two dimensionless scale parameters: the bare-entanglement cut-off $\epsilon$ and the $T\bar{T}$ length-scale $\epsilon_{T\bar{T}}$, both dimensionless. The physical dimensions can be restored by multiplying back the same target-space scale factor $R$. This setup is very reminiscent of the Hagedorn transition in standard string theories: a competition between the string scale $\sqrt{\alpha'}$ and the inverse temperature $\beta = 1/T$, which can also be viewed as a geometric scale in the world-sheet target space. The string partition function encounters an apparent singularity at the critical Hagedorn temperature $T_c \sim 1/\sqrt{\alpha'}$ due to the linear-in-energy growth of the entropy in perturbative string theory. Some descriptions of the physics beyond the critical temperature were proposed based on the condensation of string winding modes \cite{Sundborg:1984uk,Tye:1985jv,Atick:1988si}. In the perturbative regime they describe strings winding around the thermal circle. For temperatures higher than the critical value, the thermal circle becomes so small and the strings winding around it become so energetically ``cost-effective" that they tend to condense. It is therefore likely that the entanglement structure of the $T\bar{T}$-deformed theory will undergo a phase transition of the same universal class as the Hagedorn transition. In particular, fully understanding the other side $\lambda>\lambda_c$ of the transition may require one to consider the condensation of winding modes in the context of our string partition function (\ref{eq:master}). For example, one may need to relax the constraint on the mapping degree to include any integers (even the negative ones) for the path-integral of the target-space fields, and possibly re-summing the contributions in meaningful ways. Such transitions were anticipated and discussed in the context of torus partition functions \cite{Benjamin:2023nts,Dei:2024sct}. In the future it is interesting to study them in the context of entanglement structures, doing this may finally reveal the mystery of how the $T\bar{T}$ effective length-scale replaces the bare entanglement cut-off in the deformed results. 

In this paper, we have focused on a particular class of set-ups for computing the entanglement quantities: those of a single-interval sub-region in the deformed vacuum state. Making this choice facilitates our computations in many technical aspects. We make some comments about how the current framework can be extended to include more general set-ups: e.g.~multi interval sub-regions, deformed excited state, etc. 

We begin with the cases of multi interval sub-regions but still in the deformed vacuum state. We expect the following principles to remain true. Firstly we shall still take the target-space $\mathcal{M}=\tilde{\Sigma}_n$ to be the manifold obtained from the replica-trick in the un-deformed CFT, and regulated by e.g. removing small holes from each of the conical excess singularities at $\partial A$. Secondly, apart from some caveats to be discussed later, we still prefer to work in a world-sheet conformal frame $(w,\bar{w})$ whose conical excess properties match those of the target-space, in which case the one-loop determinant from the quantum fluctuations $\delta X^{\pm}$ cancel against the that from the $bc$ ghost sector. In $(w,\bar{w})$, the replica string partition function is still expected to take the general form of (\ref{eq:master}): 
\bea \label{eq:master_multi}
\text{Tr}\rho^n_A(\vec{L},\Omega_\mu)
&=&\int \mathcal{J} d\vec{\ell}\; K(\vec{L},\vec{\ell},\mu)\; \text{Tr}\rho^n_A(\vec{\ell},\Omega_{CFT})\nonumber\\
K(\vec{L},\vec{\ell},\mu) &=& \prod^{n}_{i=1}\left(\int\mathcal{D}\theta^i_1 \int \mathcal{D} \theta^i_2 \right) \sum_{X^{\pm}_{cl}(\vec{\theta}_1,\vec{\theta}_2)} e^{-S_{\tilde{\Sigma}_n}\left(X^{\pm}_{cl}(\theta_1,\theta_2),\mu,\tilde{\Sigma}'_n\right)}
\eea
where $\vec{\ell}$ includes all the conformal moduli of the world-sheet $\mathcal{M}'=\tilde{\Sigma}'_n$ with integration measure $\mathcal{J}$, while $\vec{L}$ includes the moduli parameters of the target-space $\mathcal{M}=\tilde{\Sigma}_n$. The multi-interval extension (\ref{eq:master_multi}) also describes a weighted integral over the multi-interval CFT results, whose conformal invariants of the dynamical cut-off scales are being integrated over. Analogous to the single-interval case, the kernel function $K(\vec{L},\vec{\ell},\mu)$ is computed by solving the classical on-shell action but with the boundary reparametrization modes (now there are $2n$ of them) path-integrated over. Unlike the single-interval case, the conical-excess index $n$ of the replica-trick also determines the topology of the manifold, e.g.~the manifold $\Sigma_n$ for the two-interval case is a genus $(n-1)$ surface. So regarding the properties of the world-sheet and target-space manifolds, the matching of their conical excess properties is related to the matching of their topological properties. Related to this, there is a caveat if the un-deformed CFT is non-critical, i.e. $\kappa =c-24\neq 0$. In this case the target-space CFT action contains the additional terms: 
\be 
S_{\mathcal{M}}(X^{\pm},\mu) = \left(\text{free action}\right)-\frac{\kappa}{8\pi\mu}\int_{\mathcal{M}'} d^2z R \log{\left(\partial X^+ \bar{\partial} X^-\right)} + \int_{\partial\mathcal{M}'} ds\; K \times \left(...\right)
\ee
where we do not bother with the explicit form of the boundary term except for its proportionality to $K$. In this paper, we neglected these terms because for the single-interval sub-region of our interest, the world-sheet manifold $\mathcal{M}'=\tilde{\Sigma}'_n$ is topologically an annulus with the Euler characteristics $\chi(\mathcal{M}') =2-2g-b=0$, where $b$ is the number of boundaries on $\mathcal{M}'$. According to the Gauss-Bonnet theorem: 
\be
\int_{\mathcal{M}'} R\; dA + \int_{\partial \mathcal{M}'} K\; ds = 2\pi \chi(\mathcal{M}')=0
\ee
it is consistent to find a gauge where $R=K =0$  globally and thereby neglecting the non-critical terms. For the multi-interval case, this is no longer the case. For example, for the two-interval case the corresponding Euler characteristics of $\mathcal{M'}$ is given by $\chi(\mathcal{M}') = 2-2(n-1)-4=-2n$. As a result one can no longer neglect these terms completely by setting both $R$ and $K$ to zero globally. For this reason the multi-interval extension (\ref{eq:master_multi}) may only be valid for the critical case of $c=24$, thus leaving the full treatment for the non-critical cases an open question. 

We can also consider extending the global states to include non-vacuum states. To identify the states across the original CFT and the deformed theory, we can consider states that are defined by Euclidean path-integrals in the original CFT. This includes not only the Hartle-Hawking vacuum state, but a large class of additional states such as the thermal state (represented by the Euclidean path-integral on Euclidean manifolds with a thermal circle); or primary states on a circle (represented via the state-operator correspondence by the Euclidean path-integral with local primary operators inserted asymptotically). The crucial task is to lift the Euclidean path-integrals that define the original states into a world-sheet CFT description. Take the example of the thermal state, the corresponding string partition function (\ref{eq:master}) is then formulated with the target-space CFT describing the Euclidean manifold with the thermal circle, copied and glued appropriately for the replica trick. In the future, it will be interesting to extend our analysis to include the more general set-ups just described \cite{Lai:2025xxx}. 

We end the paper by suggesting a few more things to explore in the future. Firstly, it is interesting to combine our analysis with other semi-classical limits, most notably the holographic limit of $c\to \infty$, for which the CFT results by themselves admit explicit bulk dual descriptions. In these cases the order of limits is very important, and usually the useful limit take the form of e.g. $c\to \infty, \lambda \to 0$ but keeping $(\lambda c)$ fixed, which as we have seen is the parameter entering the non-perturbative saddle-point $\ell^*_{n.p.}$ in the large $c$ limit. Secondly, the non-perturbative results we have obtained in this paper are in principle verifiable in the perturbative expansion, which can be computed in alternative ways independent of the string theory formulation. Such a connection between the non-perturbative effects and perturbation theory occurs through the resurgence analysis techniques -- if one can compute the perturbative coefficients to sufficiently large order. This can be done for the computation of the $T\bar{T}$-deformed torus partition function \cite{Gu:2024ogh} thanks to the remarkable algebraic structure of the set-up. The analogous computation for the entanglement quantities in this paper is expected to be much more difficult. Finding ways to do it, e.g.~using conformal perturbation theory, constitutes an important check not only for our non-perturbative analysis, but also for the string theory formulation as a whole.  Thirdly, in our paper we have chosen a particular way to regulate the conical excess singularities arising in performing the replica trick in the original CFT. It takes the form of a geometric cut-off with small holes removed about the singularities. This is a natural choice but not the unique one. Alternatively, one could also regulate the singularity by working with a (fixed) smoothly resolved geometry -- not as a result of the dynamical stringy back-reaction, but instead is put in by hand in the original CFT. In the future, it is also an interesting exercise to incorporate this treatment in the $T\bar{T}$-deformed theory -- the target-space manifold is now the resolved geometry which presumably is also characterized by an effective resolution length-scale, and study how it interacts with the $T\bar{T}$ deformations. Fourthly, as we have seen in section (\ref{sec:infinite}), the target-space CFT implementing the replica trick for $n>1$ is off-shell, this is the case even with the conical excess singularities removed as we did. Though for the purpose of computing the von-Neumann entropy at $n\to 1$ one may avoid dealing with the off-shell aspect of the replica trick, a proper and honest treatment of the finite $n$ Renyi entropy ultimately requires an off-shell formalism of string theory \cite{Susskind:1994sm,Rychkov:2002ni}, see recent developments in \cite{Ahmadain:2022tew,Ahmadain:2022eso}. This is a direction where further developments will make important contributions to our understanding of $T\bar{T}$-deformed entanglements, or stringy entanglements in general.  Last but not least, while in this paper we have focused on the ``stringy" sign $\mu>0$ of the $T\bar{T}$-deformation, an important follow up question is to extend the current analysis to the other sign of $\mu<0$, possibly via analytically continuing $\mu$ in the complex-plane and keeping track of the Stoke's phenomenon, as has been done for the $T\bar{T}$-deformed torus partition function in  \cite{Gu:2025tpy}. We leave this to future investigations. 

\section*{Acknowledgement}
We thank Amr Ahmadain, Luis Apolo, Pawel Caputa, Ricardo Esp\'indola, Xia Gu, Peng-Xiang Hao, Ling-Yan Hung, Hong Liu and Wei Song for helpful discussions. This work is supported by National Science Foundation of China (NSFC) grant No.\,12175238 and NSFC grant No.\,12447108.

HJW thanks the Peng Huanwu Center for Fundamental Theory, Northwestern University, Xi'an, for the hospitality during the 6th Workshop on Strings and Fields; WXL thanks the International Centre for Mathematical Sciences (ICMS) and the University of Edinburgh for the hospitality during the ``China-India-UK school in mathematical physics'', as well as the University of Tokyo for the hospitality during the ``Tokyo Holography 2025'' workshop, where part of this work was completed.

\appendix
\section{Cancellation between the one-loop determinants from \texorpdfstring{$\delta X^{\pm}$}{δX} and the \texorpdfstring{$bc$}{bc} ghosts}\label{app:cancellation}
In this appendix, we present some details regarding the computation of the total one-loop determinant from $\delta X^{\pm}$ and the ghost sector: (\ref{eq:cancellation}): 
\be\label{eq:cancellation_2}
\left[\int \mathcal{D}\delta X^{\pm}\;e^{-\frac{1}{2\pi\mu} \int d^2 w\; \delta X^+ \left(\partial_w\partial_{\bar{w}}\right) \delta X^-} \right]
\times \left[\int \mathcal{D}b\mathcal{D}c\; e^{-S^{w,\bar{w}}_{gh}(b,c,\epsilon')}\right] 
\ee
In the $(w,\bar{w})$ frame, the two modes of the target-space fluctuations $\delta X^{\pm}$ both satisfying the Dirichlet boundary condition on the boundaries of the annulus give the one-loop determinant:
\be\label{eq:target_det}
\left[\int \mathcal{D}\delta X^{\pm}\;e^{-\frac{1}{2\pi\mu} \int d^2 w\; \delta X^+ \left(\partial_w\partial_{\bar{w}}\right) \delta X^-} \right] = \frac{1}{\text{det} \left(\nabla^2_{w,\bar{w}}\right)_D}
\ee
where $\left(\nabla^2_{w,\bar{w}}\right)_D$ denotes the Laplacian operator on the annulus with Dirichlet boundary conditions on both boundaries; and $\left(\partial_w \partial_{\bar{w}}\right)_N$ will be similarly defined. The ghost action takes the standard form: 
\be
S_{gh}(b,c) = \int d^2w\; (b \hat{P}_1 c) 
\ee
where $\hat{P}_1$ is the standard differential operator that maps from spin one to (symmetric and trace-less) spin two tensors. The one-loop determinant from the ghost sector is given by the standard formula: 
\be\label{eq:ghost_det_1}
\left[\int \mathcal{D}b\mathcal{D}c\; e^{-S^{w,\bar{w}}_{gh}(b,c,\epsilon')}\right] = \sqrt{\text{det}' \hat{P}^{\dagger}_1 \hat{P}_1}
\ee
where $'$ denotes the removal of zero-modes -- they are treated separately in the path-integral. The operator $\hat{P}^{\dagger}_1 \hat{P}_1$ acts on the vector ghost field $c$, and so in this case depends on its boundary condition in (\ref{eq:ghost_bc}). We can unpack the boundary condition into more standard forms defined on the boundary $\tau=0$ of the complex coordinates $(w=\tau+ix, \bar{w}=\tau-ix)$: 
\bea
&& c_n + \bar{c}_{-n} = 0 \to  \sum_{n} c_n e^{inx} + \bar{c}_{-n} e^{inx} = c(z)+\bar{c}(\bar{z}) \propto c^\tau(z,\bar{z}) = 0 \nonumber\\
&\to & \sum_{n} n c_n e^{inx} + n\bar{c}_{-n} e^{inx} = \partial_\tau c(z)- \partial_\tau\bar{c}(\bar{z}) \propto \partial_\tau c^x(z,\bar{z}) = 0
\eea
In other words, we impose the Dirichlet boundary condition on the component $c^\tau$ while imposing the Neumman boundary condition on the component $c^x$. Based on this, the ghost determinant (\ref{eq:ghost_det_1}) can be expressed explicitly as: 
\be \label{eq:ghost_det_2}
\left[\int \mathcal{D}b\mathcal{D}c\; e^{-S^{w,\bar{w}}_{gh}(b,c,\epsilon')}\right] = \sqrt{\text{det}' \left(\nabla^2_{w,\bar{w}}\right)_{N}\text{det} \left(\nabla^2_{w,\bar{w}}\right)_{D}}
\ee
Note that $\left(\nabla^2_{w,\bar{w}}\right)_D$ does not have zero modes, so correspondingly we have written $\text{det}' \equiv \text{det}$. Combining (\ref{eq:target_det}) and (\ref{eq:ghost_det_2}), the total one-loop determinant (\ref{eq:cancellation_2}) becomes: 
\be
\left(\text{total one-loop determinant}\right) = \left[\frac{\text{det}' \left(\nabla^2_{w,\bar{w}}\right)_{N}}{\text{det} \left(\nabla^2_{w,\bar{w}}\right)_{D}}\right]^{1/2}
\ee
The Laplacian determinants on the annulus with Dirichlet and Neumann boundary conditions can be computed using the Eisenstein series, see for example \cite{OSGOOD1988148}. The (non-zero) spectrum $\lambda^{N,D}_{mn}$ of $(\nabla^2_{w,\bar{w}})_{N,D}$ are given by:
\bea\label{eq:laplace_eigenvalues}
\lambda^N_{mn} &=& \frac{\pi^2}{\tau^2}\left(4\tau^2m^2 +n^2\right),\;\;m,n \in \mathds{Z},\;\;(m,n)\neq (0,0)\nonumber\\
\lambda^D_{mn} &=& \frac{\pi^2}{\tau^2}\left(4\tau^2m^2 +n^2\right),\;\;m,n \in \mathds{Z},\;\;n>0,\;m\geq 0
\eea
where $\tau = -\frac{1}{\pi}\ln{\epsilon'}$. Using the zeta regularization, the determinants are then evaluated to take the form: 
\be 
\text{det}' \left(\nabla^2_{w,\bar{w}}\right)_N = \text{det} \left(\nabla^2_{w,\bar{w}}\right)_D \prod_{m \neq 0}(4\pi^2 m^2) = 2\tau |\eta(2i\tau)|^2 
\ee
The proportional factor is divergent and can be evaluated to unity in the zeta-function regularization:  \be 
\left[\frac{\text{det}' \left(\nabla^2_{w,\bar{w}}\right)_{N}}{\text{det}\left(\nabla^2_{w,\bar{w}}\right)_{D}}\right]^{1/2}=\prod_{m \geq 1} (4\pi^2 m^2) \to 1
\ee
This can be checked via the following manipulations involving the Riemann-zeta function $\zeta_R(s)=\sum_{m\geq 1} m^{-s}$: 
\bea 
&&\prod_{m \geq 1} (4\pi^2 m^2) = \exp{\left(-\zeta'_\Delta(0)\right)},\quad\zeta_{\Delta}(s)=\sum_{m\geq 1}(2\pi m)^{-2s} = (2\pi)^{-2s} \zeta_R(2s)\nonumber\\[.5ex]
&& \zeta'_{\Delta}(0) = 2\zeta_R'(0)-2 \ln{(2\pi)} \zeta_R(0) =0
\eea
The last equality follows from the (analytically continued) evaluations at $s=0$ regarding the Riemann-zeta function: 
\be 
\zeta_R(0)=-\frac{1}{2},\quad\zeta'_R(0) = -\frac{1}{2}\ln{(2\pi)}
\ee
These results combined thus verify the cancellation (\ref{eq:cancellation}). We remark that despite the particular choice of the zeta-function regularization, the essence of the cancellation works in a regularization-independent way.  
To see this, we note that the proportional factor comes from the eigenvalues in $\lbrace\lambda^{N}_{mn}\rbrace$ that are in addition to those of $\lbrace\lambda^D_{mn}\rbrace$, which consist of those with $(n=0, m\geq 1)$ in (\ref{eq:laplace_eigenvalues}). A key regularization-independent property of this proportional factor is that it is a constant, i.e.~not a function of the moduli $\epsilon'$ or $\tau$. Therefore the choice of regularization only affects the particular value of this constant, and does not alter any of the subsequent analysis. 

\section{Asymptotic approximations of \texorpdfstring{$F_0(\ell)$}{F0(l)}}\label{app:F0_limits}
In this section of appendix, we provide some details for computing $F_0(\ell)$ approximately in the two parametric limits of $\ell$. We discuss them separately. 

\subsubsection*{``low temperature" limit $\ell \gg 1$}
In the ``low temperature" limit of $\ell \gg 1$, the determinants $\text{det} D^{ij}_k$ exhibit a sharp transition as $k$ changes between $k<n$ and $k>n$. For $k<n$, the determinant $\text{det} D^{ij}_k$ decays exponentially with $\ell$:
\bea\label{eq:large_l_1}
\det D^{ij}_k &=& -4(n-k)\, e^{-2(n-k)\ell/n}\left((n-k)-2n e^{-(\ell-L)}\right) + \cdots \nonumber\\
&=& \begin{cases}
-4(n-k)^2\, e^{-2(n-k)\ell/n},\;\;\;\;\;\;\;\;\;\;\ell \gg L \\
4(n^2-k^2)\, e^{-2(n-k)L/n},\;\;\;\;\;\;\;\;\;\;\;\ell \approx L\\
8n(n-k)\,e^{L-(3n-2k)\ell/n},\;\;\;\;\;1\ll\ell\ll L
\end{cases}
\eea
In the case of integer $n$, there is the special case for $k=n$, at which the determinant $\text{det} D^{ij}_k$ is equal to: 
\be\label{eq:large_l_2}
\text{det} D^{ij}_{n} = \frac{4n^2}{\ell} e^{-(\ell-L)} +...
\ee
For $k>n$,  the determinant $\text{det} D^{ij}_k$ is given by: 
\be\label{eq:large_l_3}
\text{det} D^{ij}_k = 4k^2- 4k\Delta S + \frac{4n^2\left(\cosh^2{(\ell)}-2\cosh{(\ell)}\cosh{(L)}+1\right)}{\sinh^2{(\ell)}} +... 
\ee
which agrees quite well with the large $k$ approximation (\ref{eq:large_k_det}). The ... in equation (\ref{eq:large_l_1}), (\ref{eq:large_l_2}) and (\ref{eq:large_l_3}) represent terms that are further exponentially suppressed with large $\ell$.
In this limit, the threshold $k_c$ is unambiguously given by $k=\lfloor n\rfloor$, where $\lfloor n \rfloor$ denotes the integer floor of $n$. We therefore obtain that:
\bea\label{eq:highT_F_1}
&&F_0(\ell) = \frac{\lambda}{2}\ln{\left[\frac{\lambda\, e^{-\ell}}{\pi^2n}\right]}+ \frac{\lambda}{2}\,\delta_{n,\lfloor n\rfloor}\ln{\left(\frac{8n^2}{\ell} e^{-(\ell-L)}\right)}+\left(\delta F_0\right)_{finite}\nonumber\\            
&+&\frac{\lambda}{2}\sum_{k=1}^{\lfloor n\rfloor}\;
        \begin{cases}
        \ln{\left(4(n-k)^2 e^{-2(n-k)\ell/n}\right)}+i\pi,\;\;\;\;\;\;\ell \gg L \\
        \ln{\left(4(n^2-k^2) e^{-2(n-k)\ell/n}\right)},\,\;\;\;\;\;\;\;\;\;\;\;\;\ell \approx L\\
        \ln{\left(8n(n-k)e^{L-(3n-2k)\ell/n}\right)},\;\;\;\;\;\;1\ll\ell\ll L
        \end{cases} +...
\eea

At this point we need to discuss the treatment for the analytic continuation of $n$ away from their defining integer values in the replica trick. One option is to treat $n$ already as a real number in (\ref{eq:highT_F_1}), in which case it jumps as a function of $n$ due to its dependence on $\lfloor n \rfloor$; the other option is to first restrict $n$ to integer values in (\ref{eq:highT_F_1}), and then treat $n$ in the summed-result as a real number. We will follow the second option. Performing the sum gives:  
\bea\label{eq:highT_F_2}
&& F_0(\ell) = \frac{\lambda}{2}\ln{\left(\frac{8\lambda n}{\pi^2 \ell}\right)}- \frac{\lambda}{2}\left(2\ell-L\right) + \left(\delta F_0\right)_{finite}\nonumber\\
&+& \frac{\lambda}{2}
\begin{cases}
\ln{\left(4^{n-1}\Gamma(n)^2\right)}-(n-1)\ell + (n-1)i\pi,\;\;\;\;\;\;\ell \gg L\\
\ln{\left(4^{n-1}\Gamma(2n)/n\right)} - (n-1)\ell,\;\;\;\;\;\;\;\;\;\;\;\;\;\;\;\;\;\;\;\;\;\ell \approx L\\
\ln{\left((8n)^{n-1}\Gamma(n)\right)}-(n-1)(2\ell-L),\;\;\;\;\;\;1\ll \ell \ll L
\end{cases}+ ...
\label{eq:large_ell_F0}
\eea

The alert readers may notice the imaginary part $\propto (n-1)i\pi$ that shows up in the $\ell \gg L$ limit. An imaginary part in the quantum correction to the free energy indicates that there are fluctuation modes $\delta \theta_{1,2}$ about (\ref{eq:half_line_saddle}) with negative eigenvalues, and thus the instability of the classical background. This can be deduced from the expression of the $\text{det} D^{ij}_k<0$ in (\ref{eq:large_l_1}) for the limit $\ell\gg L$. It implies that for each of the $k<n$, one of the two eigenmodes has negative eigenvalue. We view the presence of negative modes for $n>1$ as reflecting the original off-shell nature of the replica background. Though we have apparently removed the conical-excess singularities by the cut-off holes, the problem did not disappear and instead found its way back as the instabilities to the fluctuating boundary reparametrizations $\theta_{1,2}$. Our attitude towards this issue is first to acknowledge that the semi-classical approach to the string partition function (\ref{eq:master}) may not be valid for general $n$, most likely not for large $n$. On the other hand, for the purpose of computing the von-Neumann entropy $S^{vN}_A = \lim_{n\to 1}S^n_A$, we shall confine ourselves to values of $n$ that are very close to $1$. In this regime, we do not expect the instability to cause problems that are as severe. For one thing, the imaginary part vanishes in the limit $n\to 1$. In addition, in taking the limit of $n\to 1$, only the $k=1$ sector may contain a negative-eigenvalue mode. For the sake of argument we have now allowed the $n\to 1$ limit to be taken at the intermediate steps. We can then further analyze for what values of $\ell$ the instability occurs. This is captured by the expression in the first line of (\ref{eq:large_l_1}) at $k=1$:
\bea
\text{det} D^{ij}_1 \approx -4(n-1)e^{-2(n-1)\ell/n}\left(n-1-2n e^{-(\ell-L)}\right)<0 
\eea
which requires $\ell$ to satisfy:
\be\label{eq:ell_unstable}
\ell > L+ \ln{\left(\frac{2n}{n-1}\right)}\; \xlongrightarrow[\ n\to 1\ ]\,\ L+\infty
\ee
We see that for the integral over $\ell$ in the string partition function (\ref{eq:master}), the instability only occurs in a region that keeps shrinking towards $\ell =\infty$ in the limit of $n\to 1$. As a result, it does not affect non-perturbative effects of the string partition function (\ref{eq:master}) that are associated with finite values of $\ell$ in the $n\to 1$ limit. Based on these, we neglect the imaginary part of (\ref{eq:highT_F_2}) and the related issue of instabilities in the subsequent analysis of the main text. We make some remarks about the instability at finite $n$. At the level of the full $\ell$-integral, the negative eigenvalues are present for values of $\ell$ satisfying (\ref{eq:ell_unstable}). In a full off-shell formalism that is beyond the scope of this work, the stringy back-reactions are expected to generate corrections (possibly in the form of counter-terms) to the target-space action and stabilize the fluctuations. At the level of the semi-classical analysis we are performing, the non-perturbative saddle-point (\ref{eq:new_saddle}) is stable against boundary-reparametrization fluctuations if it occurs well inside the instability-free regime: 
\be
\ell^*_{n.p.} = -\ln{(\lambda \kappa_n)}\lesssim L + \ln{\left(\frac{2n}{n-1}\right)}
\ee
This then sets a domain of validity for the Renyi index $n$ below which our semi-classical analysis holds. For large $c$ and small $\lambda$ this domain can be explicitly stated as:
\be
n \lesssim 1+\frac{96\pi \lambda}{c \epsilon^2}+ ...
\ee
This constraint can also be inverted to give the domain of validity for the coupling $\lambda$ at finite values of $n$. From this we see that in the semi-classical limit $\lambda \ll \epsilon^2$, the non-perturbative saddle-point is indeed controlled only in the limit $n\to 1$. 

We can now easily fix the cosmological constant counter-term $\delta F_0$ by using the $\ell = L$ limit of (\ref{eq:highT_F_2}) in conjunction with the renormalization condition (\ref{eq:ren_cond_3}):
\bea
&&\left(\delta F_0\right)_{finite}= \frac{\lambda (n+1)}{2}L + \text{(const)}\nonumber\\
&&\text{(const)}=-\frac{\lambda}{2}\ln{\left( 4^{n+2}\Gamma(2n)L/n\right)}+... 
\eea
In subsequent analysis we will disregard the constant terms that are sub-leading in the various limits we are taking. 

\subsubsection*{``high temperature" limit $\ell\ll 1$}
As it turns out, this regime does not play an important part in the non-perturbative analysis. For completeness we very briefly discuss this case. The $\ell \ll 1$ limit does not feature a sharp threshold $k_c$, instead there exists a parametric threshold $k_c \sim n\ell^{-1}$ such that the determinant $\text{det} D^{ij}_k$ behaves qualitative different across $k_c$ parametrically: 
\begin{itemize}
\item For $k\gg k_c$, it is well approximated by the large $k$ expression (\ref{eq:large_k_det}): 
\be
\text{det} D^{ij}_k = 4k^2\left(1 - \frac{\Delta S}{k}+...\right)
\ee
\item For $k\ll k_c$, it instead admits a small $k\ell$ expansion of the form: 
\be
\text{det} D^{ij}_k = \frac{4}{3}k^2 e^{L}\left(1-\frac{1}{15n^2} k^2\ell^2+\frac{2}{315 n^4}k^4\ell^4+...\right) 
\ee
\end{itemize}
The factor of $4k^2$ is present in both the $k\ll k_c$ and $k\gg k_c$ expressions, so it does not produce $\ell$-dependent terms upon taking the log determinant. We will ignore this factor. By the previous procedure, the quantum corrections to the cosmological constant in the case of $\ell \ll 1$ are given by:
\bea
F_0(\ell)&-&\left(\delta F_0\right)_{finite} \sim  \frac{\lambda}{2} \sum^{k_c}_{k=1} \ln{\left(\frac{1}{3}e^{L}\left(1-\frac{1}{15n^2}k^2 \ell^2 + \frac{2}{315n^4}k^4\ell^4+...\right)\right)}\nonumber\\
&\approx & \frac{\lambda nL}{2\ell} -\frac{n\lambda\ln{(3)}}{2\ell}+ \frac{\lambda}{2} \int^{n\ell^{-1}}_1 dk \left(-\frac{1}{15n^2}k^2\ell^2+\frac{13}{3150n^4}k^4\ell^4+...\right)\nonumber\\
&\approx & \frac{\lambda nL}{2\ell} - n\lambda\left(\frac{\text{const}}{\ell}\right),\;\;\;\text{const}= \frac{\ln{(3)}}{2}+\frac{1}{90}-\frac{13}{31500}+...
\eea
As was done before, we disregard the $\text{const}/\ell$ term that is sub-leading in large $L$.

\section{Approximate solutions to (\ref{eq:saddle_finite_2})}\label{app:saddle_approx}
In this section of appendix, we solve the saddle-point equation (\ref{eq:saddle_finite_2}) approximately in a few parametric limits of $\epsilon'$ in addition to $\epsilon \ll 1$. For reference, we reproduce the saddle-point equation below: 

\begin{eqnarray}\label{eq:EoM}
    e^{2\im n\theta(\sigma)} &=& \frac{p\sum\limits_{j=1}^\infty\frac{(-1)^j (nj)}{\sinh{(j\ell)}}\left(e^{-j\left(\ell/2+in\sigma\right)}+e^{j\left(\ell/2+in\sigma\right)}\right)-\sum\limits_{k\neq 0}\frac{(b-a)\epsilon k e^{ik\sigma} }{\sinh{(k\ell/n)}}\left(\alpha_{-k}^*+e^{k\ell/n}\alpha_{k}\right)}{p\sum\limits_{j=1}^\infty\frac{(-1)^j (nj)}{\sinh{(j\ell)}}\left(e^{-j\left(\ell/2-in\sigma\right)}+e^{j\left(\ell/2-in\sigma\right)}\right)-\sum\limits_{k\neq 0}\frac{(b-a)\epsilon k e^{-ik\sigma} }{\sinh{(k\ell/n)}}\left(\alpha_{-k}+e^{k\ell/n}\alpha_{k}^*\right)}\nonumber\\
    \alpha_{k}&=&\frac{1}{2\pi}\int^{2\pi}_0 d\sigma e^{in\theta(\sigma)} e^{-ik \sigma},\quad p=(a-b)\left(1-\epsilon\alpha_0-\epsilon\alpha_0^*\right). 
\end{eqnarray}
We can analyze the solution by considering the following limits. 
\begin{itemize}
\item For $\epsilon'=e^{-\ell/2n}\sim 1$, at the leading order in small $\epsilon$, we can neglect in (\ref{eq:EoM}) the non-linear terms (i.e. those proportional to the Fourier modes $\alpha_k$ of the solution) in both the numerator and the denominator. The equation then becomes linear and we can directly write down the solution at the leading order:
\begin{equation*}
    \begin{aligned}
        e^{2\im n\theta(\sigma)}\approx\frac{\sum\limits_{j=1}^\infty\frac{(-1)^j(nj)}{\sinh{(j\ell)}}(e^{-j(\ell/2+\im n\sigma)}+e^{j(\ell/2+\im n\sigma)})}{\sum\limits_{j=1}^\infty\frac{(-1)^j(nj)}{\sinh{(j\ell)}}(e^{-j(\ell/2-\im n\sigma)}+e^{j(\ell/2-\im n\sigma)})}\approx\frac{\sum\limits_{j=1}^\infty\frac{(-1)^jn}{l}(e^{-\im nj\sigma}+e^{\im nj\sigma})}{\sum\limits_{j=1}^\infty\frac{(-1)^jn}{l}(e^{\im nj\sigma}+e^{-\im nj\sigma})}=1.
    \end{aligned}
\end{equation*}
where in the second approximation we have taken the leading order in small $\ell$. This implies that at leading order, the solution simply takes the form: 
\be
\theta(\sigma)=0
\ee
We emphasize that even though this leading order solution appears to disobey the winding number constraint -- it is a constant, the full solution after including higher order corrections is indeed a winding number one configuration. In other words, the full solution exhibits a concentration of measure at $\sigma=0$ that becomes sharper in the $\epsilon,\ell \to 0$ limit.  

\item For $\epsilon\ll{\epsilon'}^n\ll 1$, we can still neglect the non-linear part of the equation, but this time take the large $\ell$ limit:
\begin{equation}
    \begin{aligned}
        e^{2\im n\theta(\sigma)}\approx \frac{\sum\limits_{j=1}^\infty\frac{(-1)^j(nj)}{\sinh{(j\ell)}}(e^{-j(\ell/2+\im n\sigma)}+e^{j(\ell/2+\im n\sigma)})}{\sum\limits_{j=1}^\infty\frac{(-1)^j(nj)}{\sinh{(j\ell)}}(e^{-j(\ell/2-\im n\sigma)}+e^{j(\ell/2-\im n\sigma)})}& \approx \frac{npe^{-\ell/2}e^{\im n\sigma}}{npe^{-\ell/2}e^{-\im n\sigma}}=e^{2\im n\sigma}
    \end{aligned}
\end{equation}
Thus, in this regime we have $\theta(\sigma)=\sigma$ at the leading order.

\item For ${\epsilon'}^n\ll\epsilon\ll 1$, we can instead neglect in (\ref{eq:EoM}) the ``inhomogeneous" terms, i.e. those proportional to $p$ in both the numerator and denominator. The equation then takes the following simplified yet still non-linear form: 
\begin{equation}
    e^{2\im n\theta(\sigma)}\approx\frac{\sum\limits_{k>0}ke^{\im k\sigma}\alpha_{k}}{\sum\limits_{k>0}ke^{-\im k\sigma}\alpha_{k}^*},\;\;\;\alpha_k = \frac{1}{2\pi}\int^{2\pi}_0 d\sigma e^{in\theta(\sigma)}e^{-k\sigma}
\end{equation}
In principle this equation needs to be solved self-consistently. However, one can easily observe that it can be solved self-consistently by the following solution: 
\be
e^{in\theta(\sigma)} = e^{in\sigma},\;\;\;\alpha_k = \delta_{nk}
\ee
In other words, in this regime the solution still takes the form $\theta(\sigma)=\sigma$ at the leading order. 

\item For ${\epsilon'}^n\sim\epsilon\ll 1$, we can no longer neglect in (\ref{eq:EoM}) either type of terms completely as we did before. Nonetheless, since we know the solution takes the form $\theta(\sigma)=\sigma$ at the leading order on both sides of the parametric relation between ${\epsilon'}^n$ and $\epsilon$, it is only natural to suspect it remains so at ${\epsilon'}^n\sim \epsilon$. Indeed, by taking the leading order contributions from both types of terms in (\ref{eq:EoM}), we can check that it is solved by $\theta(\sigma)=\sigma$:    
\begin{equation}
    e^{2\im n\theta(\sigma)}\left(npe^{-\im n\sigma}+(b-a)\sum\limits_{k>0}ke^{-\im k\sigma}\alpha_{k}^*\right)=\left(npe^{\im n\sigma}+(b-a)\sum\limits_{k>0}ke^{\im k\sigma}\alpha_{k}\right)
\end{equation}
\end{itemize}
In summary, for $\theta(\sigma)$ satisfying the equations of motion, its leading-order behaviors in the various regimes are given by:
\begin{equation}
    \theta(\sigma)=
    \begin{cases}
    0,\;\;\;\;\;\;\;\; {\epsilon'}^n\sim1\\
    \sigma,\;\;\;\;\;\;\;\; {\epsilon'}^n\ll 1\\
    \end{cases}
\end{equation}

\hfill
\bibliographystyle{JHEP}
\bibliography{refs.bib}

\end{document}